\documentclass[%
 reprint,
 amsmath,amssymb,
 aps,
pra,
]{revtex4-1}

\usepackage{graphicx}
\usepackage{dcolumn}
\usepackage{bm}
\usepackage{comment}
\usepackage{notoccite}
\usepackage[usenames]{color}
\usepackage{soul}

\begin{document}

\title{Crystalline clusters in mW water:\\stability, growth, and grain boundaries}

\author{Fabio Leoni}
\email{fabio.leoni@bristol.ac.uk}
\affiliation{%
School of Mathematics, University of Bristol, Bristol
BS8 1TW, United Kingdom}

 \author{Rui Shi}
 \email{rshi@iis.u-tokyo.ac.jp}
 \affiliation{Department of Fundamental Engineering, Institute of Industrial Science, \\
 University of Tokyo, 4-6-1 Komaba, Meguro-ku, Tokyo 153-8505, Japan}
 
 \author{Hajime Tanaka}
 \email{tanaka@iis.u-tokyo.ac.jp}
 \affiliation{Department of Fundamental Engineering, Institute of Industrial Science, \\
 University of Tokyo, 4-6-1 Komaba, Meguro-ku, Tokyo 153-8505, Japan}

\author{John Russo}
\email{john.russo@bristol.ac.uk}
\affiliation{%
School of Mathematics, University of Bristol, Bristol
BS8 1TW, United Kingdom}
\affiliation{Department of Fundamental Engineering, Institute of Industrial Science, \\
	University of Tokyo, 4-6-1 Komaba, Meguro-ku, Tokyo 153-8505, Japan}

\date{\today}

\begin{abstract}
With numerical simulations of the mW model of water, we investigate the energetic stability of crystalline clusters for both Ice~$I$ (cubic and hexagonal ice) and for the metastable Ice~$0$ phase as a function of the cluster size. Under a large variety of forming conditions, we find that the most stable cluster changes as a function of size: at small sizes the Ice~$0$ phase produces the most stable clusters, while at large sizes there is a crossover to Ice~$I$ clusters. 
We further investigate the growth of crystalline clusters with the seeding technique and study the growth patterns of different crystalline clusters. While energetically stable at small sizes, the growth of metastable phases (cubic and Ice~$0$) is hindered by the formation of \emph{coherent} grain boundaries. A five-fold symmetric twin boundary for cubic ice, and a newly discovered coherent grain boundary in Ice~$0$, that promotes cross nucleation of cubic ice. Our work reveals that different local structures can compete with the stable phase in mW water, and that the low energy cost of particular grain boundaries might play an important role in polymorph selection.
\end{abstract}

\maketitle

\section{Introduction}

The pioneering work of Sir Charles Frank~\cite{frank1952supercooling} revealed that the lowest energy local structure can differ from the arrangement found in the stable crystalline phases. In the Lennard-Jones fluid, for example, Frank found that the 13-atom icosahedral cluster has an energy which is $8.4\%$ lower than 13-atom clusters corresponding to nearest-neighbour environments in close-packed crystals (both the fcc and hcp crystalline phases). The idea of competing local structures has had a profound influence on our understanding of supercooled liquid states and of their non-equilibrium extension, i.e. glasses~\cite{tanaka_review,tanaka2012bond,shintani2006frustration,leocmach2012roles,royall2015role,royall2018race,russo2018glass}.

The study of locally favoured structures has also played an important
role in the understanding of the behaviour of water~\cite{russo2016crystal}. Differently from simple liquids with isotropic interactions, water is characterized by strong directional interactions that favor a local tetrahedral arrangement of particles.
It is possible to classify these tetrahedral environments in two
different groups: amorphous and crystalline local structures.

The first group is composed of amorphous local structures.
Amorphous structures are characterized by a lack of full translational
and rotational symmetry, meaning that there is a high degree of
degeneracy with respect to the relative position and/or orientation of
the tetrahedral units. One of the key properties of water, that sets
it apart from non-anomalous liquids, is that its amorphous structures
can be further grouped into two states \cite{tanaka2000simple}.
The first state, which we call the $S$-state, is a low density/energy configuration in which a central water molecule is surrounded by four tetrahedra that are translationally ordered (i.e. they form a well-defined second shell of nearest neighbours), but orientationally disordered (meaning that the relative orientation of the tetrahedra is disordered)~\cite{russo2014understanding}. The second state, which we call the $\rho$-state, is a high-density/energy configuration with a collapsed second shell of nearest neighbours. The definition of an appropriate order parameter has revealed bimodality in liquid structures up to ambient conditions, and has allowed the definition of a microscopic two-state model that accounts for all water thermodynamic anomalies~\cite{russo2014understanding}. Locally favoured structures also impose strong constraints on the dynamics of water molecules: more specifically, a water molecule has a slower diffusive dynamics the more $S$-states surround it. A hierarchical two-state model was built with this idea in mind, and was found to be able to accounts for water's dynamic anomalies~\cite{shi2018common}, while also predicting a strong-to-strong crossover at supercooled conditions~\cite{shi2018origin}.

The idea of different states playing a pivotal role in determining
water's behaviour is further supported by the behaviour of water in
supercooled and out-of-equilibrium conditions. In particular, several
simulation works have shown that water can exist in two distinct
liquid forms: the LDL (low density liquid) and HDL (high density
liquid) phases, which become indistinguishable at a liquid-liquid
critical
point~\cite{poole1992phase,Mishima1998,gallo2012ising,gallo2016water}. A
non-equilibrium counter part of this transition is also found in water
glasses, where low-density (LDA) and high density (HDA) amorphous glasses have been found (with the proposal of an additional very-high density VHDA amorphous glass~\cite{handle2017supercooled})\cite{Mishima1998,martelli2019unravelling}.  

The second group of local structures is composed of crystalline local structures.
While the anomalous behaviour of the liquid phase appears to be
strongly linked with the competition of two different amorphous local
structures, the process of ice nucleation is associated with a different type
of local structure, i.e. crystalline local structures. These
structures are defined as clusters of molecules whose
nearest-neighbour environment is the same as those found in bulk
crystals. They differ from the amorphous local structures for their
high degree of local translational and rotational order. They can thus
extend to arbitrarily large sizes, and their growth becomes
irreversible above the critical nucleus size. 
The relationship of these crystalline local structures to amorphous local structures formed in a supercooled state is one of the important factors controlling the ease of crystallization~\cite{tanaka2012bond,russo2014new,russo2016crystal,russo2018glass}.
These clusters originates from fluctuations that are very localised in space and time. For this reason, computer simulations are giving a fundamental contribution to our understanding of this important process~\cite{brukhno2008challenges,moore2011structural,gallo2016water,russo2016crystal,russo2018glass}.

Homogeneous ice nucleation occurs by the growth of the stable Ice I phase from a few nucleation sites within the supercooled liquid state. Ice I has two different polytypes: \emph{hexagonal} ice (space group $P6_3/mmc$), and \emph{cubic} ice (space group $Fd\bar{3}m$). Hexagonal ice is the thermodynamic stable form, but nucleation at supercooled conditions often proceeds through metastable phases, and deviations from six-fold rotational symmetry have been found in diffraction patterns for a vast array of crystallization scenarios (see Ref.~\cite{malkin2015stacking} for a review). These include re-crystallization from either amorphous ices~\cite{shilling2006measurements,dowell1960low,kohl2000glassy,loerting2006high,geiger2014proton}, or high-pressure ices~\cite{arnold1968neutron,kuhs1987high,hansen2008formation,hansen2008bformation}. This metastable phase was long identified with the cubic phase, but it has recently being pointed out that due to the high amount of stacking disorder~\cite{murray2007strong,carignano2007formation,pirzadeh2010understanding,rozmanov2012anisotropy,seo2012understanding,malkin2012structure,choi2014layer,quigley} this phase should be more appropriately called stacking disordered ice, ice $I_\text{sd}$, which retains a three-fold rotational symmetry and can be identified with the space group $P3m1$~\cite{kuhs2012extent,hansen2008formation}.

The formation of ice $I_\text{sd}$ can be understood in terms of cross-nucleation~\cite{desgranges2006molecular,nguyen2014cross}, which occurs when two polymorphs share a common crystallographic plane, i.e. the (111) plane for cubic ice, and the basal plane (0001) of hexagonal ice. In this case the two crystals match perfectly at the interface plane (both in symmetry and lattice constant), without discontinuities across the two phases. This causes a great reduction of the interface free energy. In the language of interphase interfaces, the two Ice I polytypes form a \emph{coherent} interface. Very recently, computer simulations of the mW model have shown that ice crystals with stacking disorder are about $14$ KJmol$^{-1}$ more stable than hexagonal crystals for sizes up to 100000 molecules at $T=230$~K~\cite{molinero2017nature}, nicely showing that the driving force for crystallization is size dependent.

Cubic ice and  ice $I_\text{sd}$ are the only metastable phases found
in laboratory experiments at ambient pressure. But there are different
candidate crystalline structures, which retain the local tetrahedral
arrangement of water molecules, and have additional properties not
found in the known forms of Ice~\cite{engel2018mapping}. We focus here on the so-called Ice~$0$ form, which is currently the only known phase (of the mW potential) that is metastable in the regime of ambient pressure and low temperatures that is relevant for homogeneous ice nucleation~\cite{russo2014new}. The unit cell of Ice $0$ contains 12 Oxygen atoms, and is tetragonal where the ratio between the long and the short sides of the cell is $c/a=1.81$.

Several simulation studies have addressed the problem of homogeneous nucleation of water, both in molecular models, such as TIP4P~\cite{matsumoto2002molecular,radhakrishnan2003nucleation,malkin2012structure} and the coarse grained mW model~\cite{molinero2008water,moore2010ice,moore2011structural,moore2011cubic}, which sacrifices orientational degrees of freedom in exchange of a considerable faster crystallization dynamics. To enhance the sampling of crystalline cluster, several techniques like Umbrella Sampling, Metadynamics, and Forward Flux Sampling have been employed~\cite{quigley2008metadynamics,li2011homogeneous,reinhardt2012local,reinhardt2013note}. An understanding of ice nucleation requires accurate studies of interfacial properties~\cite{benet2014study,benet2014computer,cheng2015solid,espinosa2016ice,espinosa2016interfacial,ambler2017solid} which are commonly used within Classical Nucleation Theory parametrizations~\cite{koop2016physically,cheng2017bridging,cheng2017gibbs}. The behaviour of water at interfaces is central in understanding heterogeneous nucleation~\cite{lupi2014heterogeneous,reinhardt2014effects,cox2015molecular,sosso2016ice,lupi2016pre}, nucleation with free surfaces~\cite{johnston2012crystallization,li2013ice,hudait2014ice,haji2014suppression,malek2015crystallization,haji2017perspective}, in nano-confinement~\cite{moore2010freezing,gonzalez2011melting}, and crystallization in solutions~\cite{bullock2013low,soria2017simulation,conde2017spontaneous}.

In the present work we focus on crystalline clusters for the mW model
of water~\cite{molinero2008water}. We consider both polytypes of
Ice~$I$, i.e. the cubic and hexagonal ice phases, and the metastable
Ice~$0$ phase. We first examine their energetic stability under a wide
variety of conditions (temperature and density). We will show that for
small clusters, the Ice~$I$ crystalline clusters are not the most
energetically stable. We instead show that small clusters with the
nearest neighbour environment of Ice~0 have a lower total energy due
to the abundance of $5$-membered rings. While Ice~I clusters (both
hexagonal and cubic forms) grow isotropoically in all directions, Ice
$0$ grows along layers composed of $5$-membered rings. We then perform seeding simulations at supercooled conditions, and examine the growth behaviour of the different clusters. In particular we highlight the importance of grain boundaries in the kinetic pathway of growth of the different crystalline clusters. We focus here on three coherent grain boundaries, whose low free energy penalties make them particularly abundant. We will estimate these free energy penalties, and try to rationalize their impact on ice nucleation.

\section{Methods}

Here we perform Monte Carlo simulations of the mW model of
water~\cite{molinero2008water}. We employ isobaric simulations at
ambient pressure, and focus on highly supercooled conditions, $T=218$~K, which is low enough to have accessible nucleation barriers, but high enough that nucleation is a rare process (not spontaneously occurring in unbiased simulations) and that dynamics can still be observed over $1000$~s of Brownian times. Ice crystals are then grown both via biased simulations, with the CNT-US method, and unbiased simulations, with the \emph{seeding} technique.

The CNT-US method is a variant of Umbrella Sampling introduced in Ref.~\cite{russo2014new}, where the expression for the free-energy barrier of Classical Nucleation Theory is used as a biasing potential in simulations. The new Hamiltonian is then written as
\begin{equation}\label{eq:cntus}
\mathcal{H}'=\mathcal{H}+|\Delta\mu|n^{2/3}(n^{1/3}-3n_c^{1/3}/2)
\end{equation}
where $\mathcal{H}$ is the given by the mW potential, $\Delta\mu$ is the chemical potential difference between Ice I and supercooled water, $n$ is the size of the largest nucleus in the system, and $n_c$ is the critical nucleus size. The advantage of this method over standard Umbrella Sampling is that the whole free energy landscape can be explored in equilibrium in one independent simulation, avoiding kinetic traps that are often incurred when restricting simulations to small sampling windows.

To distinguish the nucleation barriers between different phases,
e.g. Ice $I_h$, $I_c$, and nuclei with stacking disorder we perform
standard Umbrella Sampling simulations in which we initialize the simulations with a nuclei of the desired phase.
These simulations only preserve the biasing potential of the CNT-US scheme, but, as in traditional US schemes, the range of the order parameter (the size of the largest nucleus) is divided in many small windows, each one allowing fluctuations of the order parameter between $N_i$ and $N_i+\Delta N$. Since the initialization of each window occurs with a seed of the correct type, fluctuations to a seed with a different composition occur very rarely. We thus monitor the trajectory in each window and perform a check of the nucleus composition every 1000 Monte Carlo (MC) cycles to ensure that the nucleus has preserved the correct composition. If not, we perform a rejection move of the trajectory of the last 1000 MC time steps. We typically set $\Delta N=5$, but we have checked that the results are insensitive to this choice for $\Delta N=1$ (where the technique is formally equivalent to the Successive Umbrella Sampling scheme~\cite{rovigatti2018simulate}) and $\Delta N=10$.

We then monitor the different simulations to make sure that the original crystal structure is preserved during the sampling, and ensure that no additional stacking faults have originated.

For unbiased simulations, we use the \emph{seeding} technique~\cite{knott2012homogeneous,zimmermann2015nucleation,espinosa2016seeding,espinosa2016time}, which consists in the insertion of perfect crystalline nuclei in the melt in order to determine at which conditions the nucleus is critical, and then use Classical Nucleation Theory to estimate parameters like the surface tension and the crystallization rate. While being successful in the estimation of the crystallization rate of water at moderate supercooling, with critical nuclei composed of several thousand molecules and where the approximations of Classical Nucleation Theory are most likely to hold, potential problems with the extension of this technique to deeply supercooled conditions have been recently highlighted~\cite{lifanov2016nucleation}. On the contrary, for systems seeded with stable Ice~I crystals at deeply supercooled conditions, we will find that the critical nucleus sizes are in excellent agreement with the ones we obtain independently from Umbrella Sampling simulations.

To distinguish the different crystalline phases, and to bias the Umbrella Sampling simulations we use the standard bond-orientational order parameters. In particular we employ the $Q_{12}$ bond orientational order parameter, where a molecule is identified as having a crystal-like environment if it has more than 12 connected neighbours among its first 16 nearest neighbour. A connected neighbour of particle $i$ is defined as a neighbour $j$ for which the normalized product satisfies $Q_{12}(i)\cdot Q_{12}(j)>0.75$. See Ref.~\cite{tanaka_review} for a recent review of this methodology.

Where not differently specified, the system size of seeding simulations is $N=10000$ water molecules, while Umbrella Sampling simulations are run with $N=4000$ molecules. Since we only focus on small nuclei (up to $N=100$ molecules), no strong system size dependence of our results is expected.

\section{Results}
\label{sec:results}

\subsection{Energy of crystalline clusters}
\label{sec:results1}

\begin{figure}[!t] 
	\begin{center}
		\includegraphics[width=8.5cm]{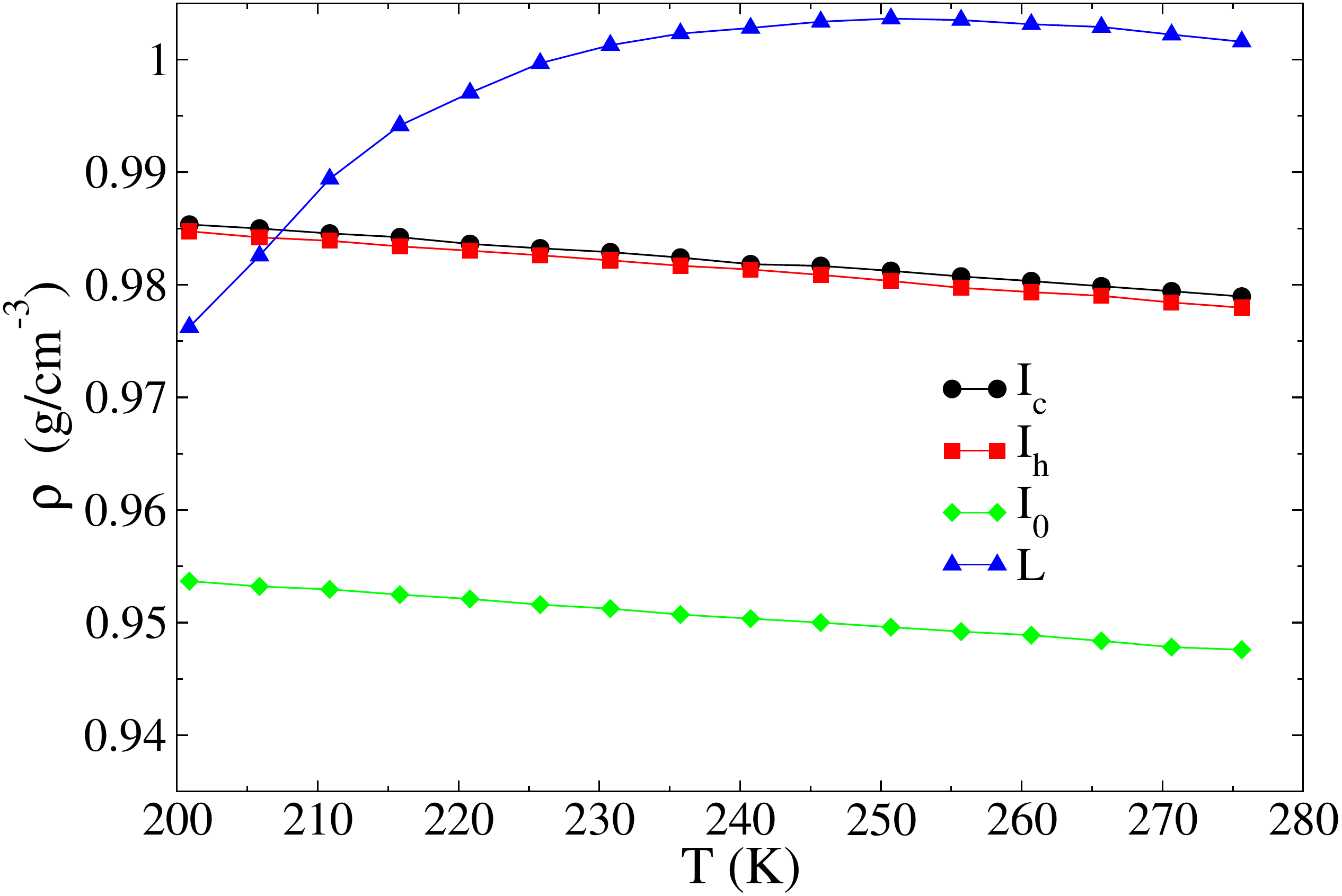}
		\caption{\label{fig:r_vs_T} $\rho$ versus $T$ relation at constant pressure $P=0$~atm, as described in the main text, for cubic ice (I$_c$, black circles), hexagonal ice (I$_h$, red squares), ice 0 (Ice 0, green rhombuses) and liquid phase (L, blue triangles).}
	\end{center}
\end{figure}

In order to understand crystal clusters competition in supercooled
water, we compare the energy of clusters of perfect crystals in three different ways to account for different scenarios of cluster formation: (i) at the equilibrium density $\rho$, corresponding to a specific temperature $T$, which
depends on the structure under consideration (see Fig.\ref{fig:r_vs_T})
; (ii) at fixed density corresponding to that of the liquid phase at
equilibrium; (iii) at fixed bond distance, in analogy with the work of Frank~\cite{frank1952supercooling}.
Afterwards, (iv) we extend the approach (i) to account for clusters equilibration in the liquid phase, and (v) we consider also the energetic contribution coming from the hydration layer.
\vspace{0.5cm}

In Fig.\ref{fig:r_vs_T} we show the $\rho$ versus $T$ relation at constant pressure $P=0$~atm for the liquid phase (L), cubic ice (I$_c$), hexagonal ice (I$_h$) and ice 0 (Ice 0) as obtained from Monte
Carlo simulations of the mW model in the {\it NPT} ensemble for $N=5376$ 
particles with periodic boundary conditions. 
We can notice the well known temperature of
maximum density (TMD) anomaly of liquid water found at around T=250~K
in the mW model at normal conditions \cite{molinero2008water,russo2018water}. 
We also note crystal densities being lower
than the liquid density above 207~K (see Fig.\ref{fig:r_vs_T}). 

Clusters of mW particles are built by adding particles to a crystalline lattice (or from a configuration of the liquid at equilibrium in the case of disordered clusters) considering the following energetic approach:
starting from any position in the crystalline lattice, we add one particle at each step choosing among the lattice positions that result in the minimum energy cluster.
This involves computing the bond energy of every particle on the surface of the cluster with its potential neighbours, and choosing to add the particle that generates the cluster with minimum
energy. In case of degenerate lattice positions, i.e. different clusters with the same size and same energy, the choice of which lattice position to fill is done randomly.
We repeat this procedure until we get a cluster of the desired size.
Since clusters will have different morphologies (i.e. different numbers of inter-particle bonds), we compute the average energy of clusters over 50 independent growth processes.

\begin{figure}[!t]
	\begin{center}
		\includegraphics[width=8.5cm]{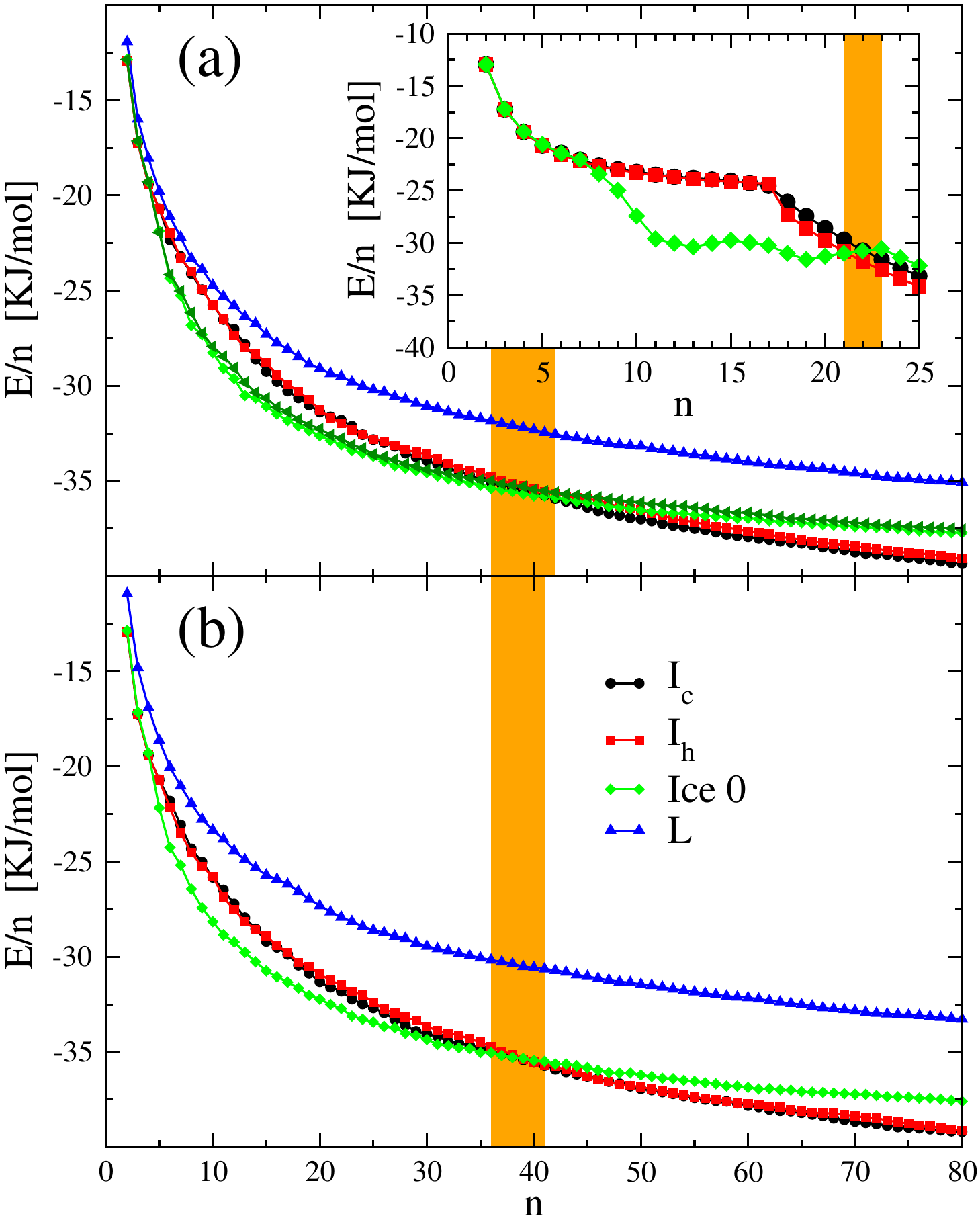}
		\caption{\label{fig:E_vs_N_Athermal} Energy per particle of clusters
			$E/n$ as a function of cluster size $n$ for cubic ice (I$_c$, black
			circles), hexagonal ice (I$_h$, red squares), ice 0 (Ice 0, green
			rhombus and dark green left-pointing triangles), and liquid phase (L, blue triangles). 
			The vertical orange band marks the crossing region between Ice 0 and I$_c$/I$_h$ curves.
			In (a) every cluster is considered at the same density of the liquid phase at equilibrium at $T=200$~K, which is $\rho=0.976$~g/cm$^3$ (method (ii) described in the main text). The shortest bond distance in each structure is $d=2.71,2.71,2.67$~{\AA} for I$_c$ (black circles), I$_h$ (red squares) and Ice 0 (green rhombus), respectively, while the shortest neighbouring distance in the case of the liquid phase L is the position of the first peak of the liquid two-point correlation function.
			In order to compare the different clusters following the method (iii), we show also Ice 0 at the density $\rho=0.937$~g/cm$^3$ (dark green left-pointing triangles) which corresponds to $d=2.71$~\AA, the same of I$_c$/I$_h$.
			In (b) every cluster is at its own equilibrium density corresponding to $T=235$~K (method (i)), which is $\rho=0.982,0.982,0.951,1.002$~g/cm$^3$ for I$_c$, I$_h$, Ice 0 and L, respectively.
			Inset in (a): Energy per particle of clusters $E/n$ as a function of cluster size $n$ using the spherical symmetric growth of clusters based on the distance from the seed, as described in the main text.
		} 
	\end{center} 
\end{figure}

In Fig.\ref{fig:E_vs_N_Athermal}(a),(b) we show the energy per particle
of clusters, $E/n$, as a function of the cluster size $n$ for the
crystals and liquid phases at the conditions described above (ii), (iii) and (i), respectively.
\vspace{0.2cm}

From Fig.\ref{fig:E_vs_N_Athermal}(a),(b)
we can notice that clusters of I$_c$ and I$_h$ have always the same
energy for any size $n$, as expected.
We also notice that Ice 0 has a lower energy respect to Ice I
for small clusters size up to the critical value $n^*\sim 40$ for the three conditions (i), (ii) and (iii)
(we computed the value of $n=n^*$ at which curves associated to Ice 0 and I$_c$/I$_h$ cross for other values of $\rho$ and $T$ following methods (ii) and (iii) and found always $n^*\sim 40$).

\begin{figure}[!t]
	\begin{center}
		\includegraphics[width=8.5cm]{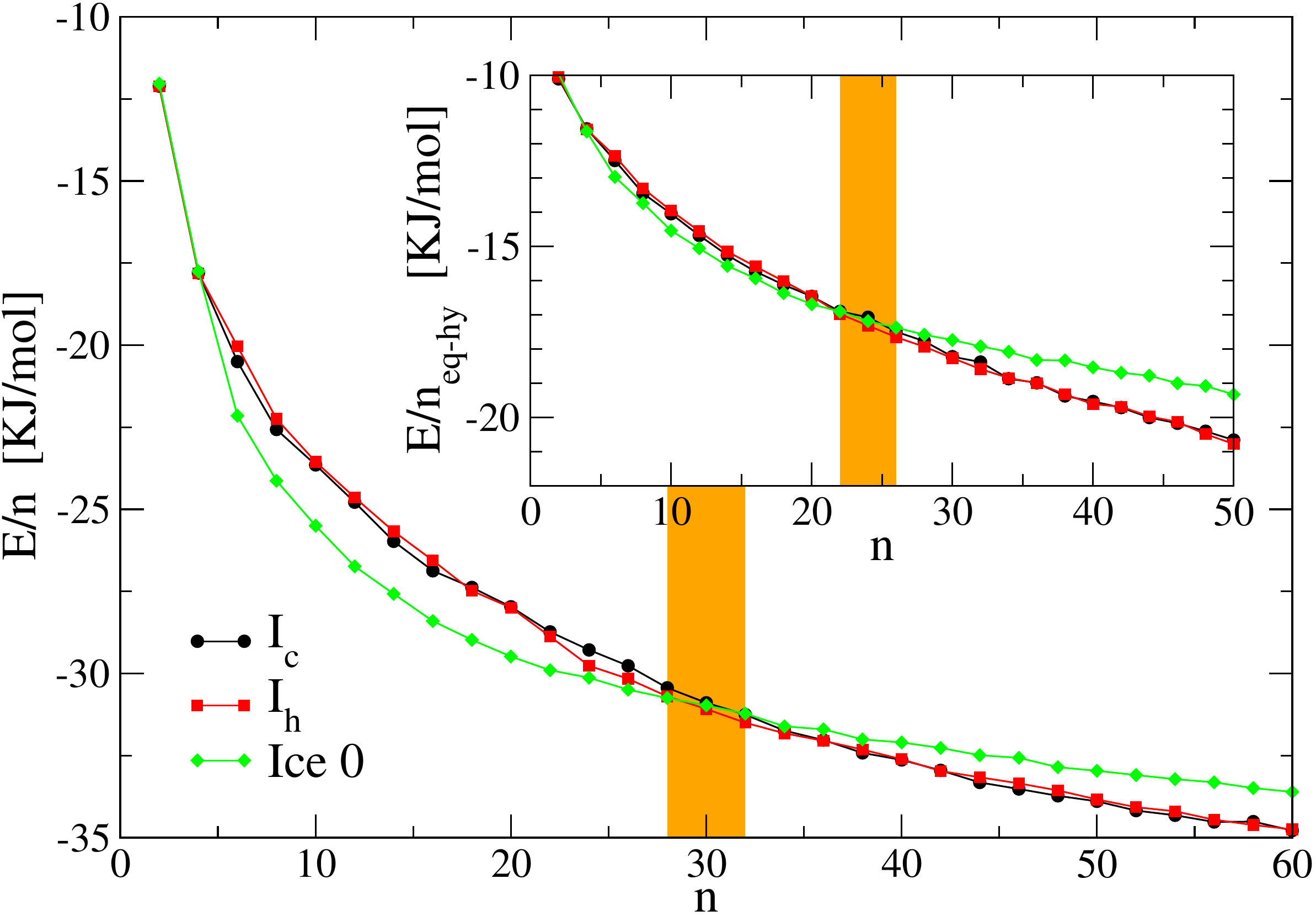}
		\caption{\label{fig:E_vs_N_Equilibrated} Energy per particle of equilibrated clusters $E/n$ as a function of cluster size $n$ for cubic ice (I$_c$, black
			circles), hexagonal ice (I$_h$, red squares), ice 0 (Ice 0, green
			rhombus). 
The vertical orange band marks the crossing region between Ice 0 and I$_c$/I$_h$ curves.
Inset: Energy per particle (including the hydration layer) of equilibrated and hydrated clusters $E/n_{eq-hy}$ as a function of cluster size $n$.} 
	\end{center} 
\end{figure}

To extend the previous results, we account for clusters equilibration in the liquid phase (method (iv)). In order to do so we perform Monte Carlo simulations sampling the dynamics according to a fixed topology structure, where the energetic neighbors of each particle of a cluster are constrained to be the same during the dynamics and their bond elongation bounded by a cutoff set to $\simeq~1.3~{\AA}$.
In Fig.\ref{fig:E_vs_N_Equilibrated} we show the energy per particle of clusters, $E/n$, as a function of the cluster size $n$ for the crystals obtained following method (iv). 
The curves are obtained averaging for each size $n$ over 10 different uncorrelated simulation times and 50 independent growth processes.
From Fig.\ref{fig:E_vs_N_Equilibrated} we can see that clusters of Ice 0 are more stable respect to clusters of ice I up to the size $n_{eq}^*\simeq 30$. 

Considering also the contribution to the energy of clusters coming from the hydration layer (method (v)), we observe (Inset of Fig.\ref{fig:E_vs_N_Equilibrated}) that the maximum size up to which clusters of Ice 0 are more stable respect to ice I is reduced further respect to the previous cases up to the value $n_{eq-hy}^*\simeq 24$.
This behavior is similar to what has been observed for the stability of 13-particles icosahedron respect to 13-particles fcc and hcp clusters interacting through the Lennard-Jones potential when embedding clusters in liquidlike environment \cite{Tarjus2003}.

\subsubsection{Clusters morphology}
\label{sec:results2}

\begin{figure}[!t] 
	\begin{center}
		\includegraphics[width=0.4cm]{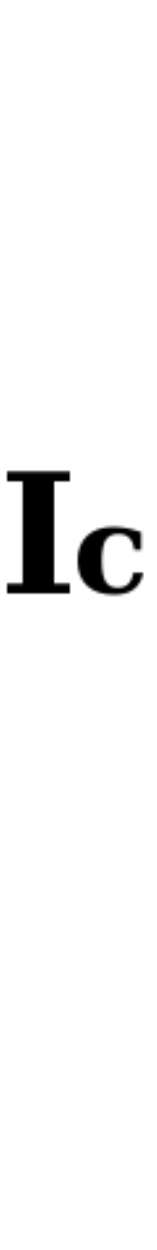}
		\includegraphics[width=3.5cm]{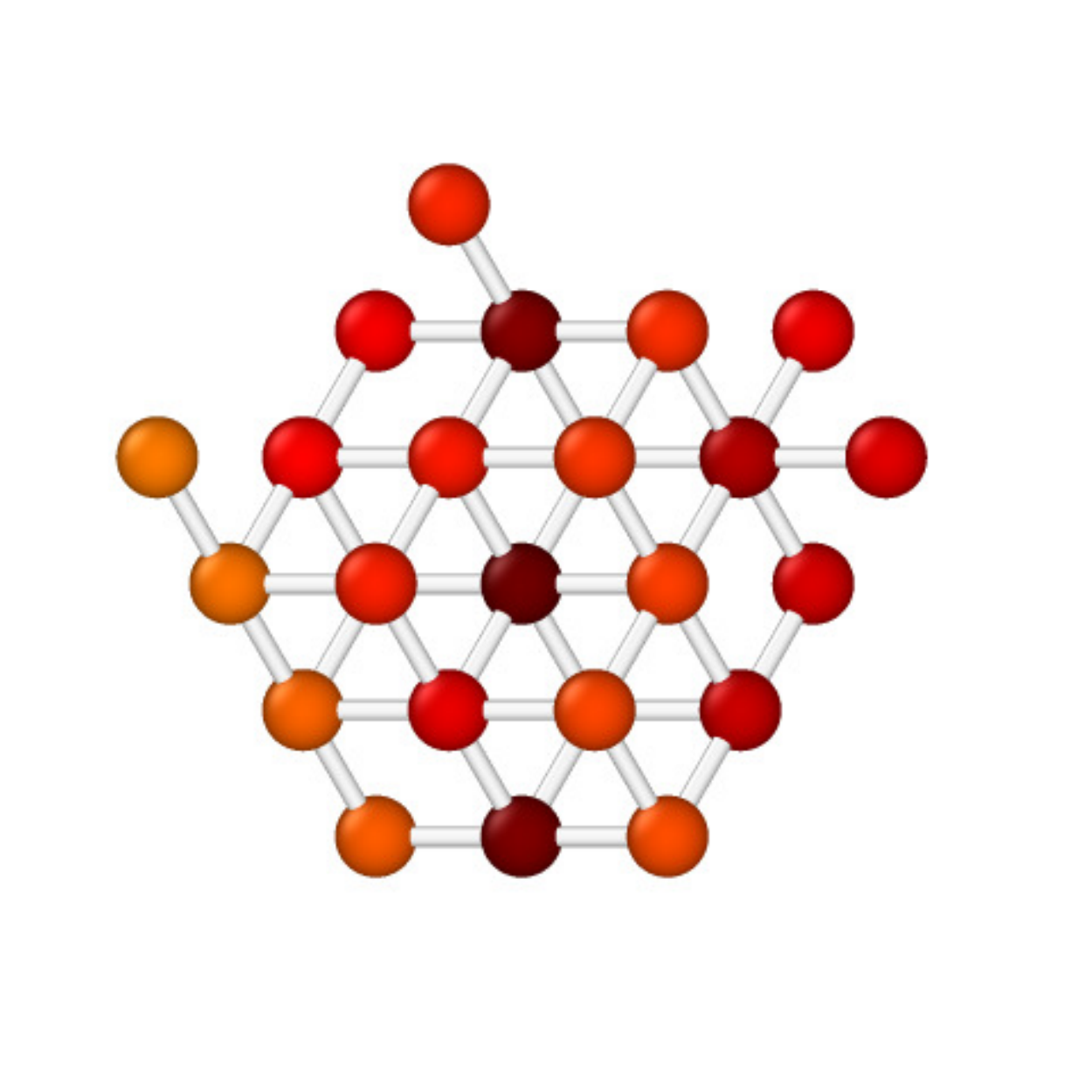}
		\includegraphics[width=3.5cm]{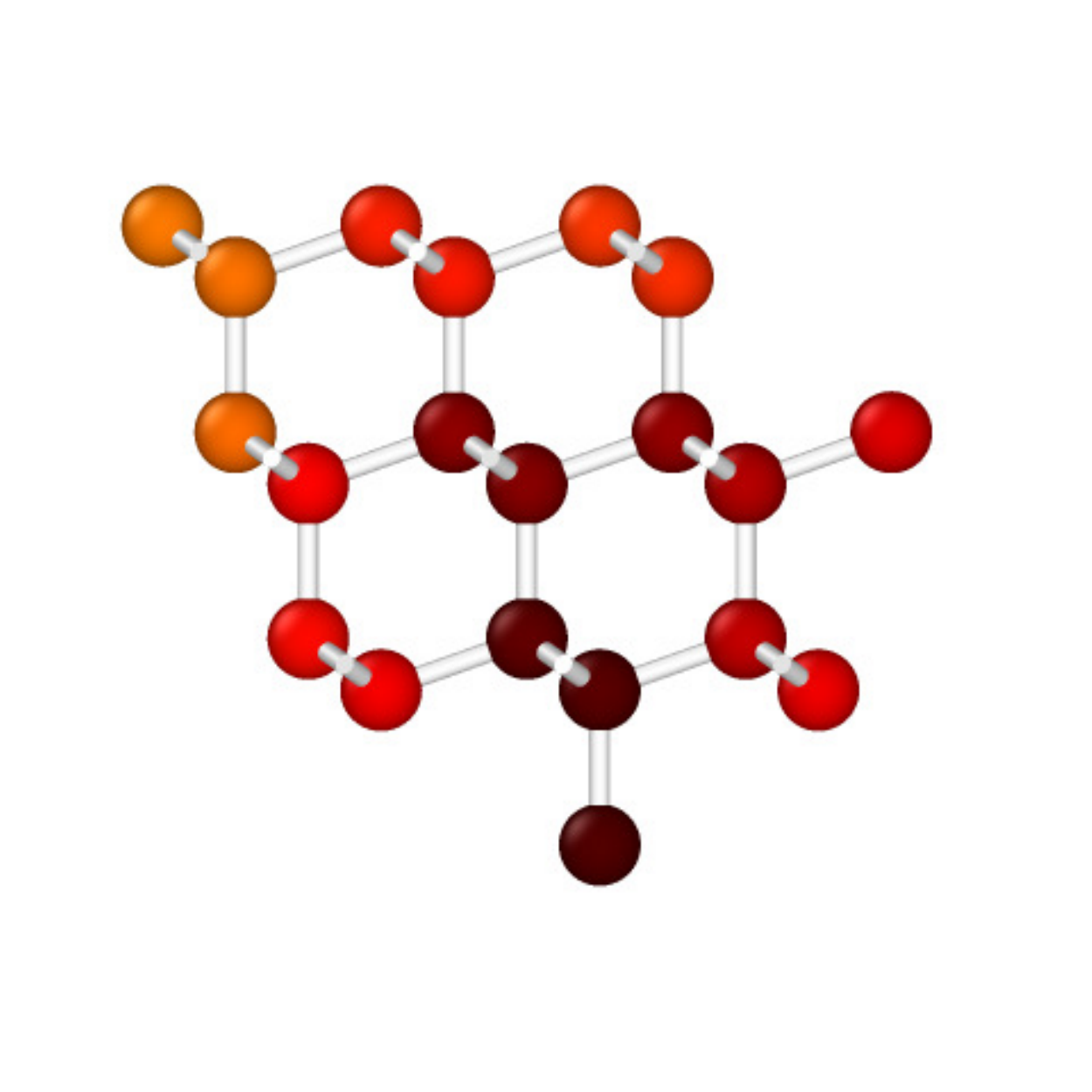}
		
		\vspace{0.1cm}
		\includegraphics[width=0.4cm]{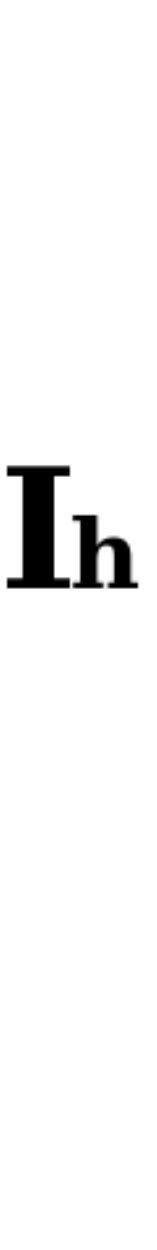}
		\includegraphics[width=3.5cm]{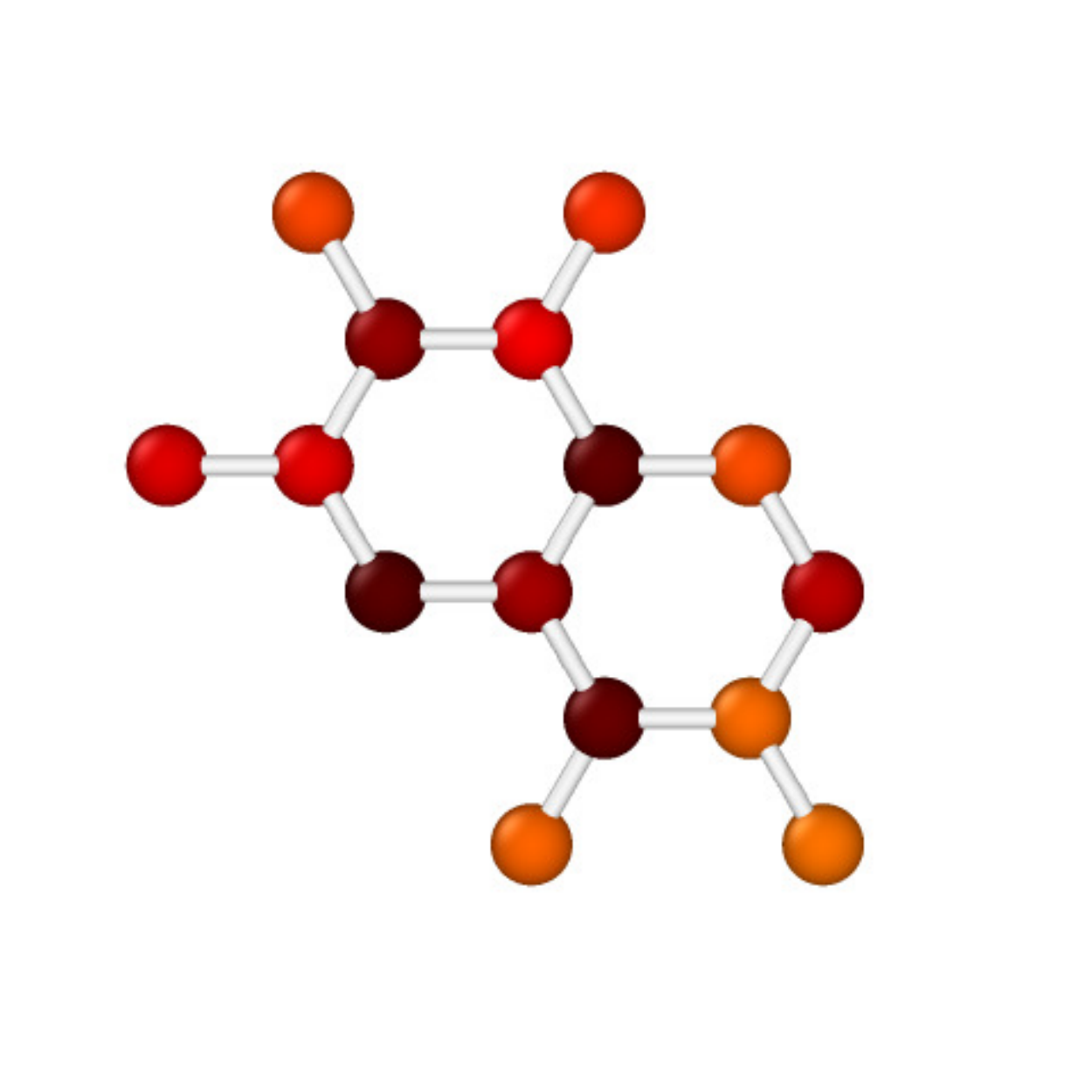}
		\includegraphics[width=3.5cm]{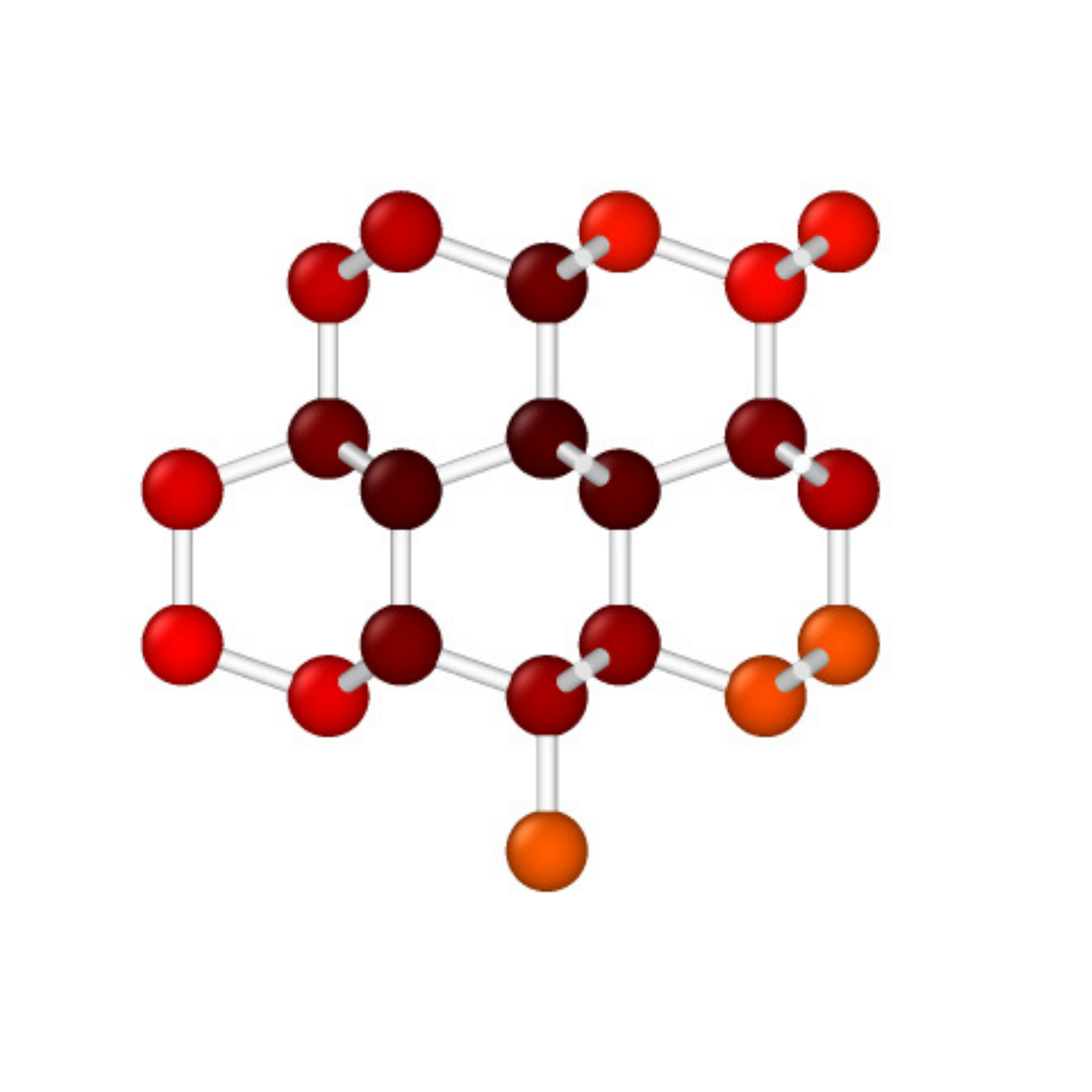}
		
		\vspace{0.1cm}
		\includegraphics[width=0.4cm]{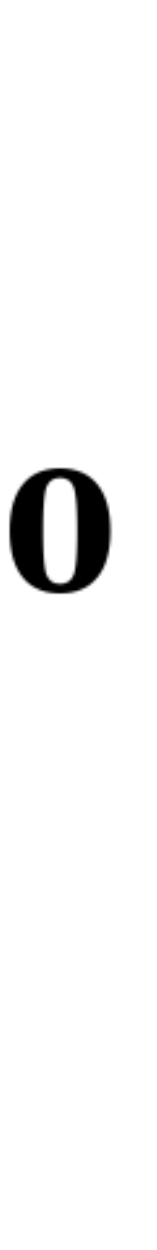}
		\includegraphics[width=3.5cm]{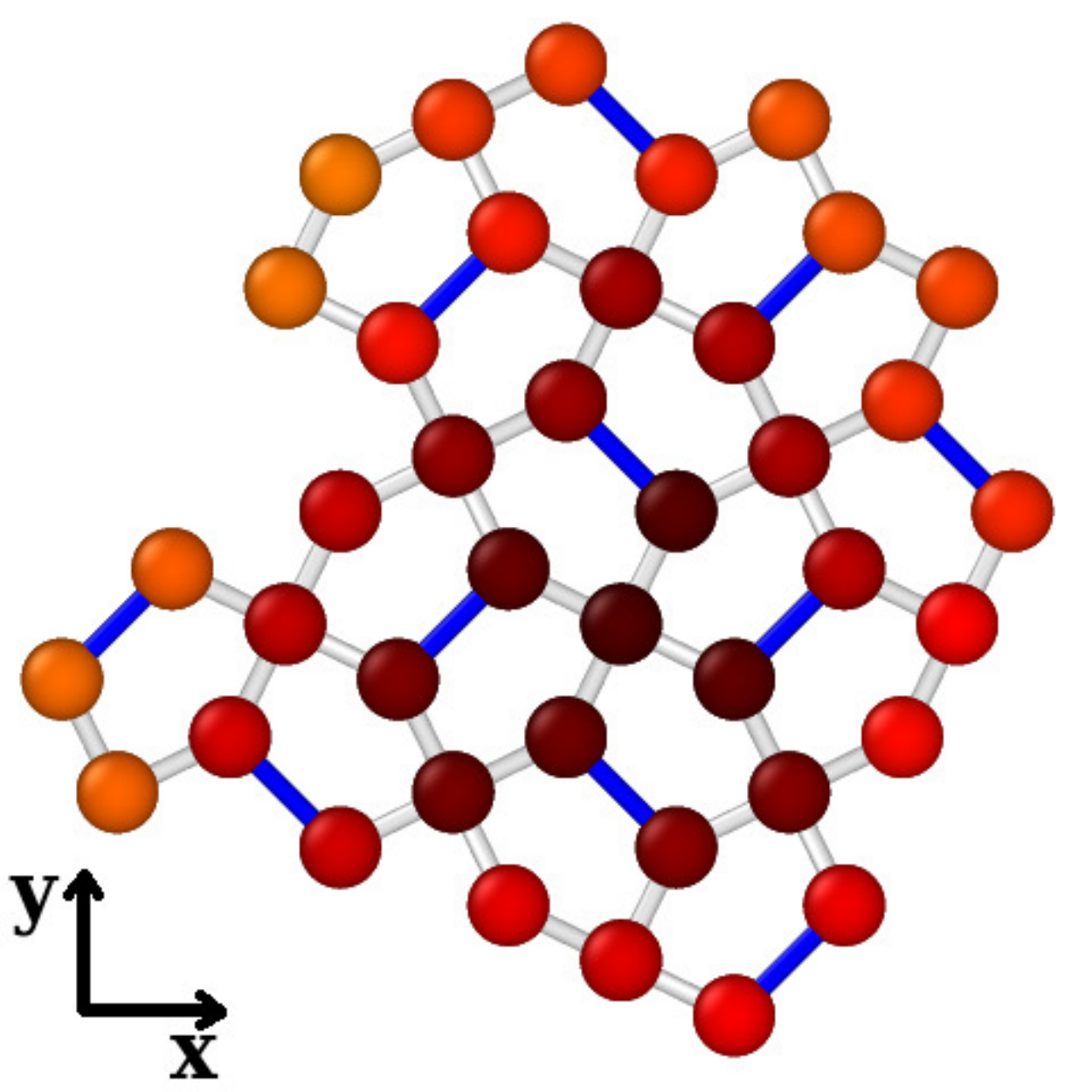}
		\includegraphics[width=3.5cm]{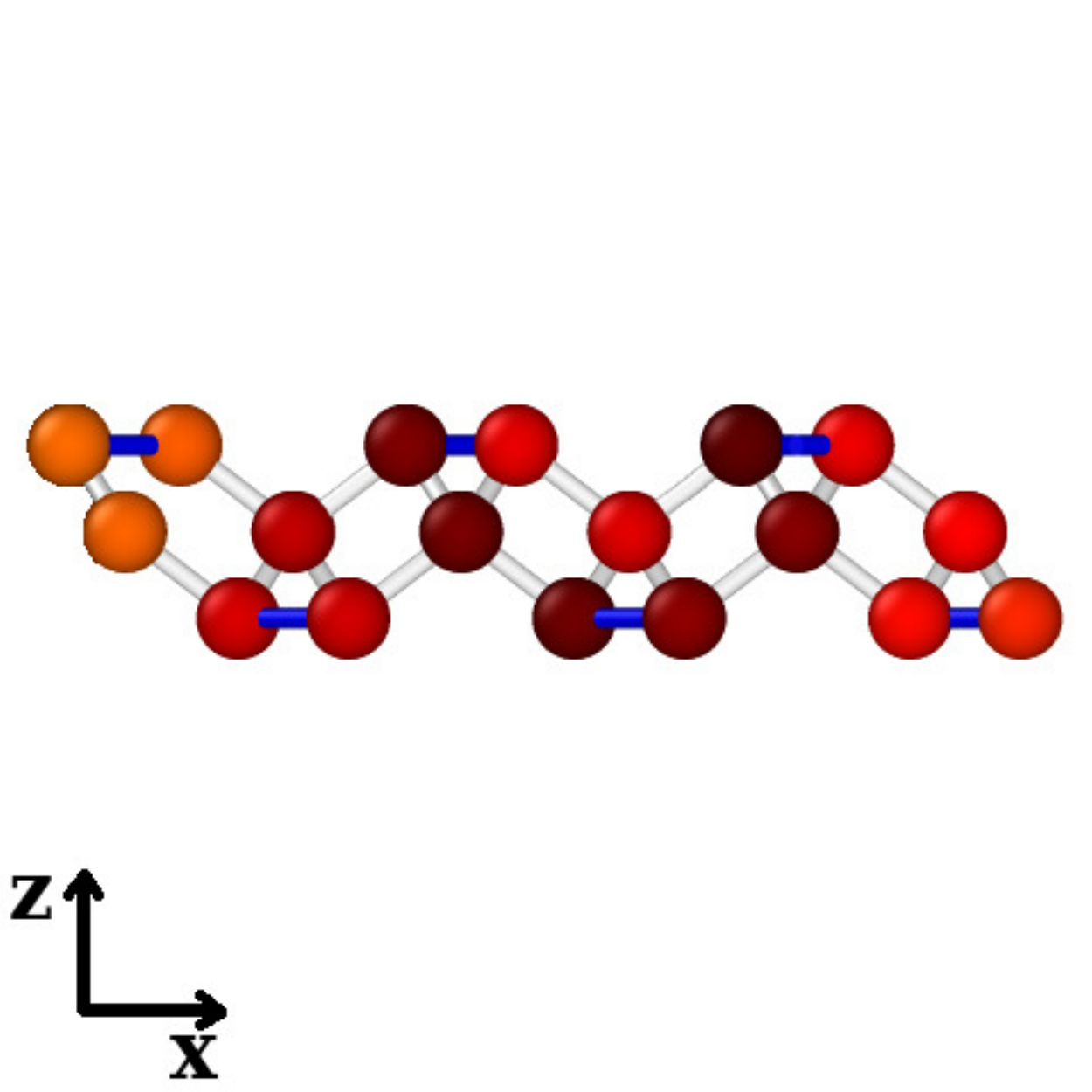}
		
		\caption{\label{fig:snapshots_N40} Snapshots of typical clusters of size $n=40$ projected in the {\it xy} (left panels) and {\it xz} (right panels) planes, for the crystalline phases I$_c$, I$_h$, and Ice 0. Bonds connect pairs of neighbouring particles. Particles are colored according to the growth sequence such that the seeding particle is the darkest one and the last particle attached to the cluster is the lightest one.}
	\end{center}
\end{figure}

In Fig.\ref{fig:snapshots_N40} we show typical clusters of Ices I$_c$, I$_h$ and  
Ice 0 projected in the {\it xy} (left panels) and {\it xz} (right panels) planes for clusters of size $n=40$ that is the typical value at which the curves of energy per particle associated to Ice 0 and I$_c$/I$_h$ cross. 
Particles are colored according to the growth sequence such that the seeding particle is the darkest one and the last particle attached to the cluster is the lightest one.
For I$_c$ the {\it xy} plane corresponds to the direction [111], while {\it xz} to [\=110]. For I$_h$ {\it xy} is the basal face corresponding to the Miller indices \{0001\}, while {\it xz} is the secondary prism face corresponding to the Miller indices \{11-20\}. For Ice 0 the long side of the unit cell $c$ is aligned along the {\it z} axis.

The first important difference between Ice 0 and I$_c$/I$_h$ and L is 
that Ice 0 grows preferentially along the (100) plane, while other clusters grow with spherical symmetry.
We understand the planar symmetry of growing clusters of Ice 0 in terms of its neighbouring bond lengths.
While I$_c$ and I$_h$ are characterized by a unique length, which is the distance between two neighbouring particles, Ice 0 is characterized by three different neighbouring distances ($d_s$, $d_i$, $d_l$), very close to each other.
Two of these three lengths ($d_s$ and $d_l$) are involved in the formation of planar structures. In the snapshots of Ice 0 in Fig.\ref{fig:snapshots_N40}, white bonds are associated to the short distance ($d_s\simeq 2.67$~\AA for $\rho=0.976$~g/cm$^3$), while blue bonds to the long distance ($d_l\simeq 2.73$~\AA for $\rho=0.976$~g/cm$^3$, which is about $2\%$ longer than $d_s$). The intermediate distance ($d_i\simeq 2.68$~\AA for $\rho=0.976$~g/cm$^3$) connects particles along the {\it z} direction forming distorted hexagonal rings and then it doesn't play any role in the formation of clusters built following the minimum energy rule, as discussed below.
From Fig.\ref{fig:snapshots_N40} we can see that clusters of Ice 0 grow forming rings of pentagons composed of $d_s$ (white) and $d_l$ (blue) bonds.  
For this reason the energy of Ice 0 is lower than I$_c$/I$_h$ already for clusters of $n=5$ particles which form pentagons in Ice 0, that is a ring, while I$_c$/I$_h$ can form the first ring only for $n=6$.
Each ring forming in the cluster gives an extra term in the calculation of its total energy.
Indeed, the number of bonds $N_b$ forming in a cluster is $N_b=n+N_r-1$, where $N_r$ is the number of rings.
Counting the number of bonds for the different growing clusters, we find that on average I$_c$ and I$_h$ can form more bonds than Ice 0 only for $n\gtrsim 40$, which corresponds to the typical value at which the curves
of energy per particle associated to Ice 0 and I$_c$/I$_h$ cross (highlighted by the orange vertical band
in Fig.\ref{fig:E_vs_N_Athermal}).

Instead of adding the strongest bonded particle to a cluster, we can add particles with increasing radial distance from a central particle. In this case we get for Ice 0 clusters that are similar to those obtained following the minimum energy rule up to a size of $n=13$, while for $n>13$ we get a spherical symmetric structure whose energy becomes in average larger than the energy of clusters of I$_c$/I$_h$ (computed with this same rule) for $n\gtrsim 22$ (see Inset of Fig.\ref{fig:E_vs_N_Athermal}(a)). This value of $n$ is smaller than the critical value of $n$ ($\gtrsim 40$) found for clusters of Ice 0 growing according to the minimum energy rule because for equal number of particles, a spherical symmetric cluster forms a smaller number of rings than a planar symmetric cluster for the Ice 0, while for I$_c$/I$_h$ the growth is always spherical symmetric following both growing rules.

The reason why the number of bonds $N_b$ can be used to compute the difference in the energy content between clusters of I$_c$/I$_h$ and Ice 0 is related to the two characteristic lengths in Ice 0, $d_s$ and $d_l$, being very close to each other (i.e. $(d_l-d_s)/d_s\ll 1$).
If the characteristic length of neighbouring particles forming a cluster is only one, as in the case of I$_c$ and I$_h$, this length will be the minimum of the potential energy (note that for perfect crystals the three-body term of the mW potential is zero). If, on the other hand, the characteristic lengths of neighbouring particles forming a cluster are more than one, in general they will be different from the length corresponding to the minimum of the potential energy and then the eventual increase in the number of bonds will not balance the increase in energy of each individual bond.
In the case of Ice 0, since the two characteristic lengths $d_s$ and $d_l$ forming clusters are close to each other and close to the distance corresponding to the minimum of the potential energy (which is 2.71~\AA for $\rho=0.976$~g/cm$^3$), the energy of each bond is very close to that of bonds in I$_c$/I$_h$ clusters such that the value of $N_b$ is a clear indication of the energy content of a cluster.

The mechanism responsible for the lower energy content of small clusters of Ice 0 respect to clusters of I$_c$/I$_h$ now described is the same causing the icosahedron to have a lower energy content than fcc and hcp clusters in simple one-component liquids with particles interacting through spherically symmetric pair-potentials, like the Lennard-Jones. 
Indeed, fcc and hcp structures are characterized by a single length, which is the minimum of the potential.
The icosahedron is characterized by two different lengths: the distance between the central particle and its 12 neighbours, and the edge of the 20 equilateral triangles forming the surface of the icosahedron where the vertices are the 12 neighbouring particles, which is about $5\%$ longer.
The total number of bonds forming the icosahedron is 42.
In the case of fcc or hcp structures, the 13 particles (central particle plus its 12 first near neighbours) composing them can form only 36 bonds.
Although the 42 bonds formed in the icosahedron have a slightly higher energy respect to the 36 bonds formed in fcc and hcp, the total energy of a 13-particles cluster of icosahedron is smaller (by 8.4$\%$ in the case of a Lennard-Jones liquid \cite{frank1952supercooling}) than that of a 13-particles cluster of fcc or hcp.
Another analogy between Ice 0 and the icosahedron is their tendency to form more clusters as the temperature is lowered \cite{steinhardt1983bond,Tarjus2003}.
On the other hand, an important difference between Ice 0 and the icosahedron is that the former can form a crystal, even if it is metastable, while the latter, due to its five-fold rotational symmetry which is not compatible with translational periodicity, cannot tile the space and form a crystal so that it leads to frustration in simple one-component liquids \cite{Tarjus2005}.


The stability range in size of clusters of Ice 0 respect to ice I$_c$/I$_h$ will depend on the particular water model adopted. For example, the mW model respect to molecular models underestimates the lattice energy of Ice 0 relative to ice I$_c$/I$_h$~\cite{quigley2014communication}. However, since in our study the stability of Ice 0 respect to ice I$_c$/I$_h$ originates from topology (that is the number of bonds formed in a cluster), we expect that a different water model should give a similar result for the energy per particle difference between Ice 0 and ice I$_c$/I$_h$ (that is Ice 0 more stable than ice I$_c$/I$_h$ up to a specific cluster size), while the energy per particle of each ice could change respect to the mW model.
 
Another aspect to take into account when considering fully atomistic realizations of clusters is
	the configuration of the hydrogen bond network. In the case of small clusters, the HB network 	rearrangements take place in the hydration layer, as verified for small aggregated domains in 	the case of an atomistic model of water in Ref.~\cite{Matsumoto2007}.

In conclusion, we find that even a tetrahedral liquid like water can be characterized by competing crystalline clusters. In our case clusters of Ice 0 are even more stable than isolated clusters of I$_c$/I$_h$ up to a critical size of $n^*\sim 40$ for clusters growing following the minimum energy rule (see Fig.\ref{fig:E_vs_N_Athermal}), while up to a critical size of $n^*\sim 22$ for clusters growing following the closest distance rule (see Fig.\ref{fig:E_vs_N_Athermal}).

\subsection{Cluster growth and grain boundaries}\label{sec:growth}

In this section we address the growth of different crystalline seeds, both from critical nuclei (with the seeding technique and Umbrella Sampling simulations), and as planar interfaces. We will show how the growth of each polymorph is associated with a specific coherent grain boundary. Ice I$_h$ is hindered by stacking disorder formed via a twin boundary between its basal plane and the (111) surface of Ice I$_c$, giving rise to stacking disordered Ice I$_\text{sd}$. Ice I$_c$ growth is instead hindered by a five-fold coherent grain boundary. Ice 0 has a coherent grain boundary between its (100) face and the (310) plane of cubic ice.

\subsubsection{Growth of hexagonal ice}

Nucleation of the stable Ice I phase in the mW model has been
extensively addressed in previous
studies~\cite{moore2011cubic,li2011homogeneous,reinhardt2012free,quigley,espinosa2016seeding,hudait2016free,molinero2017nature,cheng2018theoretical}. The
free energy difference between the two polytypes, cubic and hexagonal
ice, was found to be very small, $\Delta
G_\text{ch}=2\pm1.5\,\text{J\,mol}^{-1}$ at
$T=200$~K~\cite{quigley}. No free energy difference was indeed found
in similar studies~\cite{romano2014novel,russo2014new}. The difference
in nucleation barriers between an hexagonal and cubic nucleus at
$T=240$~K was also found to be negligibly small, $2\pm
2\,\text{J\,mol}^{-1}$~\cite{cheng2018theoretical}, which is
equivalent to a difference of about $1\,k_BT$.

\begin{figure}[!t]
	\centering
	\includegraphics[width=8cm]{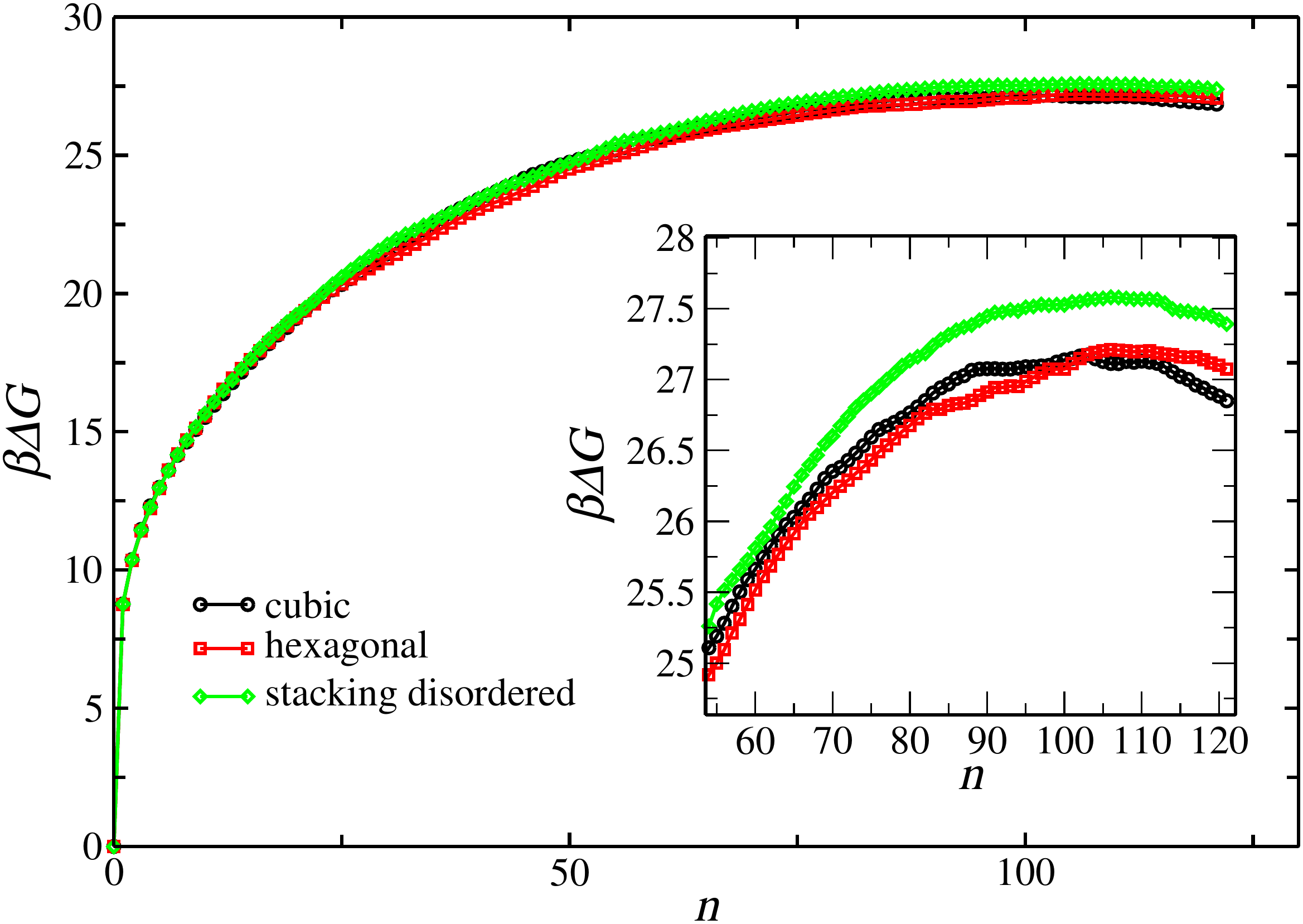}
	\caption{Free energy barriers scaled by thermal energy $\beta
		\Delta G$ as a function of nucleus size $n$ at $T=218\,K$ for the nucleation of Ice I$_h$ (red squares), I$_c$ (black circles), and a nucleus of Ice I$_\text{sd}$ with alternating layers with cubic and hexagonal stacking (green diamonds). The inset shows the barrier around its maximum.}
	\label{fig:barriers}
\end{figure}

Figure \ref{fig:barriers} shows the result for our calculations of the
nucleation barriers at $T=218$~K with the Umbrella Sampling method
detailed in the Methods section. The height of the nucleation barrier
is the same ($\beta\Delta G=27.2\pm0.25$, where $\beta=1/k_BT$) for both a nucleus of Ice I$_h$ and I$_c$. We see that the nucleation barrier difference between the two ices is indeed negligible at lower temperatures, and within the Umbrella Sampling method's precision.

Both Ice I$_h$ and I$_c$ are made of layers composed of six-membered
rings, and differ only in the way these layers are stacked. In Ice
I$_h$ the oxygen positions are the same between all layers
(parallel
	to the basal plane), while in Ice I$_c$ each layer is shifted by a distance equal to half the diameter of a ring.
Given the low free energy difference between the cubic and hexagonal polytypes, the growth of ice I$_h$ along the direction perpendicular to the basal plane is controlled by the free energy cost of introducing a twin grain boundary that changes the order of the stacking. It was recently pointed out that the presence of stacking disorder gives rise to a new crystal, ice I$_\text{sd}$ which retains only a three-fold rotational symmetry instead of the full six-fold symmetry of ice I$_h$~\cite{malkin2012structure}.
The signature of stacking disorder was found in many X-ray and neutron
diffraction patterns which were previously attributed to cubic
ice~\cite{malkin2012structure}, leading to the idea that at low
temperatures ice always nucleates as Ice~I$_\text{sd}$ rather than
cubic ice~\cite{malkin2012structure,kuhs2012extent}. We will challenge this idea in the next Section.

The free energy cost of forming a stacking fault is offset by an
increase in configurational entropy coming from the different ways to
arrange the layers. Previous studies have addressed both these
terms~\cite{quigley,molinero2017nature,cheng2018theoretical}. The most
recent estimates~\cite{cheng2018theoretical} put the stacking fault
free energy per unit area between $\gamma_\text{sf}=0\,\text{mJ
	m}^{-2}$ at $T=180$~K and $\gamma_\text{sf}=0.11\,\text{mJ m}^{-2}$
at the melting point.

We estimate the value of $\gamma_\text{sf}$ from Umbrella Sampling
simulations for an Ice I$_\text{sd}$ nucleus in which the cubic and
hexagonal layer alternate between each other (see, e.g.,
Fig.~\ref{fig:snapshots}). The barrier is shown with green diamond
symbols in Fig.~\ref{fig:barriers}. We emphasise that this is not the
nucleation barrier for Ice I$_\text{sd}$, as it does not take into
account the entropy contribution coming from different realizations of
the stacking disorder. The curve in Fig.~\ref{fig:barriers} is instead
the nucleation barrier for a single realization of stacking disorder,
from which we can estimate the cost of forming the twin boundary
defect. The free energy difference between the stacking disordered
nucleus and the hexagonal (or cubic) nucleus is shown in the inset of
Fig.~\ref{fig:barriers} and corresponds to a difference of $0.3\pm
0.1\,k_B\,T$.
The estimation of the difference in free energies between stacking disordered and cubic and hexagonal ice was done by fitting the nucleation barriers with the classical nucleation theory expression in the range $60<N<120$ around the top of the barrier. The expression is $\Delta G=a\,x+b\,x^{2/3}$. We obtain maxima at $\Delta G=27.6$ for the stacking disordered seed, and $\Delta G=27.3$ for both cubic and hexagonal seeds. The propagated error from fitting the classical nucleation theory expression is low ($~1\%$), but to account for systematic effects coming from the Umbrella Sampling method, we have increased this error to be of the same order of the deviation between barriers computed from independent trajectories, which is of the order of $0.1\,k_B\,T$.

To the value $0.3\pm
0.1\,k_B\,T$, one would need to subtract the bulk free
energy difference between the cubic and hexagonal crystals, but as one
can see from the inset of Fig.~\ref{fig:barriers}, this difference
falls below our resolution at $T=218$~K, so we simply set it to
zero. We then measure the surface tension between the hexagonal and
cubic layers during the simulation and estimate it as $\gamma_\text{sf}=0.16\pm 0.05\,\text{mJ m}^{-2}$. Ignoring the bulk free energy difference makes our value an upper bound to the true surface penalty, so overall our estimate is in agreement with the results of Ref.~\cite{cheng2018theoretical}.

\subsubsection{Growth of cubic ice}

\begin{figure}[!t]
	\centering
	\includegraphics[width=8cm]{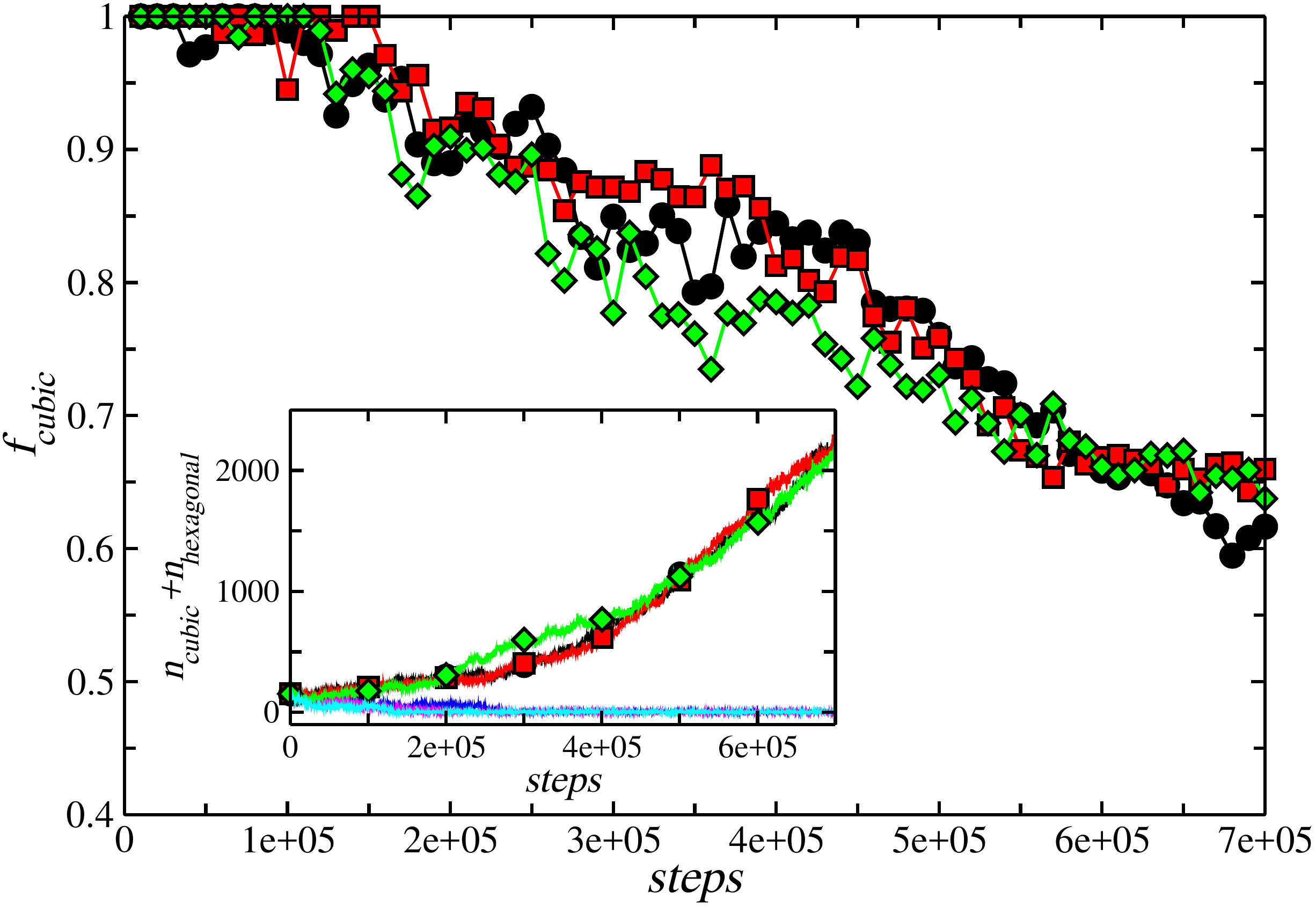}
	\caption{Fraction of molecules in a cubic ice environment as a function of simulation steps, for three independent runs at $T=218$~K. The inset shows the total amount of molecules in the largest nucleus as a function of time, both for three runs with growing nuclei (lines with symbols), and for three runs where the initial seed entirely melts (lines).}
	\label{fig:cubic_seed_analysis}
\end{figure}

We first consider unbiased simulations from cubic ice seeds following the seeding technique. We first determine the equilibrium densities of both the crystal and supercooled liquid phase, and prepare crystalline nuclei at different system size. We then run isobaric simulations of $N=10000$ water molecules at $T=218$~K and find the size that is critical, where the nucleus has the same probability to either grow or shrink. The inset of Fig.~\ref{fig:cubic_seed_analysis} shows the time evolution of different runs for seeds of size $n=110$ molecules, where half the trajectories (continuous blue lines) see the nucleus melting, and half the trajectories (lines with symbols) see the nucleus growing. We thus determine the size $n\sim 110$ to be critical, in excellent agreement with the Umbrella Sampling simulations shown in Fig.~\ref{fig:barriers}, and study the growth process of the initial ice I$_\text{c}$ seed. Via bond orientational order parameters we determine the fraction of cubic ice present in the nucleus, and compute the \emph{cubicity} parameter, defined as $f_\text{cubic}=n_\text{cubic}/(n_\text{cubic}+n_\text{hexagonal})$. The main panel in Fig.~\ref{fig:cubic_seed_analysis} plots the cubicity parameter as a function of the growth steps for the three trajectories that experience crystal growth, showing that cubicity steadily decreases during growth in all cases. These results confirm that at these sizes, crystal growth involves the formation of stacking faults.

\begin{figure}[!t]
	\centering
	\includegraphics[width=8cm]{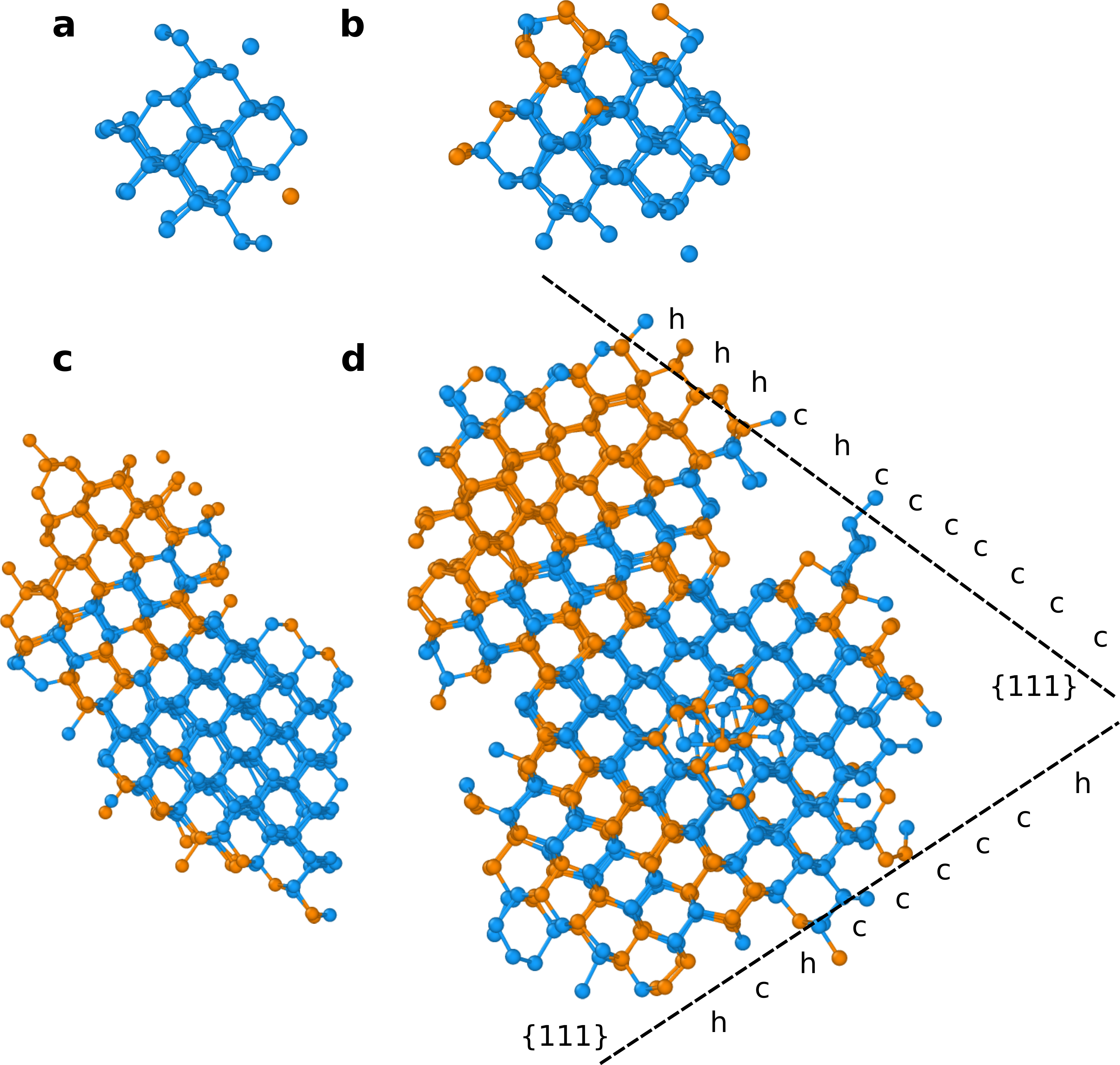}
	\caption{$<110>$ projection of a cubic ice seed composed of $48$ water molecules {\bf a}. Oxygen atoms are connected with bonds, and coloured according to their phase: blue for cubic ice, and orange for hexagonal ice.
		The different stages of growth are shown in {\bf b} ($115$ molecules), {\bf c} ($484$ molecules), and {\bf d} ($1092$ molecules). The growth occurs by random stacking of new $\{111\}$ planes (dashed lines) from the original seed. The symmetry of the stacking along two directions forming an angle of $\cos^{-1}(1/3)$ is indicated with $c$ for cubic, and $h$ for hexagonal.}
	\label{fig:cubic_seed}
\end{figure}

In order to study how the process of stacking disorder proceeds, in Fig.~\ref{fig:cubic_seed} we plot the $\left<110\right>$ projection of the crystal growth process from an initial cubic seed (panel (a)). In the growth stages (panels (b)-(d)) we notice the development of stacking disorder. A crucial difference between Ice I$_\text{sd}$ and Ice I$_\text{c}$ is that the former has stacking disorder only in one direction, while the latter, due to its cubic symmetry, has a family of equivalent planes $\{1,1,1\}=(111),(\bar{1}11),(1\bar{1}1),(11\bar{1})$. The growth along two of such planes is highlighted in Fig.~\ref{fig:cubic_seed}(d), where the order of the stacking is indicated with ``c'' for cubic arrangement, and ``h'' for hexagonal arrangement.

The appearance of random stacking along different crystallographic directions poses a problem with cubic seeds trying to grow. As seen in Fig.~\ref{fig:cubic_seed}, the interface between the two stacking directions is incompatible with crystalline order, i.e. a grain boundary forms. We observe the formation of such grain boundaries in all 10 trajectories that are started with a cubic seed.

\begin{figure}[!t]
	\centering
	\includegraphics[width=8cm]{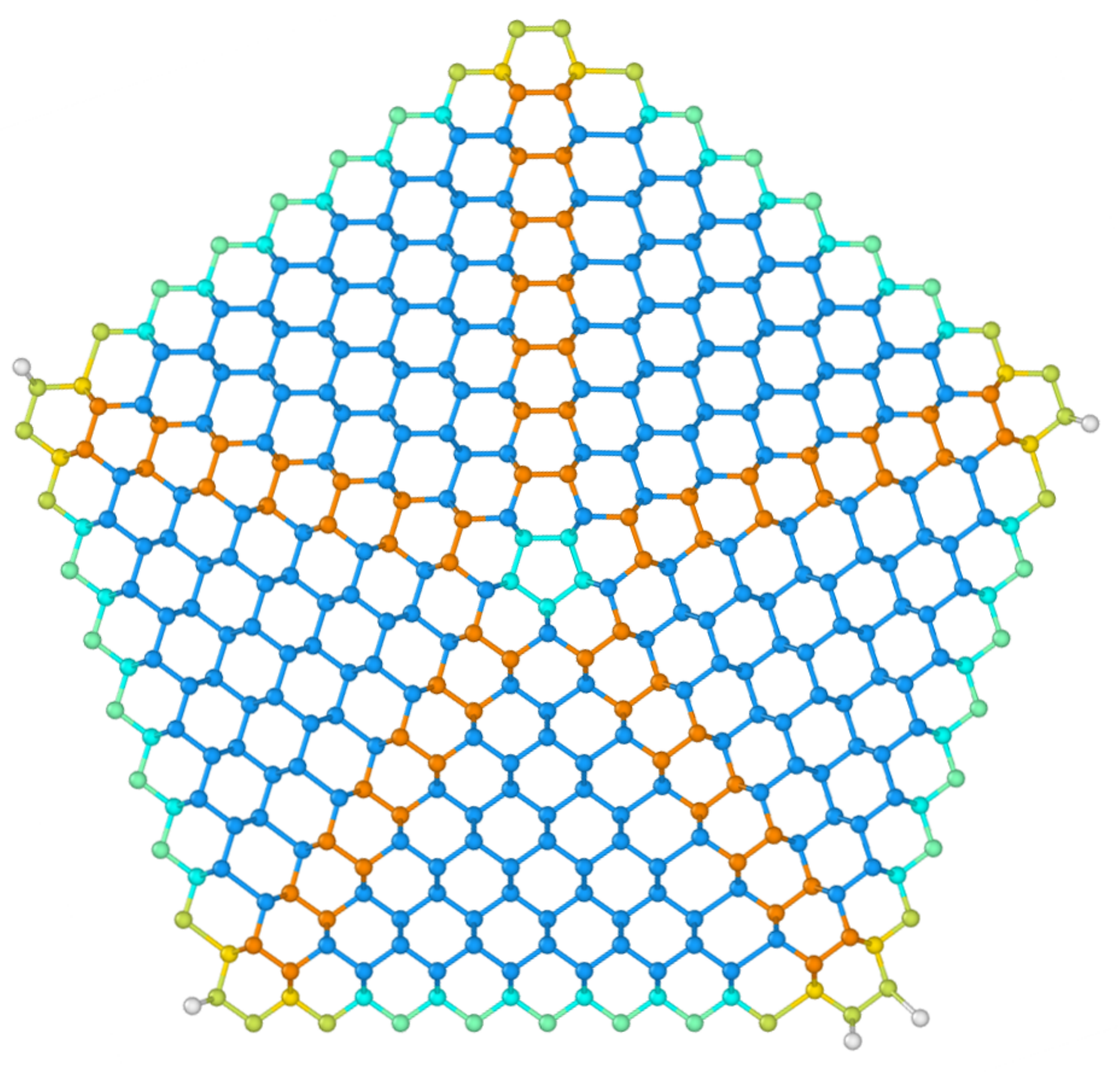}
	\caption{Five-fold symmetric coherent grain boundary formed from cubic ice (blue atoms) with hexagonal stacking (orange atoms) on co-planar \{111\} planes.}
	\label{fig:flower}
\end{figure}

While Ice I$_\text{sd}$ only pays a free energy penalty for stacking, ice I$_\text{c}$ must additional incur in the free energy penalty associated with the formation of grain boundaries. For incoherent boundaries, this penalty would be too high, and cubic ice would then play no role in crystal nucleation. Instead we always observe the formation of a coherent grain boundary with five-fold symmetry. This particular structure was first noted in Ref.~\cite{li2011homogeneous}.
We can already get a hint of this process by considering the angle $\theta$ at which the $\{1,1,1\}$ planes meet in Fig.~\ref{fig:cubic_seed}d, which is $\theta=\cos^{-1}(1/3)$. We notice that this angle, $\theta\sim 70.5^\circ$, is very close to the value $72^\circ$ that corresponds to five-fold symmetry.
The five-fold grain boundary is depicted in Fig.~\ref{fig:flower}. It is a defect in which five cubic ice grains with hexagonal stacking on co-planar \{111\} planes merge together in a coherent way.

It is important to notice that this particular grain only appears as a
consequence of stacking disorder of ice I$_c$, and cannot originate
when nucleation of Ice I$_h$ or I$_\text{sd}$ occurs instead (since
there is only one stacking direction in these forms of ice). The
formation of this grain boundary is thus an indirect piece of evidence
towards nucleation of ice I$_c$ compared to Ice I$_h$ or
I$_\text{sd}$. In the following we will show that nucleation
simulations indicate the presence of this structure in at least $50\%$ of trajectories.

\begin{figure*}[!t]
	\centering
	\includegraphics[width=16cm]{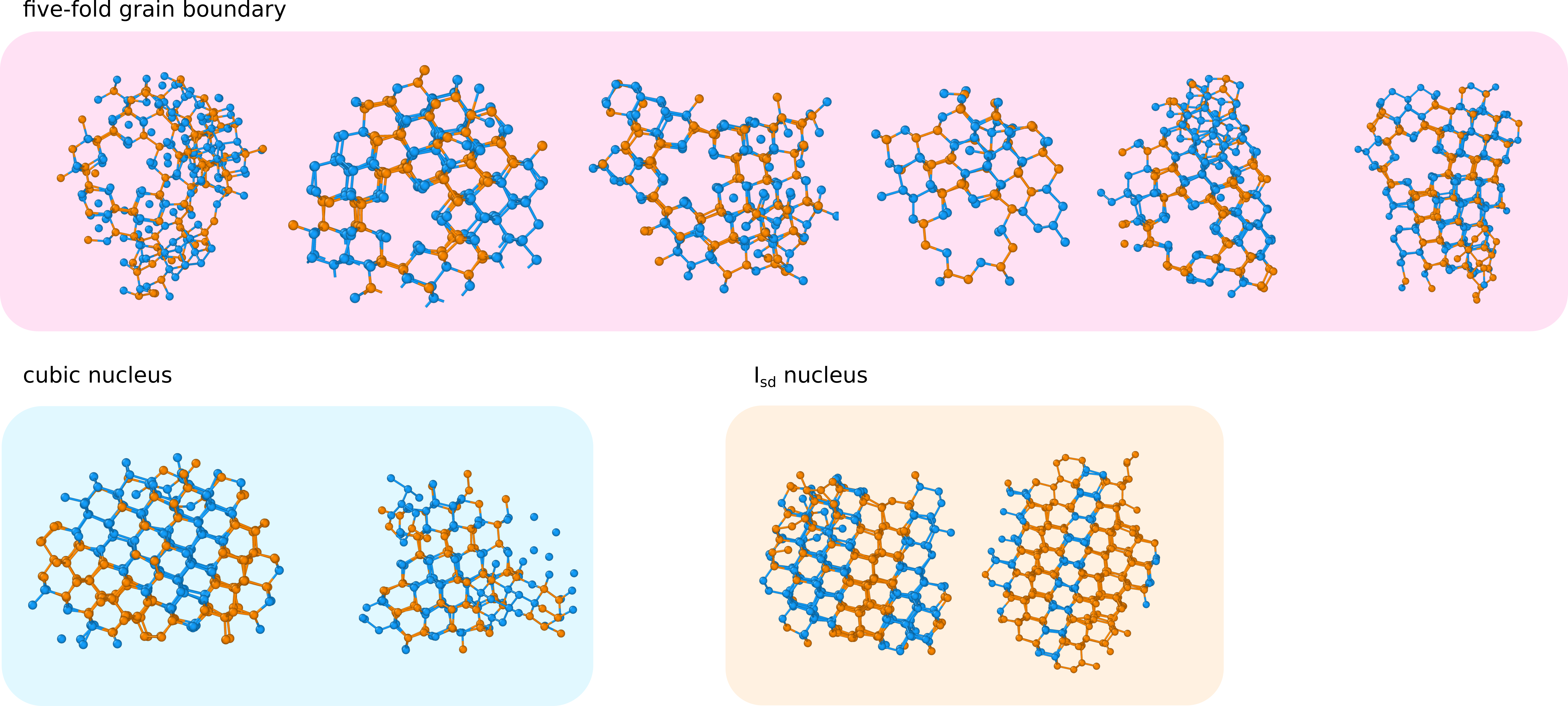}
	\caption{Results from 10 independent runs, for ice nuclei grown in CNT-US simulations at $T=218$~K. Nuclei correspond to sizes $n\sim 300$. The output are classified in three groups: 1) nuclei with five-fold symmetry grain; 2) cubic ice nuclei, in which stacking occurs in at least two-different directions; 3) ice I$_\text{sd}$ nuclei, in which stacking occurs only in one direction (i.e. the basal plane).}
	\label{fig:snapshots}
\end{figure*}

To study nucleation events without seeding, we use CNT-US simulations at $T=218$~K. According to Eq.~\ref{eq:cntus}, the two unknown parameters that need to be determined are the driving force for crystallization, $\Delta\mu$, and the critical nucleus size, $n_c$. The first one is obtained via thermodynamic integration from the melting point of mW ice~I, $T=275.6$~K. 
The parameter $n_c$ is instead determined as follows: several simulations at different values of $n_c$ are run independently, and the fluctuations of the largest nucleus size is monitored. When $n_c$ is close to the true value of the critical size, one observes large fluctuations of the size $n$, with the system sampling from small to large crystalline sizes in the course of the same simulation. From these simulations one can reconstruct the free energy barrier, and thus obtain the true value of $n_c$. We run $10$ independent trajectories at the pre-determined values of $\beta\Delta\mu$ and $n_c$.

Figure \ref{fig:snapshots} shows representative configurations in which only nuclei of size $n\sim 300$ are plotted, and colored according to their phase (orange for the hexagonal crystal, and blue for the cubic crystal). The results reveal three possible outcomes of nucleation. In the top row we see the formation of a five-fold symmetric grain boundary of Fig.~\ref{fig:flower}. In the bottom row (left panel) we see nuclei with stacking disorder in at least two directions, which also originates from the cubic phase. Finally, the bottom row (right panel) shows nuclei with stacking in only one direction, which corresponds to Ice I$_\text{sd}$ nucleation.

The abundance of stacking disorder in two or more directions, and in particular the presence of the five-fold symmetric grain, provides indirect evidence that nucleation indeed involves also the cubic phase. The cubic phase is able to avoid paying the full free-energy cost of an incoherent grain boundary by forming coherent five-fold symmetric grain boundaries, in which the hydrogen bond network is only distorted and not disrupted. 
The five-fold symmetric grain boundary forms a coherent interface at the expense of some elastic energy. We will now attempt a first free energy estimate of the grain boundary cost.

We tried obtaining a free energy barrier for crystals containing the
five-fold grain boundary, but the nucleus fluctuations always favored growth on one of the cubic sides, until eventually the grain boundary disappeared. This could be remedied by introducing additional biasing fields that stabilize the grain boundary, but here instead we use a simpler approach, and leave a more detailed estimate for future studies.

We decrease the temperature to $T=204$~K, in a regime where nucleation occurs spontaneously in the system. We then estimate the free energy barrier of nucleation via mean first passage theory~\cite{wedekind2008kinetic,russo2013interplay}. The mean first passage time $t_\text{fp}(n)$ is defined as the average time elapsed until the appearance of a nucleus of size $n$ in the system. For homogeneous nucleation in the steady-state, the mean first passage time is given by

\begin{equation}\label{eq:mfpt}
t_\text{fp}(n)=\frac{1}{2kV}\left\lbrace 1+\text{erf}\left[c\,(n-n_c)\right]\right\rbrace
\end{equation}

where $k$ is the nucleation rate, $n_c$ is the critical nucleus size, erf is the error function, and $c$ is related to the curvature of the free energy barrier at its maximum (i.e. the Zeldovich factor), $c=\sqrt{\Delta G''(n_c)/k_B\,T}$.

\begin{figure}[!t]
	\centering
	\includegraphics[width=8cm]{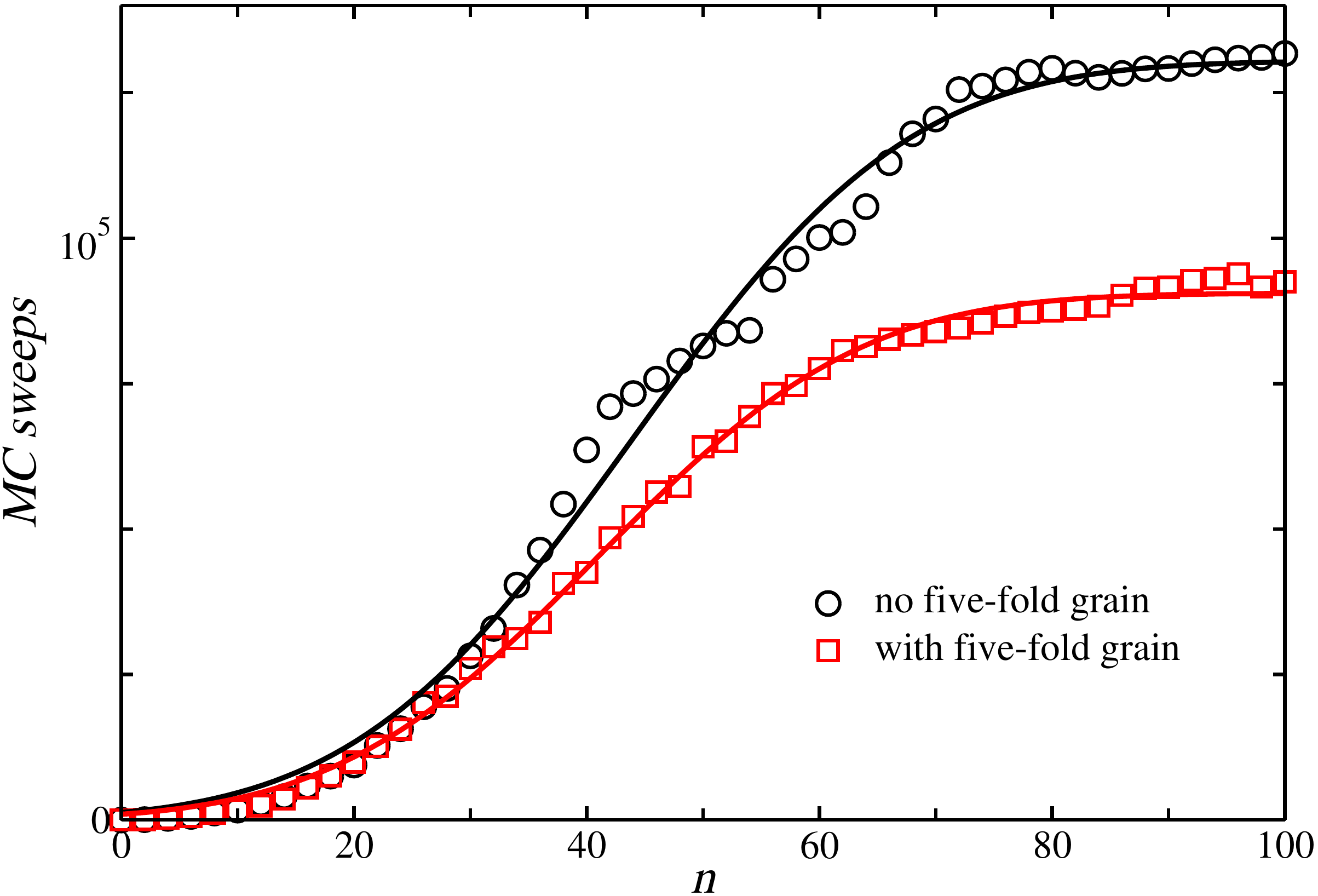}
	\caption{Mean first passage time as a function of nucleus size, $n$. The time is in arbitrary Monte Carlo Sweeps. Symbols are measured averaging mean first passage times over 100 trajectories of homogeneous nucleation at $T=204$~K, divided in two groups depending on whether they form the five-fold grain boundary or not. The lines are fits according to Eq.~\ref{eq:mfpt}.}
	\label{fig:mfpt}
\end{figure}

We run $100$ homogeneous nucleation trajectories at $T=204$~K until
the appearance of a super-critical nucleus. We then divide the
trajectories in two sets depending on whether the nucleus has formed a
five-fold symmetric grain boundary or not. To identify the grain
boundary we use bond orientational order parameters to detect the
atoms in the inner five-membered ring of the defect (depicted in cyan
in Fig.~\ref{fig:flower}). These atoms are identified when the
following two conditions are satisfied: $Q_4<0.1$ and
$Q_{12}>0.2$. Both $Q_4$ and $Q_{12}$ refer to coarse-grained bond
orientational parameters~\cite{lechner2008accurate} (see Ref.~\cite{tanaka_review} for a review of these methods).

The computed mean first passage times for nuclei of size up to $n=100$
are plotted in Fig.~\ref{fig:mfpt}, distinguishing between nuclei with
the five-fold grain (red squares) and without it (black circles). We
notice a very clear distinction between the two different sets, with
nuclei with the five-fold grain having shorter mean first passage
times. This already reveals that the nucleation barrier for these
states is lower compared to the nuclei without the grain. By using
Eq.~\ref{eq:mfpt} we fit both curves, and display the results as
continuous lines in Fig.~\ref{fig:mfpt}. The results for the fit are
(the superscript $5f$ refers to structures with the five-fold grain
boundary): $n_c^{5f}=41$, $n_c=44$, $k^{5f}/k=1.44$, $c^{5f}=0.040$,
and $c=0.038$. We notice that both the critical size and the free
energy curvature are about the same in both cases, while the main
difference is in the nucleation rate. We can thus estimate the free
energy difference between the nuclei with and without a five-fold
grain boundary $\Delta G^{5f}$ via the following relation:
$\beta\Delta G^{5f}=-\log (k^{5f}/k)=-0.37$, or $\Delta G^{5f}=-0.6
KJ/mol$. As discussed above, for small nuclei (here $n_c\sim 40$) the
energetic cost of forming the five-fold grain boundary is very small,
so the free energy advantage comes from the entropic gain in the
coherent stacking in multiple directions. For much bigger nuclei, we
expect the strain on the bonds far from the grain core to make $\Delta G^{5f}$ big and positive.

Overall, the abundance of five-fold grain boundaries in both direct nucleation and umbrella sampling simulations, and the absence of a significant free-energy penalty for small nuclei, shows that the nucleation of the cubic phase occurs distinctly and at least with the same probability of the nucleation of Ice I$_h$ and I$_\text{sd}$. The free-energy cost of the five-fold grain boundary is expected to grow considerably with increasing nucleus size, due to the strain in the bonds far from the grain core, and this is likely to stop the stacking of cubic ice in multiple directions. It should thus be more advantageous, for cubic seeds above a certain size, to have stacking disorder in only one direction, thus converting them into Ice I$_\text{sd}$.
This suggests another mechanism, besides just the difference in bulk free energy between the cubic and hexagonal phases, for the conversion of cubic ice in Ice I$_\text{sd}$, that is governed by the size dependence of the free energy cost of the five-fold grain boundary.

\subsubsection{Growth of Ice 0}

\begin{figure}[!t]
	\centering
	\includegraphics[width=8cm]{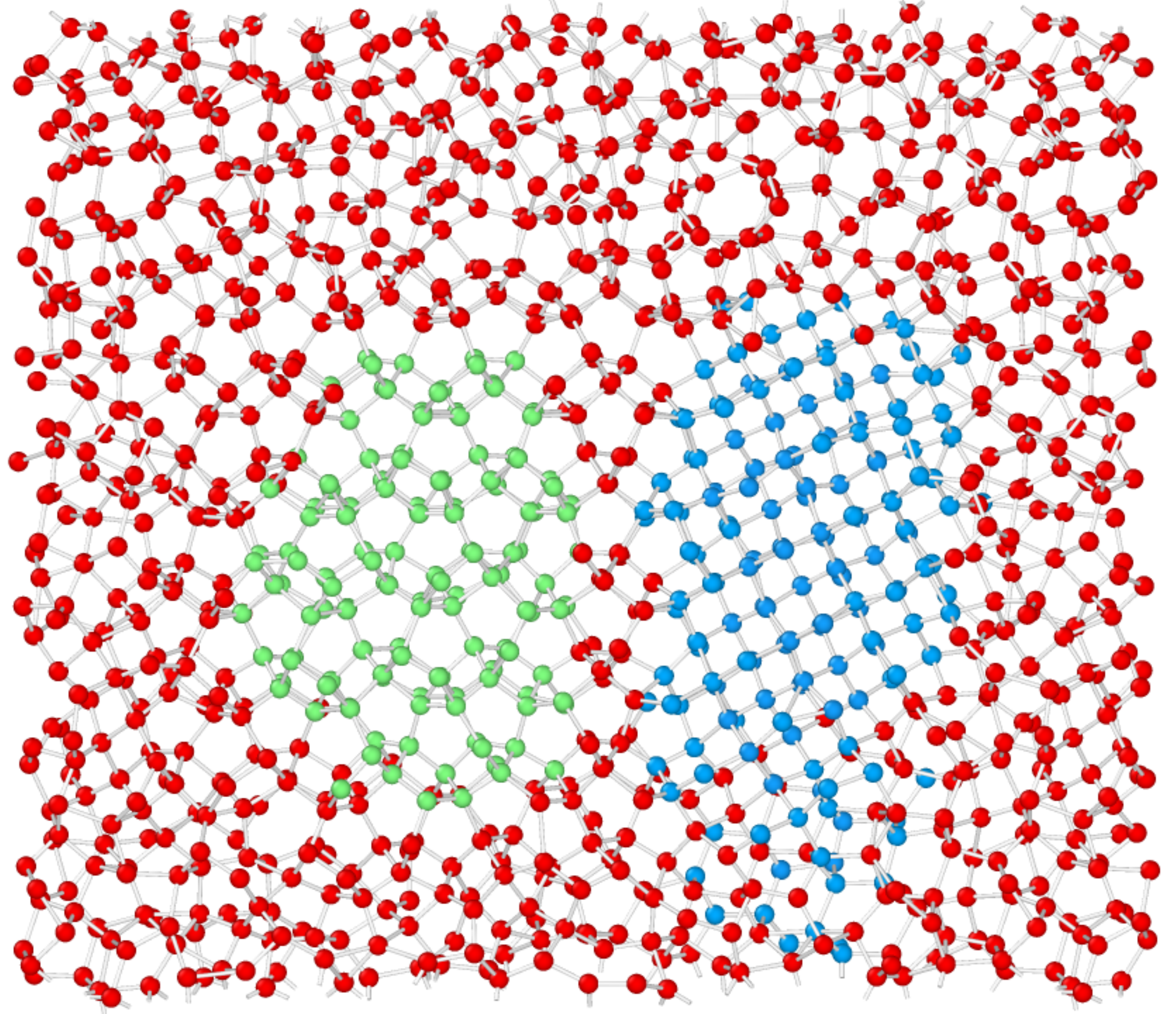}
	\caption{Seeding of an ice $0$ nucleus at $T=208$~K. Ice~0 particles are colored in green, and the cubic ice that cross nucleates on the surface of ice~$0$ is colored in blue.}
	\label{fig:ice0_seeding}
\end{figure}

In this section we consider the growth of Ice 0 nuclei with the seeding technique at $T=208$~K. We run simulations with an Ice 0 seed and determine the critical size to be around $n_c=400$ molecules. We then run 100 independent simulations and focus on the trajectories where the size of the nucleus grows. We observe two different scenarios, occurring with similar probability (number of trajectories that follow one growth mechanism or the other): i) growth of a defect free ice 0 crystal; ii) cross-nucleation of ice I$_c$ on the ice $0$ seed. A snapshot from a configuration following scenario ii) is shown in Fig.~\ref{fig:ice0_seeding}. While the first scenario is the expected growth mode for a metastable nucleus, the observation of cross nucleation of cubic ice poses some interesting questions. Firstly, cross nucleation at such small sizes implies that the surface penalty between ice $0$ and cubic ice should be considerably low, which is unexpected given the differences between the two different crystalline structures. Secondly, the question remains why only the cubic form is cross-nucleated, and not the hexagonal phase.

\begin{figure}[!t]
	\centering
	\includegraphics[width=8cm]{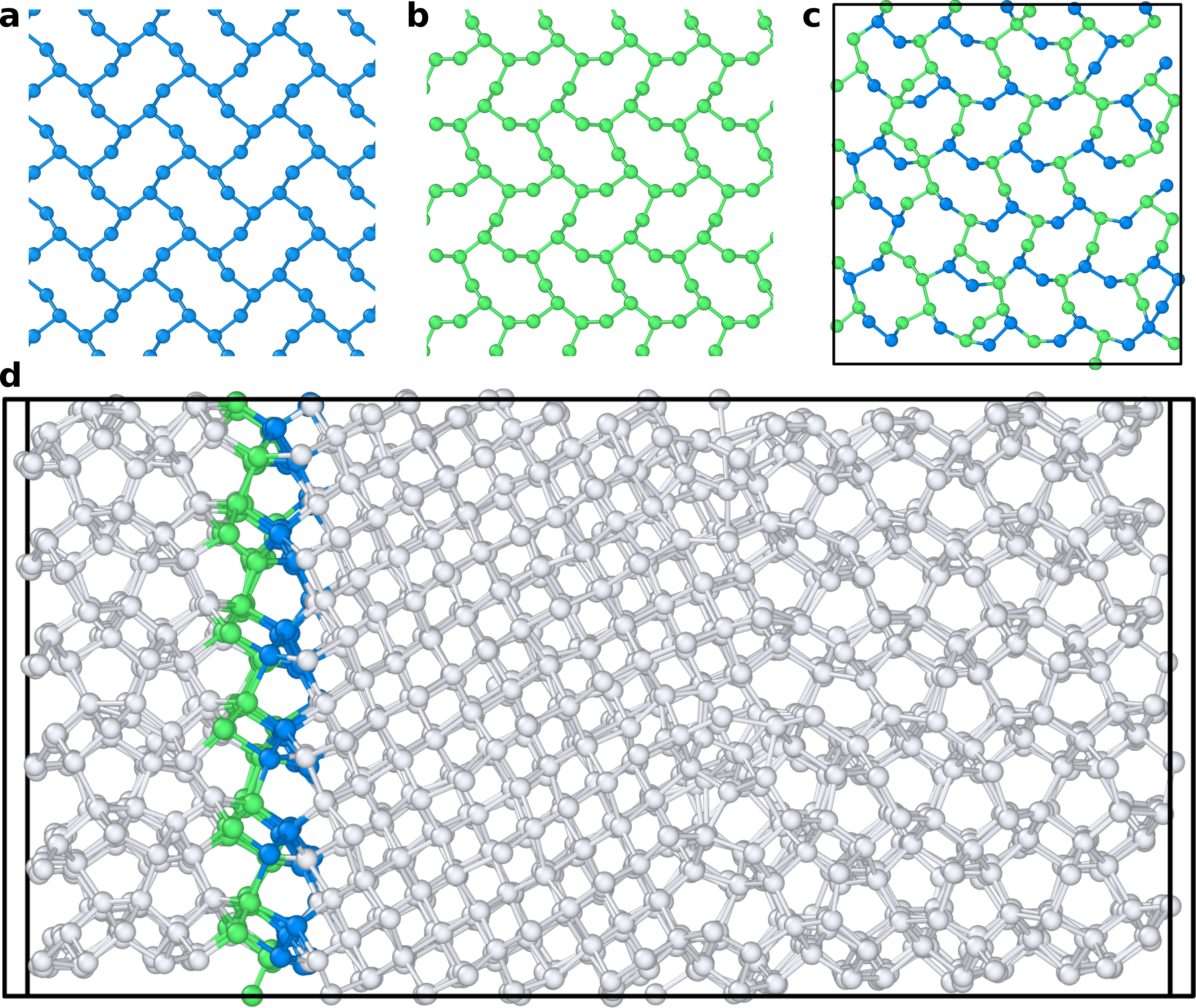}
	\caption{Ice0/cubic interface. Surface projections of slabs of thickness $4.3${\AA} for the (310) plane of cubic ice ({\bf a}) and the (100) plane of Ice~$0$ ({\bf b}). Panel {\bf c} shows the coherent interface between the two crystals, and {\bf d} the direct coexistence box. The slab thickness corresponds to the range of the mW potential.}
	\label{fig:interface}
\end{figure}

To answers these questions, we run growth simulations with the direct coexistence method, where a slab of Ice 0 is prepared in contact with the supercooled liquid phase. Also in this case we observe the cross-nucleation of defect free cubic ice from the (100) plane of Ice 0. The (100) plane is the same that minimizes the cluster energy, as we described in Fig.~\ref{fig:snapshots_N40}.

In Fig.~\ref{fig:interface} we plot a snapshot from such configurations. Panel (d) shows that a coherent grain boundary is formed between the face (100) of Ice 0 (panel (b)) and the (310) face of cubic ice (panel (a)). The two faces join coherently (without breaking of hydrogen bonds), as shown for example in panel (c). From these equilibrium simulations we measure a very small strain (around $2\%$) of the cubic unit cell compared to its equilibrium value close to the interface. This signals a very low surface free energy cost of the defect. The cost of coherent grain boundaries between two solid surfaces of the same material (mW) originates almost entirely from the strain energy. We can estimate this cost as
$$
\Delta E_\text{c0}=e_I-\frac{e_0+e_c}{2}
$$
where $e_I$ is the energy of an interface atom, while $e_0$ and $e_c$ are respectively the energy of atoms in the bulk Ice 0 and I$_c$ phases. We use our simulations to estimate these energies as: $e_I=94.71$~kJ/mol, $e_0=96.23$~kJ/mol, $e_c=97.18$~kJ/mol, for $\Delta E_\text{c0}=1.99$~kJ/mol.
$e_I$ is obtained by measuring the average energy of the molecules at the interface between ice 0 and cubic ice from the direct-coexistence simulations . Since the interface is coherent, it is very sharp (one particle width), and there is no ambiguity in the selection of the correct particles.
{$\gamma_\text{c0}$ is obtained by assuming (as usually done for coherent interfaces) that the dominant contribution to the free energy is energetic (given by $\Delta E_\text{c0}$), and then dividing the total surface energy for the area of the surface.
This leads to a surface energy of stacking between the (100) plane of Ice 0 and the (310) plane of cubic ice of $\gamma_\text{c0}=29.67$~mJ/m$^2$. This energy is much larger than the stacking surface energy of cubic and hexagonal layers in Ice I (that we estimated earlier as $\gamma_\text{sf}\sim 0.16$~mJ/m$^2$ at similar thermodynamic conditions), but is well below the limit of $200$~mJ/m$^2$ that is conventionally used to separate coherent from semi-coherent grains.

We observe that the low cost of the grain boundary also makes it difficult to obtain free energy barriers of Ice 0 (as shown for the other phases in Fig.~\ref{fig:barriers}), as we always observe the appearance of cubic ice in the Umbrella Sampling simulations.

One important consequence of this grain boundary is the fact that the (100) face of Ice 0 appears to be an ideal template for the cross nucleation of Ice I$_c$. Our simulations always show the absence of hexagonal stacking, and the formation of a crystal with cubicity of $c=1$. The reason for this is that the grain boundary involves a high-order face of cubic ice, the (310) face, and so the new cubic phase grows along a direction incompatible with stacking faults.

\section{Conclusions}

In this article we have focused on crystalline clusters in the mW model of water. In the spirit of Frank's original argument, we have shown that there is a competition between different crystalline clusters, and that the most energetic clusters at small sizes include five-membered rings. Here we have demonstrated this with the Ice 0 structure, which is a metastable crystal for the mW potential. Five-membered rings allow small clusters to increase the ratio between number of bonds and number of particles in the cluster, without significant penalties to the bond energy. As the clusters grow in size, six-membered rings (as found in the stable Ice I forms) become favourable due to their optimal bond energy.
Frustration effects between local structures can suppress the nucleation of crystals, and are thus linked to glass-forming ability~\cite{ronceray2017suppression,russo2018glass,wei2018assessing,tanaka_review}. The competition between different local structures also plays an important role in polymorph selection, i.e. which metastable phase is nucleated.

To study the growth of crystalline clusters in supercooled conditions
we have used both direct simulations, seeding, and umbrella sampling
biasing techniques. In particular we have highlighted the role of
grain boundaries in the growth process of the clusters. The lowest
energy grain boundary is the stacking fault between hexagonal and
cubic layers that can appear both on the basal plane of Ice I$_h$ and
the (111) plane of Ice I$_c$, which we estimate as
$\gamma_\text{sf}=0.16\pm 0.05$mJm$^{-2}$ at $T=218$~K. Given the negligible
bulk free energy difference between I$_c$ and I$_h$, this predicts, at
low temperatures, the formation of Ice I$_\text{sd}$ as the first
stage of nucleation, favored by the configurational entropy associated
with its disordered stacking sequence. But our results show also that cubic ice
still plays an important role in ice nucleation, thanks to another
coherent grain boundary. Differently from Ice
I$_h$ and I$_\text{sd}$, cubic ices can generate stacking faults in
four different directions. The associated cost of intersecting these
stacking faults, which would otherwise suppress the nucleation of
small cubic crystallites, is instead accommodated by the formation of
a coherent five-fold symmetric grain boundary. At deep supercooled
conditions ($T=208$~K) we estimate the free energy difference between
nuclei with and without the five-fold grain boundary to be of the
order of $0.37\,k_BT$ in favor of the first ones. The abundance of configurations with the five-fold grain boundary is an indirect measure of the importance of the nucleation of cubic ice. Due to its five-fold symmetric nature, the grain boundary cost will increase rapidly with nucleus size, due to the bond strain far from the core of the boundary. We can thus expect that the grain boundary will also be a contributing factor in selecting one growth direction for cubic ice, which would then gradually transform in Ice I$_\text{sd}$, in accordance with Ostwald's step rule.

We have also considered seeding simulations of Ice 0 nuclei, and found that its growth is also controlled by a coherent grain boundary. The new grain boundary is formed between the (100) face of Ice 0 and the (310) face of cubic ice. We estimated the surface free energy of the grain boundary to be $\gamma_\text{c0}=29.67$~mJ/m$^2$. The fact that the grain boundary involves a high order face of cubic ice, makes the (100) face of Ice 0 an ideal template for the cross-nucleation of pure cubic ice. Indeed our simulations show growth of cubic ice without stacking faults.

\vspace{0.3cm}
\noindent
\textbf{Acknowledgements}\linebreak 
J.R. and F.L. acknowledge support from the European Research Council
Grant DLV-759187. J.R. acknowledges support from the Royal Society
University Research Fellowship and a travel support from Institute
of Industrial Science (IIS), University of Tokyo for a
short stay at IIS. This study was partially supported by Grants-in-Aid for Specially Promoted Research (Grant No. JP25000002) and Scientific Research (A) (18H03675) from the Japan Society for the Promotion of Science (JSPS).


\begin{thebibliography}{100}%
	\makeatletter
	\providecommand \@ifxundefined [1]{%
		\@ifx{#1\undefined}
	}%
	\providecommand \@ifnum [1]{%
		\ifnum #1\expandafter \@firstoftwo
		\else \expandafter \@secondoftwo
		\fi
	}%
	\providecommand \@ifx [1]{%
		\ifx #1\expandafter \@firstoftwo
		\else \expandafter \@secondoftwo
		\fi
	}%
	\providecommand \natexlab [1]{#1}%
	\providecommand \enquote  [1]{``#1''}%
	\providecommand \bibnamefont  [1]{#1}%
	\providecommand \bibfnamefont [1]{#1}%
	\providecommand \citenamefont [1]{#1}%
	\providecommand \href@noop [0]{\@secondoftwo}%
	\providecommand \href [0]{\begingroup \@sanitize@url \@href}%
	\providecommand \@href[1]{\@@startlink{#1}\@@href}%
	\providecommand \@@href[1]{\endgroup#1\@@endlink}%
	\providecommand \@sanitize@url [0]{\catcode `\\12\catcode `\$12\catcode
		`\&12\catcode `\#12\catcode `\^12\catcode `\_12\catcode `\%12\relax}%
	\providecommand \@@startlink[1]{}%
	\providecommand \@@endlink[0]{}%
	\providecommand \url  [0]{\begingroup\@sanitize@url \@url }%
	\providecommand \@url [1]{\endgroup\@href {#1}{\urlprefix }}%
	\providecommand \urlprefix  [0]{URL }%
	\providecommand \Eprint [0]{\href }%
	\providecommand \doibase [0]{http://dx.doi.org/}%
	\providecommand \selectlanguage [0]{\@gobble}%
	\providecommand \bibinfo  [0]{\@secondoftwo}%
	\providecommand \bibfield  [0]{\@secondoftwo}%
	\providecommand \translation [1]{[#1]}%
	\providecommand \BibitemOpen [0]{}%
	\providecommand \bibitemStop [0]{}%
	\providecommand \bibitemNoStop [0]{.\EOS\space}%
	\providecommand \EOS [0]{\spacefactor3000\relax}%
	\providecommand \BibitemShut  [1]{\csname bibitem#1\endcsname}%
	\let\auto@bib@innerbib\@empty
	\bibitem [{\citenamefont {Frank}(1952)}]{frank1952supercooling}%
	\BibitemOpen
	\bibfield  {author} {\bibinfo {author} {\bibfnamefont {F.~C.}\ \bibnamefont
			{Frank}},\ }\href@noop {} {\bibfield  {journal} {\bibinfo  {journal}
			{Proceedings of the Royal Society of London. Series A, Mathematical and
				Physical Sciences}\ ,\ \bibinfo {pages} {43}} (\bibinfo {year}
		{1952})}\BibitemShut {NoStop}%
	\bibitem [{\citenamefont {Tanaka}\ \emph {et~al.}(2019)\citenamefont {Tanaka},
		\citenamefont {Shi}, \citenamefont {Tong},\ and\ \citenamefont
		{Russo}}]{tanaka_review}%
	\BibitemOpen
	\bibfield  {author} {\bibinfo {author} {\bibfnamefont {H.}~\bibnamefont
			{Tanaka}}, \bibinfo {author} {\bibfnamefont {R.}~\bibnamefont {Shi}},
		\bibinfo {author} {\bibfnamefont {H.}~\bibnamefont {Tong}}, \ and\ \bibinfo
		{author} {\bibfnamefont {J.}~\bibnamefont {Russo}},\ }\href@noop {}
	{\bibfield  {journal} {\bibinfo  {journal} {Nature Review Physics}\ }\textbf
		{\bibinfo {volume} {1}},\ \bibinfo {pages} {333} (\bibinfo {year}
		{2019})}\BibitemShut {NoStop}%
	\bibitem [{\citenamefont {Tanaka}(2012)}]{tanaka2012bond}%
	\BibitemOpen
	\bibfield  {author} {\bibinfo {author} {\bibfnamefont {H.}~\bibnamefont
			{Tanaka}},\ }\href@noop {} {\bibfield  {journal} {\bibinfo  {journal} {Eur.
				Phys. J E}\ }\textbf {\bibinfo {volume} {35}},\ \bibinfo {pages} {113}
		(\bibinfo {year} {2012})}\BibitemShut {NoStop}%
	\bibitem [{\citenamefont {Shintani}\ and\ \citenamefont
		{Tanaka}(2006)}]{shintani2006frustration}%
	\BibitemOpen
	\bibfield  {author} {\bibinfo {author} {\bibfnamefont {H.}~\bibnamefont
			{Shintani}}\ and\ \bibinfo {author} {\bibfnamefont {H.}~\bibnamefont
			{Tanaka}},\ }\href@noop {} {\bibfield  {journal} {\bibinfo  {journal} {Nat.
				Phys.}\ }\textbf {\bibinfo {volume} {2}},\ \bibinfo {pages} {200} (\bibinfo
		{year} {2006})}\BibitemShut {NoStop}%
	\bibitem [{\citenamefont {Leocmach}\ and\ \citenamefont
		{Tanaka}(2012)}]{leocmach2012roles}%
	\BibitemOpen
	\bibfield  {author} {\bibinfo {author} {\bibfnamefont {M.}~\bibnamefont
			{Leocmach}}\ and\ \bibinfo {author} {\bibfnamefont {H.}~\bibnamefont
			{Tanaka}},\ }\href@noop {} {\bibfield  {journal} {\bibinfo  {journal} {Nat.
				Commun.}\ }\textbf {\bibinfo {volume} {3}},\ \bibinfo {pages} {974} (\bibinfo
		{year} {2012})}\BibitemShut {NoStop}%
	\bibitem [{\citenamefont {Royall}\ and\ \citenamefont
		{Williams}(2015)}]{royall2015role}%
	\BibitemOpen
	\bibfield  {author} {\bibinfo {author} {\bibfnamefont {C.~P.}\ \bibnamefont
			{Royall}}\ and\ \bibinfo {author} {\bibfnamefont {S.~R.}\ \bibnamefont
			{Williams}},\ }\href@noop {} {\bibfield  {journal} {\bibinfo  {journal}
			{Physics Reports}\ }\textbf {\bibinfo {volume} {560}},\ \bibinfo {pages} {1}
		(\bibinfo {year} {2015})}\BibitemShut {NoStop}%
	\bibitem [{\citenamefont {Royall}\ \emph {et~al.}(2018)\citenamefont {Royall},
		\citenamefont {Turci}, \citenamefont {Tatsumi}, \citenamefont {Russo},\ and\
		\citenamefont {Robinson}}]{royall2018race}%
	\BibitemOpen
	\bibfield  {author} {\bibinfo {author} {\bibfnamefont {C.~P.}\ \bibnamefont
			{Royall}}, \bibinfo {author} {\bibfnamefont {F.}~\bibnamefont {Turci}},
		\bibinfo {author} {\bibfnamefont {S.}~\bibnamefont {Tatsumi}}, \bibinfo
		{author} {\bibfnamefont {J.}~\bibnamefont {Russo}}, \ and\ \bibinfo {author}
		{\bibfnamefont {J.}~\bibnamefont {Robinson}},\ }\href@noop {} {\bibfield
		{journal} {\bibinfo  {journal} {Journal of Physics: Condensed Matter}\
		}\textbf {\bibinfo {volume} {30}},\ \bibinfo {pages} {363001} (\bibinfo
		{year} {2018})}\BibitemShut {NoStop}%
	\bibitem [{\citenamefont {Russo}\ \emph
		{et~al.}(2018{\natexlab{a}})\citenamefont {Russo}, \citenamefont {Romano},\
		and\ \citenamefont {Tanaka}}]{russo2018glass}%
	\BibitemOpen
	\bibfield  {author} {\bibinfo {author} {\bibfnamefont {J.}~\bibnamefont
			{Russo}}, \bibinfo {author} {\bibfnamefont {F.}~\bibnamefont {Romano}}, \
		and\ \bibinfo {author} {\bibfnamefont {H.}~\bibnamefont {Tanaka}},\
	}\href@noop {} {\bibfield  {journal} {\bibinfo  {journal} {Physical Review
			X}\ }\textbf {\bibinfo {volume} {8}},\ \bibinfo {pages} {021040} (\bibinfo
	{year} {2018}{\natexlab{a}})}\BibitemShut {NoStop}%
\bibitem [{\citenamefont {Russo}\ and\ \citenamefont
	{Tanaka}(2016)}]{russo2016crystal}%
\BibitemOpen
\bibfield  {author} {\bibinfo {author} {\bibfnamefont {J.}~\bibnamefont
		{Russo}}\ and\ \bibinfo {author} {\bibfnamefont {H.}~\bibnamefont {Tanaka}},\
}\href@noop {} {\bibfield  {journal} {\bibinfo  {journal} {The Journal of
		Chemical Physics}\ }\textbf {\bibinfo {volume} {145}},\ \bibinfo {pages}
{211801} (\bibinfo {year} {2016})}\BibitemShut {NoStop}%
\bibitem [{\citenamefont {Tanaka}(2000)}]{tanaka2000simple}%
\BibitemOpen
\bibfield  {author} {\bibinfo {author} {\bibfnamefont {H.}~\bibnamefont
		{Tanaka}},\ }\href {\doibase 10.1063/1.480609} {\bibfield  {journal}
	{\bibinfo  {journal} {J. Chem. Phys.}\ }\textbf {\bibinfo {volume} {112}},\
	\bibinfo {pages} {799} (\bibinfo {year} {2000})}\BibitemShut {NoStop}%
\bibitem [{\citenamefont {Russo}\ and\ \citenamefont
	{Tanaka}(2014)}]{russo2014understanding}%
\BibitemOpen
\bibfield  {author} {\bibinfo {author} {\bibfnamefont {J.}~\bibnamefont
		{Russo}}\ and\ \bibinfo {author} {\bibfnamefont {H.}~\bibnamefont {Tanaka}},\
}\href@noop {} {\bibfield  {journal} {\bibinfo  {journal} {Nature
		communications}\ }\textbf {\bibinfo {volume} {5}},\ \bibinfo {pages} {3556}
(\bibinfo {year} {2014})}\BibitemShut {NoStop}%
\bibitem [{\citenamefont {Shi}\ \emph {et~al.}(2018{\natexlab{a}})\citenamefont
	{Shi}, \citenamefont {Russo},\ and\ \citenamefont {Tanaka}}]{shi2018common}%
\BibitemOpen
\bibfield  {author} {\bibinfo {author} {\bibfnamefont {R.}~\bibnamefont
		{Shi}}, \bibinfo {author} {\bibfnamefont {J.}~\bibnamefont {Russo}}, \ and\
	\bibinfo {author} {\bibfnamefont {H.}~\bibnamefont {Tanaka}},\ }\href@noop {}
{\bibfield  {journal} {\bibinfo  {journal} {The Journal of chemical physics}\
	}\textbf {\bibinfo {volume} {149}},\ \bibinfo {pages} {224502} (\bibinfo
	{year} {2018}{\natexlab{a}})}\BibitemShut {NoStop}%
\bibitem [{\citenamefont {Shi}\ \emph {et~al.}(2018{\natexlab{b}})\citenamefont
	{Shi}, \citenamefont {Russo},\ and\ \citenamefont {Tanaka}}]{shi2018origin}%
\BibitemOpen
\bibfield  {author} {\bibinfo {author} {\bibfnamefont {R.}~\bibnamefont
		{Shi}}, \bibinfo {author} {\bibfnamefont {J.}~\bibnamefont {Russo}}, \ and\
	\bibinfo {author} {\bibfnamefont {H.}~\bibnamefont {Tanaka}},\ }\href@noop {}
{\bibfield  {journal} {\bibinfo  {journal} {Proceedings of the National
			Academy of Sciences}\ }\textbf {\bibinfo {volume} {115}},\ \bibinfo {pages}
	{9444} (\bibinfo {year} {2018}{\natexlab{b}})}\BibitemShut {NoStop}%
\bibitem [{\citenamefont {Poole}\ \emph {et~al.}(1992)\citenamefont {Poole},
	\citenamefont {Sciortino}, \citenamefont {Essmann},\ and\ \citenamefont
	{Stanley}}]{poole1992phase}%
\BibitemOpen
\bibfield  {author} {\bibinfo {author} {\bibfnamefont {P.~H.}\ \bibnamefont
		{Poole}}, \bibinfo {author} {\bibfnamefont {F.}~\bibnamefont {Sciortino}},
	\bibinfo {author} {\bibfnamefont {U.}~\bibnamefont {Essmann}}, \ and\
	\bibinfo {author} {\bibfnamefont {H.~E.}\ \bibnamefont {Stanley}},\
}\href@noop {} {\bibfield  {journal} {\bibinfo  {journal} {Nature}\ }\textbf
{\bibinfo {volume} {360}},\ \bibinfo {pages} {324} (\bibinfo {year}
{1992})}\BibitemShut {NoStop}%
\bibitem [{\citenamefont {Mishima}\ and\ \citenamefont
	{Stanley}(1998)}]{Mishima1998}%
\BibitemOpen
\bibfield  {author} {\bibinfo {author} {\bibfnamefont {O.}~\bibnamefont
		{Mishima}}\ and\ \bibinfo {author} {\bibfnamefont {H.~E.}\ \bibnamefont
		{Stanley}},\ }\href {\doibase 10.1038/24540} {\bibfield  {journal} {\bibinfo
		{journal} {Nature}\ }\textbf {\bibinfo {volume} {396}},\ \bibinfo {pages}
	{329} (\bibinfo {year} {1998})}\BibitemShut {NoStop}%
\bibitem [{\citenamefont {Gallo}\ and\ \citenamefont
	{Sciortino}(2012)}]{gallo2012ising}%
\BibitemOpen
\bibfield  {author} {\bibinfo {author} {\bibfnamefont {P.}~\bibnamefont
		{Gallo}}\ and\ \bibinfo {author} {\bibfnamefont {F.}~\bibnamefont
		{Sciortino}},\ }\href@noop {} {\bibfield  {journal} {\bibinfo  {journal}
		{Physical review letters}\ }\textbf {\bibinfo {volume} {109}},\ \bibinfo
	{pages} {177801} (\bibinfo {year} {2012})}\BibitemShut {NoStop}%
\bibitem [{\citenamefont {Gallo}\ \emph {et~al.}(2016)\citenamefont {Gallo},
	\citenamefont {Amann-Winkel}, \citenamefont {Angell}, \citenamefont
	{Anisimov}, \citenamefont {Caupin}, \citenamefont {Chakravarty},
	\citenamefont {Lascaris}, \citenamefont {Loerting}, \citenamefont
	{Panagiotopoulos}, \citenamefont {Russo} \emph {et~al.}}]{gallo2016water}%
\BibitemOpen
\bibfield  {author} {\bibinfo {author} {\bibfnamefont {P.}~\bibnamefont
		{Gallo}}, \bibinfo {author} {\bibfnamefont {K.}~\bibnamefont {Amann-Winkel}},
	\bibinfo {author} {\bibfnamefont {C.~A.}\ \bibnamefont {Angell}}, \bibinfo
	{author} {\bibfnamefont {M.~A.}\ \bibnamefont {Anisimov}}, \bibinfo {author}
	{\bibfnamefont {F.}~\bibnamefont {Caupin}}, \bibinfo {author} {\bibfnamefont
		{C.}~\bibnamefont {Chakravarty}}, \bibinfo {author} {\bibfnamefont
		{E.}~\bibnamefont {Lascaris}}, \bibinfo {author} {\bibfnamefont
		{T.}~\bibnamefont {Loerting}}, \bibinfo {author} {\bibfnamefont {A.~Z.}\
		\bibnamefont {Panagiotopoulos}}, \bibinfo {author} {\bibfnamefont
		{J.}~\bibnamefont {Russo}},  \emph {et~al.},\ }\href@noop {} {\bibfield
	{journal} {\bibinfo  {journal} {Chemical reviews}\ }\textbf {\bibinfo
		{volume} {116}},\ \bibinfo {pages} {7463} (\bibinfo {year}
	{2016})}\BibitemShut {NoStop}%
\bibitem [{\citenamefont {Handle}\ \emph {et~al.}(2017)\citenamefont {Handle},
	\citenamefont {Loerting},\ and\ \citenamefont
	{Sciortino}}]{handle2017supercooled}%
\BibitemOpen
\bibfield  {author} {\bibinfo {author} {\bibfnamefont {P.~H.}\ \bibnamefont
		{Handle}}, \bibinfo {author} {\bibfnamefont {T.}~\bibnamefont {Loerting}}, \
	and\ \bibinfo {author} {\bibfnamefont {F.}~\bibnamefont {Sciortino}},\
}\href@noop {} {\bibfield  {journal} {\bibinfo  {journal} {Proceedings of the
		National Academy of Sciences}\ }\textbf {\bibinfo {volume} {114}},\ \bibinfo
{pages} {13336} (\bibinfo {year} {2017})}\BibitemShut {NoStop}%
\bibitem [{\citenamefont {Martelli}(2019)}]{martelli2019unravelling}%
\BibitemOpen
\bibfield  {author} {\bibinfo {author} {\bibfnamefont {F.}~\bibnamefont
		{Martelli}},\ }\href@noop {} {\bibfield  {journal} {\bibinfo  {journal} {The
			Journal of chemical physics}\ }\textbf {\bibinfo {volume} {150}},\ \bibinfo
	{pages} {094506} (\bibinfo {year} {2019})}\BibitemShut {NoStop}%
\bibitem [{\citenamefont {Russo}\ \emph {et~al.}(2014)\citenamefont {Russo},
	\citenamefont {Romano},\ and\ \citenamefont {Tanaka}}]{russo2014new}%
\BibitemOpen
\bibfield  {author} {\bibinfo {author} {\bibfnamefont {J.}~\bibnamefont
		{Russo}}, \bibinfo {author} {\bibfnamefont {F.}~\bibnamefont {Romano}}, \
	and\ \bibinfo {author} {\bibfnamefont {H.}~\bibnamefont {Tanaka}},\
}\href@noop {} {\bibfield  {journal} {\bibinfo  {journal} {Nature materials}\
}\textbf {\bibinfo {volume} {13}},\ \bibinfo {pages} {733} (\bibinfo {year}
{2014})}\BibitemShut {NoStop}%
\bibitem [{\citenamefont {Brukhno}\ \emph {et~al.}(2008)\citenamefont
	{Brukhno}, \citenamefont {Anwar}, \citenamefont {Davidchack},\ and\
	\citenamefont {Handel}}]{brukhno2008challenges}%
\BibitemOpen
\bibfield  {author} {\bibinfo {author} {\bibfnamefont {A.~V.}\ \bibnamefont
		{Brukhno}}, \bibinfo {author} {\bibfnamefont {J.}~\bibnamefont {Anwar}},
	\bibinfo {author} {\bibfnamefont {R.}~\bibnamefont {Davidchack}}, \ and\
	\bibinfo {author} {\bibfnamefont {R.}~\bibnamefont {Handel}},\ }\href@noop {}
{\bibfield  {journal} {\bibinfo  {journal} {Journal of Physics: Condensed
			Matter}\ }\textbf {\bibinfo {volume} {20}},\ \bibinfo {pages} {494243}
	(\bibinfo {year} {2008})}\BibitemShut {NoStop}%
\bibitem [{\citenamefont {Moore}\ and\ \citenamefont
	{Molinero}(2011{\natexlab{a}})}]{moore2011structural}%
\BibitemOpen
\bibfield  {author} {\bibinfo {author} {\bibfnamefont {E.~B.}\ \bibnamefont
		{Moore}}\ and\ \bibinfo {author} {\bibfnamefont {V.}~\bibnamefont
		{Molinero}},\ }\href@noop {} {\bibfield  {journal} {\bibinfo  {journal}
		{Nature}\ }\textbf {\bibinfo {volume} {479}},\ \bibinfo {pages} {506}
	(\bibinfo {year} {2011}{\natexlab{a}})}\BibitemShut {NoStop}%
\bibitem [{\citenamefont {Malkin}\ \emph {et~al.}(2015)\citenamefont {Malkin},
	\citenamefont {Murray}, \citenamefont {Salzmann}, \citenamefont {Molinero},
	\citenamefont {Pickering},\ and\ \citenamefont {Whale}}]{malkin2015stacking}%
\BibitemOpen
\bibfield  {author} {\bibinfo {author} {\bibfnamefont {T.~L.}\ \bibnamefont
		{Malkin}}, \bibinfo {author} {\bibfnamefont {B.~J.}\ \bibnamefont {Murray}},
	\bibinfo {author} {\bibfnamefont {C.~G.}\ \bibnamefont {Salzmann}}, \bibinfo
	{author} {\bibfnamefont {V.}~\bibnamefont {Molinero}}, \bibinfo {author}
	{\bibfnamefont {S.~J.}\ \bibnamefont {Pickering}}, \ and\ \bibinfo {author}
	{\bibfnamefont {T.~F.}\ \bibnamefont {Whale}},\ }\href@noop {} {\bibfield
	{journal} {\bibinfo  {journal} {Physical Chemistry Chemical Physics}\
	}\textbf {\bibinfo {volume} {17}},\ \bibinfo {pages} {60} (\bibinfo {year}
	{2015})}\BibitemShut {NoStop}%
\bibitem [{\citenamefont {Shilling}\ \emph {et~al.}(2006)\citenamefont
	{Shilling}, \citenamefont {Tolbert}, \citenamefont {Toon}, \citenamefont
	{Jensen}, \citenamefont {Murray},\ and\ \citenamefont
	{Bertram}}]{shilling2006measurements}%
\BibitemOpen
\bibfield  {author} {\bibinfo {author} {\bibfnamefont {J.}~\bibnamefont
		{Shilling}}, \bibinfo {author} {\bibfnamefont {M.}~\bibnamefont {Tolbert}},
	\bibinfo {author} {\bibfnamefont {O.}~\bibnamefont {Toon}}, \bibinfo {author}
	{\bibfnamefont {E.}~\bibnamefont {Jensen}}, \bibinfo {author} {\bibfnamefont
		{B.~J.}\ \bibnamefont {Murray}}, \ and\ \bibinfo {author} {\bibfnamefont
		{A.~K.}\ \bibnamefont {Bertram}},\ }\href@noop {} {\bibfield  {journal}
	{\bibinfo  {journal} {Geophysical research letters}\ }\textbf {\bibinfo
		{volume} {33}} (\bibinfo {year} {2006})}\BibitemShut {NoStop}%
\bibitem [{\citenamefont {Dowell}\ and\ \citenamefont
	{Rinfret}(1960)}]{dowell1960low}%
\BibitemOpen
\bibfield  {author} {\bibinfo {author} {\bibfnamefont {L.}~\bibnamefont
		{Dowell}}\ and\ \bibinfo {author} {\bibfnamefont {A.}~\bibnamefont
		{Rinfret}},\ }\href@noop {} {\bibfield  {journal} {\bibinfo  {journal}
		{Nature}\ }\textbf {\bibinfo {volume} {188}},\ \bibinfo {pages} {1144}
	(\bibinfo {year} {1960})}\BibitemShut {NoStop}%
\bibitem [{\citenamefont {Kohl}\ \emph {et~al.}(2000)\citenamefont {Kohl},
	\citenamefont {Mayer},\ and\ \citenamefont {Hallbrucker}}]{kohl2000glassy}%
\BibitemOpen
\bibfield  {author} {\bibinfo {author} {\bibfnamefont {I.}~\bibnamefont
		{Kohl}}, \bibinfo {author} {\bibfnamefont {E.}~\bibnamefont {Mayer}}, \ and\
	\bibinfo {author} {\bibfnamefont {A.}~\bibnamefont {Hallbrucker}},\
}\href@noop {} {\bibfield  {journal} {\bibinfo  {journal} {Physical Chemistry
		Chemical Physics}\ }\textbf {\bibinfo {volume} {2}},\ \bibinfo {pages} {1579}
(\bibinfo {year} {2000})}\BibitemShut {NoStop}%
\bibitem [{\citenamefont {Loerting}\ \emph {et~al.}(2006)\citenamefont
	{Loerting}, \citenamefont {Kohl}, \citenamefont {Schustereder}, \citenamefont
	{Winkel},\ and\ \citenamefont {Mayer}}]{loerting2006high}%
\BibitemOpen
\bibfield  {author} {\bibinfo {author} {\bibfnamefont {T.}~\bibnamefont
		{Loerting}}, \bibinfo {author} {\bibfnamefont {I.}~\bibnamefont {Kohl}},
	\bibinfo {author} {\bibfnamefont {W.}~\bibnamefont {Schustereder}}, \bibinfo
	{author} {\bibfnamefont {K.}~\bibnamefont {Winkel}}, \ and\ \bibinfo {author}
	{\bibfnamefont {E.}~\bibnamefont {Mayer}},\ }\href@noop {} {\bibfield
	{journal} {\bibinfo  {journal} {ChemPhysChem}\ }\textbf {\bibinfo {volume}
		{7}},\ \bibinfo {pages} {1203} (\bibinfo {year} {2006})}\BibitemShut
{NoStop}%
\bibitem [{\citenamefont {Geiger}\ \emph {et~al.}(2014)\citenamefont {Geiger},
	\citenamefont {Dellago}, \citenamefont {Macher}, \citenamefont {Franchini},
	\citenamefont {Kresse}, \citenamefont {Bernard}, \citenamefont {Stern},\ and\
	\citenamefont {Loerting}}]{geiger2014proton}%
\BibitemOpen
\bibfield  {author} {\bibinfo {author} {\bibfnamefont {P.}~\bibnamefont
		{Geiger}}, \bibinfo {author} {\bibfnamefont {C.}~\bibnamefont {Dellago}},
	\bibinfo {author} {\bibfnamefont {M.}~\bibnamefont {Macher}}, \bibinfo
	{author} {\bibfnamefont {C.}~\bibnamefont {Franchini}}, \bibinfo {author}
	{\bibfnamefont {G.}~\bibnamefont {Kresse}}, \bibinfo {author} {\bibfnamefont
		{J.}~\bibnamefont {Bernard}}, \bibinfo {author} {\bibfnamefont {J.~N.}\
		\bibnamefont {Stern}}, \ and\ \bibinfo {author} {\bibfnamefont
		{T.}~\bibnamefont {Loerting}},\ }\href@noop {} {\bibfield  {journal}
	{\bibinfo  {journal} {The Journal of Physical Chemistry C}\ }\textbf
	{\bibinfo {volume} {118}},\ \bibinfo {pages} {10989} (\bibinfo {year}
	{2014})}\BibitemShut {NoStop}%
\bibitem [{\citenamefont {Arnold}\ \emph {et~al.}(1968)\citenamefont {Arnold},
	\citenamefont {Finch}, \citenamefont {Rabideau},\ and\ \citenamefont
	{Wenzel}}]{arnold1968neutron}%
\BibitemOpen
\bibfield  {author} {\bibinfo {author} {\bibfnamefont {G.}~\bibnamefont
		{Arnold}}, \bibinfo {author} {\bibfnamefont {E.}~\bibnamefont {Finch}},
	\bibinfo {author} {\bibfnamefont {S.}~\bibnamefont {Rabideau}}, \ and\
	\bibinfo {author} {\bibfnamefont {R.}~\bibnamefont {Wenzel}},\ }\href@noop {}
{\bibfield  {journal} {\bibinfo  {journal} {The Journal of Chemical Physics}\
	}\textbf {\bibinfo {volume} {49}},\ \bibinfo {pages} {4365} (\bibinfo {year}
	{1968})}\BibitemShut {NoStop}%
\bibitem [{\citenamefont {Kuhs}\ \emph {et~al.}(1987)\citenamefont {Kuhs},
	\citenamefont {Bliss},\ and\ \citenamefont {Finney}}]{kuhs1987high}%
\BibitemOpen
\bibfield  {author} {\bibinfo {author} {\bibfnamefont {W.}~\bibnamefont
		{Kuhs}}, \bibinfo {author} {\bibfnamefont {D.}~\bibnamefont {Bliss}}, \ and\
	\bibinfo {author} {\bibfnamefont {J.}~\bibnamefont {Finney}},\ }\href@noop {}
{\bibfield  {journal} {\bibinfo  {journal} {Le Journal de Physique
			Colloques}\ }\textbf {\bibinfo {volume} {48}},\ \bibinfo {pages} {C1}
	(\bibinfo {year} {1987})}\BibitemShut {NoStop}%
\bibitem [{\citenamefont {Hansen}\ \emph
	{et~al.}(2008{\natexlab{a}})\citenamefont {Hansen}, \citenamefont {Koza},\
	and\ \citenamefont {Kuhs}}]{hansen2008formation}%
\BibitemOpen
\bibfield  {author} {\bibinfo {author} {\bibfnamefont {T.}~\bibnamefont
		{Hansen}}, \bibinfo {author} {\bibfnamefont {M.}~\bibnamefont {Koza}}, \ and\
	\bibinfo {author} {\bibfnamefont {W.}~\bibnamefont {Kuhs}},\ }\href@noop {}
{\bibfield  {journal} {\bibinfo  {journal} {Journal of Physics: Condensed
			Matter}\ }\textbf {\bibinfo {volume} {20}},\ \bibinfo {pages} {285104}
	(\bibinfo {year} {2008}{\natexlab{a}})}\BibitemShut {NoStop}%
\bibitem [{\citenamefont {Hansen}\ \emph
	{et~al.}(2008{\natexlab{b}})\citenamefont {Hansen}, \citenamefont {Koza},
	\citenamefont {Lindner},\ and\ \citenamefont {Kuhs}}]{hansen2008bformation}%
\BibitemOpen
\bibfield  {author} {\bibinfo {author} {\bibfnamefont {T.}~\bibnamefont
		{Hansen}}, \bibinfo {author} {\bibfnamefont {M.}~\bibnamefont {Koza}},
	\bibinfo {author} {\bibfnamefont {P.}~\bibnamefont {Lindner}}, \ and\
	\bibinfo {author} {\bibfnamefont {W.}~\bibnamefont {Kuhs}},\ }\href@noop {}
{\bibfield  {journal} {\bibinfo  {journal} {Journal of Physics: Condensed
			Matter}\ }\textbf {\bibinfo {volume} {20}},\ \bibinfo {pages} {285105}
	(\bibinfo {year} {2008}{\natexlab{b}})}\BibitemShut {NoStop}%
\bibitem [{\citenamefont {Murray}\ and\ \citenamefont
	{Bertram}(2007)}]{murray2007strong}%
\BibitemOpen
\bibfield  {author} {\bibinfo {author} {\bibfnamefont {B.~J.}\ \bibnamefont
		{Murray}}\ and\ \bibinfo {author} {\bibfnamefont {A.~K.}\ \bibnamefont
		{Bertram}},\ }\href@noop {} {\bibfield  {journal} {\bibinfo  {journal}
		{Geophysical Research Letters}\ }\textbf {\bibinfo {volume} {34}} (\bibinfo
	{year} {2007})}\BibitemShut {NoStop}%
\bibitem [{\citenamefont {Carignano}(2007)}]{carignano2007formation}%
\BibitemOpen
\bibfield  {author} {\bibinfo {author} {\bibfnamefont {M.~A.}\ \bibnamefont
		{Carignano}},\ }\href@noop {} {\bibfield  {journal} {\bibinfo  {journal} {The
			Journal of Physical Chemistry C}\ }\textbf {\bibinfo {volume} {111}},\
	\bibinfo {pages} {501} (\bibinfo {year} {2007})}\BibitemShut {NoStop}%
\bibitem [{\citenamefont {Pirzadeh}\ and\ \citenamefont
	{Kusalik}(2010)}]{pirzadeh2010understanding}%
\BibitemOpen
\bibfield  {author} {\bibinfo {author} {\bibfnamefont {P.}~\bibnamefont
		{Pirzadeh}}\ and\ \bibinfo {author} {\bibfnamefont {P.~G.}\ \bibnamefont
		{Kusalik}},\ }\href@noop {} {\bibfield  {journal} {\bibinfo  {journal}
		{Journal of the American Chemical Society}\ }\textbf {\bibinfo {volume}
		{133}},\ \bibinfo {pages} {704} (\bibinfo {year} {2010})}\BibitemShut
{NoStop}%
\bibitem [{\citenamefont {Rozmanov}\ and\ \citenamefont
	{Kusalik}(2012)}]{rozmanov2012anisotropy}%
\BibitemOpen
\bibfield  {author} {\bibinfo {author} {\bibfnamefont {D.}~\bibnamefont
		{Rozmanov}}\ and\ \bibinfo {author} {\bibfnamefont {P.~G.}\ \bibnamefont
		{Kusalik}},\ }\href@noop {} {\bibfield  {journal} {\bibinfo  {journal} {The
			Journal of chemical physics}\ }\textbf {\bibinfo {volume} {137}},\ \bibinfo
	{pages} {094702} (\bibinfo {year} {2012})}\BibitemShut {NoStop}%
\bibitem [{\citenamefont {Seo}\ \emph {et~al.}(2012)\citenamefont {Seo},
	\citenamefont {Jang}, \citenamefont {Kim}, \citenamefont {Choi},\ and\
	\citenamefont {Kim}}]{seo2012understanding}%
\BibitemOpen
\bibfield  {author} {\bibinfo {author} {\bibfnamefont {M.}~\bibnamefont
		{Seo}}, \bibinfo {author} {\bibfnamefont {E.}~\bibnamefont {Jang}}, \bibinfo
	{author} {\bibfnamefont {K.}~\bibnamefont {Kim}}, \bibinfo {author}
	{\bibfnamefont {S.}~\bibnamefont {Choi}}, \ and\ \bibinfo {author}
	{\bibfnamefont {J.~S.}\ \bibnamefont {Kim}},\ }\href@noop {} {\bibfield
	{journal} {\bibinfo  {journal} {The Journal of chemical physics}\ }\textbf
	{\bibinfo {volume} {137}},\ \bibinfo {pages} {154503} (\bibinfo {year}
	{2012})}\BibitemShut {NoStop}%
\bibitem [{\citenamefont {Malkin}\ \emph {et~al.}(2012)\citenamefont {Malkin},
	\citenamefont {Murray}, \citenamefont {Brukhno}, \citenamefont {Anwar},\ and\
	\citenamefont {Salzmann}}]{malkin2012structure}%
\BibitemOpen
\bibfield  {author} {\bibinfo {author} {\bibfnamefont {T.~L.}\ \bibnamefont
		{Malkin}}, \bibinfo {author} {\bibfnamefont {B.~J.}\ \bibnamefont {Murray}},
	\bibinfo {author} {\bibfnamefont {A.~V.}\ \bibnamefont {Brukhno}}, \bibinfo
	{author} {\bibfnamefont {J.}~\bibnamefont {Anwar}}, \ and\ \bibinfo {author}
	{\bibfnamefont {C.~G.}\ \bibnamefont {Salzmann}},\ }\href@noop {} {\bibfield
	{journal} {\bibinfo  {journal} {Proceedings of the National Academy of
			Sciences}\ }\textbf {\bibinfo {volume} {109}},\ \bibinfo {pages} {1041}
	(\bibinfo {year} {2012})}\BibitemShut {NoStop}%
\bibitem [{\citenamefont {Choi}\ \emph {et~al.}(2014)\citenamefont {Choi},
	\citenamefont {Jang},\ and\ \citenamefont {Kim}}]{choi2014layer}%
\BibitemOpen
\bibfield  {author} {\bibinfo {author} {\bibfnamefont {S.}~\bibnamefont
		{Choi}}, \bibinfo {author} {\bibfnamefont {E.}~\bibnamefont {Jang}}, \ and\
	\bibinfo {author} {\bibfnamefont {J.~S.}\ \bibnamefont {Kim}},\ }\href@noop
{} {\bibfield  {journal} {\bibinfo  {journal} {The Journal of chemical
			physics}\ }\textbf {\bibinfo {volume} {140}},\ \bibinfo {pages} {014701}
	(\bibinfo {year} {2014})}\BibitemShut {NoStop}%
\bibitem [{\citenamefont {Quigley}(2014)}]{quigley}%
\BibitemOpen
\bibfield  {author} {\bibinfo {author} {\bibfnamefont {D.}~\bibnamefont
		{Quigley}},\ }\href {\doibase 10.1063/1.4896376} {\bibfield  {journal}
	{\bibinfo  {journal} {The Journal of Chemical Physics}\ }\textbf {\bibinfo
		{volume} {141}},\ \bibinfo {pages} {121101} (\bibinfo {year}
	{2014})}\BibitemShut {NoStop}%
\bibitem [{\citenamefont {Kuhs}\ \emph {et~al.}(2012)\citenamefont {Kuhs},
	\citenamefont {Sippel}, \citenamefont {Falenty},\ and\ \citenamefont
	{Hansen}}]{kuhs2012extent}%
\BibitemOpen
\bibfield  {author} {\bibinfo {author} {\bibfnamefont {W.~F.}\ \bibnamefont
		{Kuhs}}, \bibinfo {author} {\bibfnamefont {C.}~\bibnamefont {Sippel}},
	\bibinfo {author} {\bibfnamefont {A.}~\bibnamefont {Falenty}}, \ and\
	\bibinfo {author} {\bibfnamefont {T.~C.}\ \bibnamefont {Hansen}},\
}\href@noop {} {\bibfield  {journal} {\bibinfo  {journal} {Proceedings of the
		National Academy of Sciences}\ }\textbf {\bibinfo {volume} {109}},\ \bibinfo
{pages} {21259} (\bibinfo {year} {2012})}\BibitemShut {NoStop}%
\bibitem [{\citenamefont {Desgranges}\ and\ \citenamefont
	{Delhommelle}(2006)}]{desgranges2006molecular}%
\BibitemOpen
\bibfield  {author} {\bibinfo {author} {\bibfnamefont {C.}~\bibnamefont
		{Desgranges}}\ and\ \bibinfo {author} {\bibfnamefont {J.}~\bibnamefont
		{Delhommelle}},\ }\href@noop {} {\bibfield  {journal} {\bibinfo  {journal}
		{Journal of the American Chemical Society}\ }\textbf {\bibinfo {volume}
		{128}},\ \bibinfo {pages} {10368} (\bibinfo {year} {2006})}\BibitemShut
{NoStop}%
\bibitem [{\citenamefont {Nguyen}\ and\ \citenamefont
	{Molinero}(2014)}]{nguyen2014cross}%
\BibitemOpen
\bibfield  {author} {\bibinfo {author} {\bibfnamefont {A.~H.}\ \bibnamefont
		{Nguyen}}\ and\ \bibinfo {author} {\bibfnamefont {V.}~\bibnamefont
		{Molinero}},\ }\href@noop {} {\bibfield  {journal} {\bibinfo  {journal} {The
			Journal of chemical physics}\ }\textbf {\bibinfo {volume} {140}},\ \bibinfo
	{pages} {084506} (\bibinfo {year} {2014})}\BibitemShut {NoStop}%
\bibitem [{\citenamefont {Lupi}\ \emph {et~al.}(2017)\citenamefont {Lupi},
	\citenamefont {Hudait}, \citenamefont {Peters}, \citenamefont {Gr{\"u}nwald},
	\citenamefont {Gotchy~Mullen}, \citenamefont {Nguyen},\ and\ \citenamefont
	{Molinero}}]{molinero2017nature}%
\BibitemOpen
\bibfield  {author} {\bibinfo {author} {\bibfnamefont {L.}~\bibnamefont
		{Lupi}}, \bibinfo {author} {\bibfnamefont {A.}~\bibnamefont {Hudait}},
	\bibinfo {author} {\bibfnamefont {B.}~\bibnamefont {Peters}}, \bibinfo
	{author} {\bibfnamefont {M.}~\bibnamefont {Gr{\"u}nwald}}, \bibinfo {author}
	{\bibfnamefont {R.}~\bibnamefont {Gotchy~Mullen}}, \bibinfo {author}
	{\bibfnamefont {A.~H.}\ \bibnamefont {Nguyen}}, \ and\ \bibinfo {author}
	{\bibfnamefont {V.}~\bibnamefont {Molinero}},\ }\href@noop {} {\bibfield
	{journal} {\bibinfo  {journal} {Nature}\ }\textbf {\bibinfo {volume} {551}},\
	\bibinfo {pages} {218–222} (\bibinfo {year} {2017})}\BibitemShut {NoStop}%
\bibitem [{\citenamefont {Engel}\ \emph {et~al.}(2018)\citenamefont {Engel},
	\citenamefont {Anelli}, \citenamefont {Ceriotti}, \citenamefont {Pickard},\
	and\ \citenamefont {Needs}}]{engel2018mapping}%
\BibitemOpen
\bibfield  {author} {\bibinfo {author} {\bibfnamefont {E.~A.}\ \bibnamefont
		{Engel}}, \bibinfo {author} {\bibfnamefont {A.}~\bibnamefont {Anelli}},
	\bibinfo {author} {\bibfnamefont {M.}~\bibnamefont {Ceriotti}}, \bibinfo
	{author} {\bibfnamefont {C.~J.}\ \bibnamefont {Pickard}}, \ and\ \bibinfo
	{author} {\bibfnamefont {R.~J.}\ \bibnamefont {Needs}},\ }\href@noop {}
{\bibfield  {journal} {\bibinfo  {journal} {Nature communications}\ }\textbf
	{\bibinfo {volume} {9}},\ \bibinfo {pages} {2173} (\bibinfo {year}
	{2018})}\BibitemShut {NoStop}%
\bibitem [{\citenamefont {Matsumoto}\ \emph {et~al.}(2002)\citenamefont
	{Matsumoto}, \citenamefont {Saito},\ and\ \citenamefont
	{Ohmine}}]{matsumoto2002molecular}%
\BibitemOpen
\bibfield  {author} {\bibinfo {author} {\bibfnamefont {M.}~\bibnamefont
		{Matsumoto}}, \bibinfo {author} {\bibfnamefont {S.}~\bibnamefont {Saito}}, \
	and\ \bibinfo {author} {\bibfnamefont {I.}~\bibnamefont {Ohmine}},\
}\href@noop {} {\bibfield  {journal} {\bibinfo  {journal} {Nature}\ }\textbf
{\bibinfo {volume} {416}},\ \bibinfo {pages} {409} (\bibinfo {year}
{2002})}\BibitemShut {NoStop}%
\bibitem [{\citenamefont {Radhakrishnan}\ and\ \citenamefont
	{Trout}(2003)}]{radhakrishnan2003nucleation}%
\BibitemOpen
\bibfield  {author} {\bibinfo {author} {\bibfnamefont {R.}~\bibnamefont
		{Radhakrishnan}}\ and\ \bibinfo {author} {\bibfnamefont {B.~L.}\ \bibnamefont
		{Trout}},\ }\href@noop {} {\bibfield  {journal} {\bibinfo  {journal}
		{Physical review letters}\ }\textbf {\bibinfo {volume} {90}},\ \bibinfo
	{pages} {158301} (\bibinfo {year} {2003})}\BibitemShut {NoStop}%
\bibitem [{\citenamefont {Molinero}\ and\ \citenamefont
	{Moore}(2008)}]{molinero2008water}%
\BibitemOpen
\bibfield  {author} {\bibinfo {author} {\bibfnamefont {V.}~\bibnamefont
		{Molinero}}\ and\ \bibinfo {author} {\bibfnamefont {E.~B.}\ \bibnamefont
		{Moore}},\ }\href@noop {} {\bibfield  {journal} {\bibinfo  {journal} {The
			Journal of Physical Chemistry B}\ }\textbf {\bibinfo {volume} {113}},\
	\bibinfo {pages} {4008} (\bibinfo {year} {2008})}\BibitemShut {NoStop}%
\bibitem [{\citenamefont {Moore}\ and\ \citenamefont
	{Molinero}(2010)}]{moore2010ice}%
\BibitemOpen
\bibfield  {author} {\bibinfo {author} {\bibfnamefont {E.~B.}\ \bibnamefont
		{Moore}}\ and\ \bibinfo {author} {\bibfnamefont {V.}~\bibnamefont
		{Molinero}},\ }\href@noop {} {\bibfield  {journal} {\bibinfo  {journal} {The
			Journal of chemical physics}\ }\textbf {\bibinfo {volume} {132}},\ \bibinfo
	{pages} {244504} (\bibinfo {year} {2010})}\BibitemShut {NoStop}%
\bibitem [{\citenamefont {Moore}\ and\ \citenamefont
	{Molinero}(2011{\natexlab{b}})}]{moore2011cubic}%
\BibitemOpen
\bibfield  {author} {\bibinfo {author} {\bibfnamefont {E.~B.}\ \bibnamefont
		{Moore}}\ and\ \bibinfo {author} {\bibfnamefont {V.}~\bibnamefont
		{Molinero}},\ }\href@noop {} {\bibfield  {journal} {\bibinfo  {journal}
		{Physical Chemistry Chemical Physics}\ }\textbf {\bibinfo {volume} {13}},\
	\bibinfo {pages} {20008} (\bibinfo {year} {2011}{\natexlab{b}})}\BibitemShut
{NoStop}%
\bibitem [{\citenamefont {Quigley}\ and\ \citenamefont
	{Rodger}(2008)}]{quigley2008metadynamics}%
\BibitemOpen
\bibfield  {author} {\bibinfo {author} {\bibfnamefont {D.}~\bibnamefont
		{Quigley}}\ and\ \bibinfo {author} {\bibfnamefont {P.~M.}\ \bibnamefont
		{Rodger}},\ }\href@noop {} {\bibfield  {journal} {\bibinfo  {journal} {The
			Journal of chemical physics}\ }\textbf {\bibinfo {volume} {128}},\ \bibinfo
	{pages} {154518} (\bibinfo {year} {2008})}\BibitemShut {NoStop}%
\bibitem [{\citenamefont {Li}\ \emph {et~al.}(2011)\citenamefont {Li},
	\citenamefont {Donadio}, \citenamefont {Russo},\ and\ \citenamefont
	{Galli}}]{li2011homogeneous}%
\BibitemOpen
\bibfield  {author} {\bibinfo {author} {\bibfnamefont {T.}~\bibnamefont
		{Li}}, \bibinfo {author} {\bibfnamefont {D.}~\bibnamefont {Donadio}},
	\bibinfo {author} {\bibfnamefont {G.}~\bibnamefont {Russo}}, \ and\ \bibinfo
	{author} {\bibfnamefont {G.}~\bibnamefont {Galli}},\ }\href@noop {}
{\bibfield  {journal} {\bibinfo  {journal} {Physical Chemistry Chemical
			Physics}\ }\textbf {\bibinfo {volume} {13}},\ \bibinfo {pages} {19807}
	(\bibinfo {year} {2011})}\BibitemShut {NoStop}%
\bibitem [{\citenamefont {Reinhardt}\ \emph {et~al.}(2012)\citenamefont
	{Reinhardt}, \citenamefont {Doye}, \citenamefont {Noya},\ and\ \citenamefont
	{Vega}}]{reinhardt2012local}%
\BibitemOpen
\bibfield  {author} {\bibinfo {author} {\bibfnamefont {A.}~\bibnamefont
		{Reinhardt}}, \bibinfo {author} {\bibfnamefont {J.~P.}\ \bibnamefont {Doye}},
	\bibinfo {author} {\bibfnamefont {E.~G.}\ \bibnamefont {Noya}}, \ and\
	\bibinfo {author} {\bibfnamefont {C.}~\bibnamefont {Vega}},\ }\href@noop {}
{\bibfield  {journal} {\bibinfo  {journal} {The Journal of chemical physics}\
	}\textbf {\bibinfo {volume} {137}},\ \bibinfo {pages} {194504} (\bibinfo
	{year} {2012})}\BibitemShut {NoStop}%
\bibitem [{\citenamefont {Reinhardt}\ and\ \citenamefont
	{Doye}(2013)}]{reinhardt2013note}%
\BibitemOpen
\bibfield  {author} {\bibinfo {author} {\bibfnamefont {A.}~\bibnamefont
		{Reinhardt}}\ and\ \bibinfo {author} {\bibfnamefont {J.~P.}\ \bibnamefont
		{Doye}},\ }\href@noop {} {\bibfield  {journal} {\bibinfo  {journal} {The
			Journal of chemical physics}\ }\textbf {\bibinfo {volume} {139}},\ \bibinfo
	{pages} {096102} (\bibinfo {year} {2013})}\BibitemShut {NoStop}%
\bibitem [{\citenamefont {Benet}\ \emph
	{et~al.}(2014{\natexlab{a}})\citenamefont {Benet}, \citenamefont
	{MacDowell},\ and\ \citenamefont {Sanz}}]{benet2014study}%
\BibitemOpen
\bibfield  {author} {\bibinfo {author} {\bibfnamefont {J.}~\bibnamefont
		{Benet}}, \bibinfo {author} {\bibfnamefont {L.~G.}\ \bibnamefont
		{MacDowell}}, \ and\ \bibinfo {author} {\bibfnamefont {E.}~\bibnamefont
		{Sanz}},\ }\href@noop {} {\bibfield  {journal} {\bibinfo  {journal} {Physical
			Chemistry Chemical Physics}\ }\textbf {\bibinfo {volume} {16}},\ \bibinfo
	{pages} {22159} (\bibinfo {year} {2014}{\natexlab{a}})}\BibitemShut {NoStop}%
\bibitem [{\citenamefont {Benet}\ \emph
	{et~al.}(2014{\natexlab{b}})\citenamefont {Benet}, \citenamefont
	{MacDowell},\ and\ \citenamefont {Sanz}}]{benet2014computer}%
\BibitemOpen
\bibfield  {author} {\bibinfo {author} {\bibfnamefont {J.}~\bibnamefont
		{Benet}}, \bibinfo {author} {\bibfnamefont {L.~G.}\ \bibnamefont
		{MacDowell}}, \ and\ \bibinfo {author} {\bibfnamefont {E.}~\bibnamefont
		{Sanz}},\ }\href@noop {} {\bibfield  {journal} {\bibinfo  {journal} {The
			Journal of chemical physics}\ }\textbf {\bibinfo {volume} {141}},\ \bibinfo
	{pages} {034701} (\bibinfo {year} {2014}{\natexlab{b}})}\BibitemShut
{NoStop}%
\bibitem [{\citenamefont {Cheng}\ \emph {et~al.}(2015)\citenamefont {Cheng},
	\citenamefont {Tribello},\ and\ \citenamefont {Ceriotti}}]{cheng2015solid}%
\BibitemOpen
\bibfield  {author} {\bibinfo {author} {\bibfnamefont {B.}~\bibnamefont
		{Cheng}}, \bibinfo {author} {\bibfnamefont {G.~A.}\ \bibnamefont {Tribello}},
	\ and\ \bibinfo {author} {\bibfnamefont {M.}~\bibnamefont {Ceriotti}},\
}\href@noop {} {\bibfield  {journal} {\bibinfo  {journal} {Physical Review
		B}\ }\textbf {\bibinfo {volume} {92}},\ \bibinfo {pages} {180102} (\bibinfo
{year} {2015})}\BibitemShut {NoStop}%
\bibitem [{\citenamefont {Espinosa}\ \emph
	{et~al.}(2016{\natexlab{a}})\citenamefont {Espinosa}, \citenamefont {Vega},\
	and\ \citenamefont {Sanz}}]{espinosa2016ice}%
\BibitemOpen
\bibfield  {author} {\bibinfo {author} {\bibfnamefont {J.~R.}\ \bibnamefont
		{Espinosa}}, \bibinfo {author} {\bibfnamefont {C.}~\bibnamefont {Vega}}, \
	and\ \bibinfo {author} {\bibfnamefont {E.}~\bibnamefont {Sanz}},\ }\href@noop
{} {\bibfield  {journal} {\bibinfo  {journal} {The Journal of Physical
			Chemistry C}\ }\textbf {\bibinfo {volume} {120}},\ \bibinfo {pages} {8068}
	(\bibinfo {year} {2016}{\natexlab{a}})}\BibitemShut {NoStop}%
\bibitem [{\citenamefont {Espinosa}\ \emph
	{et~al.}(2016{\natexlab{b}})\citenamefont {Espinosa}, \citenamefont
	{Zaragoza}, \citenamefont {Rosales-Pelaez}, \citenamefont {Navarro},
	\citenamefont {Valeriani}, \citenamefont {Vega},\ and\ \citenamefont
	{Sanz}}]{espinosa2016interfacial}%
\BibitemOpen
\bibfield  {author} {\bibinfo {author} {\bibfnamefont {J.~R.}\ \bibnamefont
		{Espinosa}}, \bibinfo {author} {\bibfnamefont {A.}~\bibnamefont {Zaragoza}},
	\bibinfo {author} {\bibfnamefont {P.}~\bibnamefont {Rosales-Pelaez}},
	\bibinfo {author} {\bibfnamefont {C.}~\bibnamefont {Navarro}}, \bibinfo
	{author} {\bibfnamefont {C.}~\bibnamefont {Valeriani}}, \bibinfo {author}
	{\bibfnamefont {C.}~\bibnamefont {Vega}}, \ and\ \bibinfo {author}
	{\bibfnamefont {E.}~\bibnamefont {Sanz}},\ }\href@noop {} {\bibfield
	{journal} {\bibinfo  {journal} {Physical review letters}\ }\textbf {\bibinfo
		{volume} {117}},\ \bibinfo {pages} {135702} (\bibinfo {year}
	{2016}{\natexlab{b}})}\BibitemShut {NoStop}%
\bibitem [{\citenamefont {Ambler}\ \emph {et~al.}(2017)\citenamefont {Ambler},
	\citenamefont {Vorselaars}, \citenamefont {Allen},\ and\ \citenamefont
	{Quigley}}]{ambler2017solid}%
\BibitemOpen
\bibfield  {author} {\bibinfo {author} {\bibfnamefont {M.}~\bibnamefont
		{Ambler}}, \bibinfo {author} {\bibfnamefont {B.}~\bibnamefont {Vorselaars}},
	\bibinfo {author} {\bibfnamefont {M.~P.}\ \bibnamefont {Allen}}, \ and\
	\bibinfo {author} {\bibfnamefont {D.}~\bibnamefont {Quigley}},\ }\href@noop
{} {\bibfield  {journal} {\bibinfo  {journal} {The Journal of Chemical
			Physics}\ }\textbf {\bibinfo {volume} {146}},\ \bibinfo {pages} {074701}
	(\bibinfo {year} {2017})}\BibitemShut {NoStop}%
\bibitem [{\citenamefont {Koop}\ and\ \citenamefont
	{Murray}(2016)}]{koop2016physically}%
\BibitemOpen
\bibfield  {author} {\bibinfo {author} {\bibfnamefont {T.}~\bibnamefont
		{Koop}}\ and\ \bibinfo {author} {\bibfnamefont {B.~J.}\ \bibnamefont
		{Murray}},\ }\href@noop {} {\bibfield  {journal} {\bibinfo  {journal} {The
			Journal of Chemical Physics}\ }\textbf {\bibinfo {volume} {145}},\ \bibinfo
	{pages} {211915} (\bibinfo {year} {2016})}\BibitemShut {NoStop}%
\bibitem [{\citenamefont {Cheng}\ and\ \citenamefont
	{Ceriotti}(2017)}]{cheng2017bridging}%
\BibitemOpen
\bibfield  {author} {\bibinfo {author} {\bibfnamefont {B.}~\bibnamefont
		{Cheng}}\ and\ \bibinfo {author} {\bibfnamefont {M.}~\bibnamefont
		{Ceriotti}},\ }\href@noop {} {\bibfield  {journal} {\bibinfo  {journal} {The
			Journal of chemical physics}\ }\textbf {\bibinfo {volume} {146}},\ \bibinfo
	{pages} {034106} (\bibinfo {year} {2017})}\BibitemShut {NoStop}%
\bibitem [{\citenamefont {Cheng}\ \emph {et~al.}(2017)\citenamefont {Cheng},
	\citenamefont {Tribello},\ and\ \citenamefont {Ceriotti}}]{cheng2017gibbs}%
\BibitemOpen
\bibfield  {author} {\bibinfo {author} {\bibfnamefont {B.}~\bibnamefont
		{Cheng}}, \bibinfo {author} {\bibfnamefont {G.~A.}\ \bibnamefont {Tribello}},
	\ and\ \bibinfo {author} {\bibfnamefont {M.}~\bibnamefont {Ceriotti}},\
}\href@noop {} {\bibfield  {journal} {\bibinfo  {journal} {arXiv preprint
		arXiv:1703.06062}\ } (\bibinfo {year} {2017})}\BibitemShut {NoStop}%
\bibitem [{\citenamefont {Lupi}\ \emph {et~al.}(2014)\citenamefont {Lupi},
	\citenamefont {Hudait},\ and\ \citenamefont
	{Molinero}}]{lupi2014heterogeneous}%
\BibitemOpen
\bibfield  {author} {\bibinfo {author} {\bibfnamefont {L.}~\bibnamefont
		{Lupi}}, \bibinfo {author} {\bibfnamefont {A.}~\bibnamefont {Hudait}}, \ and\
	\bibinfo {author} {\bibfnamefont {V.}~\bibnamefont {Molinero}},\ }\href@noop
{} {\bibfield  {journal} {\bibinfo  {journal} {Journal of the American
			Chemical Society}\ }\textbf {\bibinfo {volume} {136}},\ \bibinfo {pages}
	{3156} (\bibinfo {year} {2014})}\BibitemShut {NoStop}%
\bibitem [{\citenamefont {Reinhardt}\ and\ \citenamefont
	{Doye}(2014)}]{reinhardt2014effects}%
\BibitemOpen
\bibfield  {author} {\bibinfo {author} {\bibfnamefont {A.}~\bibnamefont
		{Reinhardt}}\ and\ \bibinfo {author} {\bibfnamefont {J.~P.}\ \bibnamefont
		{Doye}},\ }\href@noop {} {\bibfield  {journal} {\bibinfo  {journal} {The
			Journal of chemical physics}\ }\textbf {\bibinfo {volume} {141}},\ \bibinfo
	{pages} {084501} (\bibinfo {year} {2014})}\BibitemShut {NoStop}%
\bibitem [{\citenamefont {Cox}\ \emph {et~al.}(2015)\citenamefont {Cox},
	\citenamefont {Kathmann}, \citenamefont {Slater},\ and\ \citenamefont
	{Michaelides}}]{cox2015molecular}%
\BibitemOpen
\bibfield  {author} {\bibinfo {author} {\bibfnamefont {S.~J.}\ \bibnamefont
		{Cox}}, \bibinfo {author} {\bibfnamefont {S.~M.}\ \bibnamefont {Kathmann}},
	\bibinfo {author} {\bibfnamefont {B.}~\bibnamefont {Slater}}, \ and\ \bibinfo
	{author} {\bibfnamefont {A.}~\bibnamefont {Michaelides}},\ }\href@noop {}
{\bibfield  {journal} {\bibinfo  {journal} {The Journal of chemical physics}\
	}\textbf {\bibinfo {volume} {142}},\ \bibinfo {pages} {184704} (\bibinfo
	{year} {2015})}\BibitemShut {NoStop}%
\bibitem [{\citenamefont {Sosso}\ \emph {et~al.}(2016)\citenamefont {Sosso},
	\citenamefont {Tribello}, \citenamefont {Zen}, \citenamefont {Pedevilla},\
	and\ \citenamefont {Michaelides}}]{sosso2016ice}%
\BibitemOpen
\bibfield  {author} {\bibinfo {author} {\bibfnamefont {G.~C.}\ \bibnamefont
		{Sosso}}, \bibinfo {author} {\bibfnamefont {G.~A.}\ \bibnamefont {Tribello}},
	\bibinfo {author} {\bibfnamefont {A.}~\bibnamefont {Zen}}, \bibinfo {author}
	{\bibfnamefont {P.}~\bibnamefont {Pedevilla}}, \ and\ \bibinfo {author}
	{\bibfnamefont {A.}~\bibnamefont {Michaelides}},\ }\href@noop {} {\bibfield
	{journal} {\bibinfo  {journal} {The Journal of Chemical Physics}\ }\textbf
	{\bibinfo {volume} {145}},\ \bibinfo {pages} {211927} (\bibinfo {year}
	{2016})}\BibitemShut {NoStop}%
\bibitem [{\citenamefont {Lupi}\ \emph {et~al.}(2016)\citenamefont {Lupi},
	\citenamefont {Peters},\ and\ \citenamefont {Molinero}}]{lupi2016pre}%
\BibitemOpen
\bibfield  {author} {\bibinfo {author} {\bibfnamefont {L.}~\bibnamefont
		{Lupi}}, \bibinfo {author} {\bibfnamefont {B.}~\bibnamefont {Peters}}, \ and\
	\bibinfo {author} {\bibfnamefont {V.}~\bibnamefont {Molinero}},\ }\href@noop
{} {\bibfield  {journal} {\bibinfo  {journal} {The Journal of Chemical
			Physics}\ }\textbf {\bibinfo {volume} {145}},\ \bibinfo {pages} {211910}
	(\bibinfo {year} {2016})}\BibitemShut {NoStop}%
\bibitem [{\citenamefont {Johnston}\ and\ \citenamefont
	{Molinero}(2012)}]{johnston2012crystallization}%
\BibitemOpen
\bibfield  {author} {\bibinfo {author} {\bibfnamefont {J.~C.}\ \bibnamefont
		{Johnston}}\ and\ \bibinfo {author} {\bibfnamefont {V.}~\bibnamefont
		{Molinero}},\ }\href@noop {} {\bibfield  {journal} {\bibinfo  {journal}
		{Journal of the American Chemical Society}\ }\textbf {\bibinfo {volume}
		{134}},\ \bibinfo {pages} {6650} (\bibinfo {year} {2012})}\BibitemShut
{NoStop}%
\bibitem [{\citenamefont {Li}\ \emph {et~al.}(2013)\citenamefont {Li},
	\citenamefont {Donadio},\ and\ \citenamefont {Galli}}]{li2013ice}%
\BibitemOpen
\bibfield  {author} {\bibinfo {author} {\bibfnamefont {T.}~\bibnamefont
		{Li}}, \bibinfo {author} {\bibfnamefont {D.}~\bibnamefont {Donadio}}, \ and\
	\bibinfo {author} {\bibfnamefont {G.}~\bibnamefont {Galli}},\ }\href@noop {}
{\bibfield  {journal} {\bibinfo  {journal} {Nature communications}\ }\textbf
	{\bibinfo {volume} {4}},\ \bibinfo {pages} {1887} (\bibinfo {year}
	{2013})}\BibitemShut {NoStop}%
\bibitem [{\citenamefont {Hudait}\ and\ \citenamefont
	{Molinero}(2014)}]{hudait2014ice}%
\BibitemOpen
\bibfield  {author} {\bibinfo {author} {\bibfnamefont {A.}~\bibnamefont
		{Hudait}}\ and\ \bibinfo {author} {\bibfnamefont {V.}~\bibnamefont
		{Molinero}},\ }\href@noop {} {\bibfield  {journal} {\bibinfo  {journal}
		{Journal of the American Chemical Society}\ }\textbf {\bibinfo {volume}
		{136}},\ \bibinfo {pages} {8081} (\bibinfo {year} {2014})}\BibitemShut
{NoStop}%
\bibitem [{\citenamefont {Haji-Akbari}\ \emph {et~al.}(2014)\citenamefont
	{Haji-Akbari}, \citenamefont {DeFever}, \citenamefont {Sarupria},\ and\
	\citenamefont {Debenedetti}}]{haji2014suppression}%
\BibitemOpen
\bibfield  {author} {\bibinfo {author} {\bibfnamefont {A.}~\bibnamefont
		{Haji-Akbari}}, \bibinfo {author} {\bibfnamefont {R.~S.}\ \bibnamefont
		{DeFever}}, \bibinfo {author} {\bibfnamefont {S.}~\bibnamefont {Sarupria}}, \
	and\ \bibinfo {author} {\bibfnamefont {P.~G.}\ \bibnamefont {Debenedetti}},\
}\href@noop {} {\bibfield  {journal} {\bibinfo  {journal} {Physical Chemistry
		Chemical Physics}\ }\textbf {\bibinfo {volume} {16}},\ \bibinfo {pages}
{25916} (\bibinfo {year} {2014})}\BibitemShut {NoStop}%
\bibitem [{\citenamefont {Malek}\ \emph {et~al.}(2015)\citenamefont {Malek},
	\citenamefont {Morrow},\ and\ \citenamefont
	{Saika-Voivod}}]{malek2015crystallization}%
\BibitemOpen
\bibfield  {author} {\bibinfo {author} {\bibfnamefont {S.~M.}\ \bibnamefont
		{Malek}}, \bibinfo {author} {\bibfnamefont {G.~P.}\ \bibnamefont {Morrow}}, \
	and\ \bibinfo {author} {\bibfnamefont {I.}~\bibnamefont {Saika-Voivod}},\
}\href@noop {} {\bibfield  {journal} {\bibinfo  {journal} {The Journal of
		chemical physics}\ }\textbf {\bibinfo {volume} {142}},\ \bibinfo {pages}
{124506} (\bibinfo {year} {2015})}\BibitemShut {NoStop}%
\bibitem [{\citenamefont {Haji-Akbari}\ and\ \citenamefont
	{Debenedetti}(2017)}]{haji2017perspective}%
\BibitemOpen
\bibfield  {author} {\bibinfo {author} {\bibfnamefont {A.}~\bibnamefont
		{Haji-Akbari}}\ and\ \bibinfo {author} {\bibfnamefont {P.~G.}\ \bibnamefont
		{Debenedetti}},\ }\href@noop {} {\bibfield  {journal} {\bibinfo  {journal}
		{The Journal of Chemical Physics}\ }\textbf {\bibinfo {volume} {147}},\
	\bibinfo {pages} {060901} (\bibinfo {year} {2017})}\BibitemShut {NoStop}%
\bibitem [{\citenamefont {Moore}\ \emph {et~al.}(2010)\citenamefont {Moore},
	\citenamefont {De~La~Llave}, \citenamefont {Welke}, \citenamefont
	{Scherlis},\ and\ \citenamefont {Molinero}}]{moore2010freezing}%
\BibitemOpen
\bibfield  {author} {\bibinfo {author} {\bibfnamefont {E.~B.}\ \bibnamefont
		{Moore}}, \bibinfo {author} {\bibfnamefont {E.}~\bibnamefont {De~La~Llave}},
	\bibinfo {author} {\bibfnamefont {K.}~\bibnamefont {Welke}}, \bibinfo
	{author} {\bibfnamefont {D.~A.}\ \bibnamefont {Scherlis}}, \ and\ \bibinfo
	{author} {\bibfnamefont {V.}~\bibnamefont {Molinero}},\ }\href@noop {}
{\bibfield  {journal} {\bibinfo  {journal} {Physical Chemistry Chemical
			Physics}\ }\textbf {\bibinfo {volume} {12}},\ \bibinfo {pages} {4124}
	(\bibinfo {year} {2010})}\BibitemShut {NoStop}%
\bibitem [{\citenamefont {Gonzalez~Solveyra}\ \emph {et~al.}(2011)\citenamefont
	{Gonzalez~Solveyra}, \citenamefont {De~La~Llave}, \citenamefont {Scherlis},\
	and\ \citenamefont {Molinero}}]{gonzalez2011melting}%
\BibitemOpen
\bibfield  {author} {\bibinfo {author} {\bibfnamefont {E.}~\bibnamefont
		{Gonzalez~Solveyra}}, \bibinfo {author} {\bibfnamefont {E.}~\bibnamefont
		{De~La~Llave}}, \bibinfo {author} {\bibfnamefont {D.~A.}\ \bibnamefont
		{Scherlis}}, \ and\ \bibinfo {author} {\bibfnamefont {V.}~\bibnamefont
		{Molinero}},\ }\href@noop {} {\bibfield  {journal} {\bibinfo  {journal} {The
			Journal of Physical Chemistry B}\ }\textbf {\bibinfo {volume} {115}},\
	\bibinfo {pages} {14196} (\bibinfo {year} {2011})}\BibitemShut {NoStop}%
\bibitem [{\citenamefont {Bullock}\ and\ \citenamefont
	{Molinero}(2013)}]{bullock2013low}%
\BibitemOpen
\bibfield  {author} {\bibinfo {author} {\bibfnamefont {G.}~\bibnamefont
		{Bullock}}\ and\ \bibinfo {author} {\bibfnamefont {V.}~\bibnamefont
		{Molinero}},\ }\href@noop {} {\bibfield  {journal} {\bibinfo  {journal}
		{Faraday discussions}\ }\textbf {\bibinfo {volume} {167}},\ \bibinfo {pages}
	{371} (\bibinfo {year} {2013})}\BibitemShut {NoStop}%
\bibitem [{\citenamefont {Soria}\ \emph {et~al.}(2017)\citenamefont {Soria},
	\citenamefont {Espinosa}, \citenamefont {Ramirez}, \citenamefont {Valeriani},
	\citenamefont {Vega},\ and\ \citenamefont {Sanz}}]{soria2017simulation}%
\BibitemOpen
\bibfield  {author} {\bibinfo {author} {\bibfnamefont {G.~D.}\ \bibnamefont
		{Soria}}, \bibinfo {author} {\bibfnamefont {J.~R.}\ \bibnamefont {Espinosa}},
	\bibinfo {author} {\bibfnamefont {J.}~\bibnamefont {Ramirez}}, \bibinfo
	{author} {\bibfnamefont {C.}~\bibnamefont {Valeriani}}, \bibinfo {author}
	{\bibfnamefont {C.}~\bibnamefont {Vega}}, \ and\ \bibinfo {author}
	{\bibfnamefont {E.}~\bibnamefont {Sanz}},\ }\href@noop {} {\bibfield
	{journal} {\bibinfo  {journal} {arXiv preprint arXiv:1709.00619}\ } (\bibinfo
	{year} {2017})}\BibitemShut {NoStop}%
\bibitem [{\citenamefont {Conde}\ \emph {et~al.}(2017)\citenamefont {Conde},
	\citenamefont {Rovere},\ and\ \citenamefont {Gallo}}]{conde2017spontaneous}%
\BibitemOpen
\bibfield  {author} {\bibinfo {author} {\bibfnamefont {M.}~\bibnamefont
		{Conde}}, \bibinfo {author} {\bibfnamefont {M.}~\bibnamefont {Rovere}}, \
	and\ \bibinfo {author} {\bibfnamefont {P.}~\bibnamefont {Gallo}},\
}\href@noop {} {\bibfield  {journal} {\bibinfo  {journal} {Physical Chemistry
		Chemical Physics}\ }\textbf {\bibinfo {volume} {19}},\ \bibinfo {pages}
{9566} (\bibinfo {year} {2017})}\BibitemShut {NoStop}%
\bibitem [{\citenamefont {Rovigatti}\ \emph {et~al.}(2018)\citenamefont
	{Rovigatti}, \citenamefont {Russo},\ and\ \citenamefont
	{Romano}}]{rovigatti2018simulate}%
\BibitemOpen
\bibfield  {author} {\bibinfo {author} {\bibfnamefont {L.}~\bibnamefont
		{Rovigatti}}, \bibinfo {author} {\bibfnamefont {J.}~\bibnamefont {Russo}}, \
	and\ \bibinfo {author} {\bibfnamefont {F.}~\bibnamefont {Romano}},\
}\href@noop {} {\bibfield  {journal} {\bibinfo  {journal} {The European
		Physical Journal E}\ }\textbf {\bibinfo {volume} {41}},\ \bibinfo {pages}
{59} (\bibinfo {year} {2018})}\BibitemShut {NoStop}%
\bibitem [{\citenamefont {Knott}\ \emph {et~al.}(2012)\citenamefont {Knott},
	\citenamefont {Molinero}, \citenamefont {Doherty},\ and\ \citenamefont
	{Peters}}]{knott2012homogeneous}%
\BibitemOpen
\bibfield  {author} {\bibinfo {author} {\bibfnamefont {B.~C.}\ \bibnamefont
		{Knott}}, \bibinfo {author} {\bibfnamefont {V.}~\bibnamefont {Molinero}},
	\bibinfo {author} {\bibfnamefont {M.~F.}\ \bibnamefont {Doherty}}, \ and\
	\bibinfo {author} {\bibfnamefont {B.}~\bibnamefont {Peters}},\ }\href@noop {}
{\bibfield  {journal} {\bibinfo  {journal} {Journal of the American Chemical
			Society}\ }\textbf {\bibinfo {volume} {134}},\ \bibinfo {pages} {19544}
	(\bibinfo {year} {2012})}\BibitemShut {NoStop}%
\bibitem [{\citenamefont {Zimmermann}\ \emph {et~al.}(2015)\citenamefont
	{Zimmermann}, \citenamefont {Vorselaars}, \citenamefont {Quigley},\ and\
	\citenamefont {Peters}}]{zimmermann2015nucleation}%
\BibitemOpen
\bibfield  {author} {\bibinfo {author} {\bibfnamefont {N.~E.}\ \bibnamefont
		{Zimmermann}}, \bibinfo {author} {\bibfnamefont {B.}~\bibnamefont
		{Vorselaars}}, \bibinfo {author} {\bibfnamefont {D.}~\bibnamefont {Quigley}},
	\ and\ \bibinfo {author} {\bibfnamefont {B.}~\bibnamefont {Peters}},\
}\href@noop {} {\bibfield  {journal} {\bibinfo  {journal} {Journal of the
		American Chemical Society}\ }\textbf {\bibinfo {volume} {137}},\ \bibinfo
{pages} {13352} (\bibinfo {year} {2015})}\BibitemShut {NoStop}%
\bibitem [{\citenamefont {Espinosa}\ \emph
	{et~al.}(2016{\natexlab{c}})\citenamefont {Espinosa}, \citenamefont {Vega},
	\citenamefont {Valeriani},\ and\ \citenamefont {Sanz}}]{espinosa2016seeding}%
\BibitemOpen
\bibfield  {author} {\bibinfo {author} {\bibfnamefont {J.~R.}\ \bibnamefont
		{Espinosa}}, \bibinfo {author} {\bibfnamefont {C.}~\bibnamefont {Vega}},
	\bibinfo {author} {\bibfnamefont {C.}~\bibnamefont {Valeriani}}, \ and\
	\bibinfo {author} {\bibfnamefont {E.}~\bibnamefont {Sanz}},\ }\href@noop {}
{\bibfield  {journal} {\bibinfo  {journal} {The Journal of chemical physics}\
	}\textbf {\bibinfo {volume} {144}},\ \bibinfo {pages} {034501} (\bibinfo
	{year} {2016}{\natexlab{c}})}\BibitemShut {NoStop}%
\bibitem [{\citenamefont {Espinosa}\ \emph
	{et~al.}(2016{\natexlab{d}})\citenamefont {Espinosa}, \citenamefont
	{Navarro}, \citenamefont {Sanz}, \citenamefont {Valeriani},\ and\
	\citenamefont {Vega}}]{espinosa2016time}%
\BibitemOpen
\bibfield  {author} {\bibinfo {author} {\bibfnamefont {J.}~\bibnamefont
		{Espinosa}}, \bibinfo {author} {\bibfnamefont {C.}~\bibnamefont {Navarro}},
	\bibinfo {author} {\bibfnamefont {E.}~\bibnamefont {Sanz}}, \bibinfo {author}
	{\bibfnamefont {C.}~\bibnamefont {Valeriani}}, \ and\ \bibinfo {author}
	{\bibfnamefont {C.}~\bibnamefont {Vega}},\ }\href@noop {} {\bibfield
	{journal} {\bibinfo  {journal} {The Journal of Chemical Physics}\ }\textbf
	{\bibinfo {volume} {145}},\ \bibinfo {pages} {211922} (\bibinfo {year}
	{2016}{\natexlab{d}})}\BibitemShut {NoStop}%
\bibitem [{\citenamefont {Lifanov}\ \emph {et~al.}(2016)\citenamefont
	{Lifanov}, \citenamefont {Vorselaars},\ and\ \citenamefont
	{Quigley}}]{lifanov2016nucleation}%
\BibitemOpen
\bibfield  {author} {\bibinfo {author} {\bibfnamefont {Y.}~\bibnamefont
		{Lifanov}}, \bibinfo {author} {\bibfnamefont {B.}~\bibnamefont {Vorselaars}},
	\ and\ \bibinfo {author} {\bibfnamefont {D.}~\bibnamefont {Quigley}},\
}\href@noop {} {\bibfield  {journal} {\bibinfo  {journal} {The Journal of
		Chemical Physics}\ }\textbf {\bibinfo {volume} {145}},\ \bibinfo {pages}
{211912} (\bibinfo {year} {2016})}\BibitemShut {NoStop}%
\bibitem [{\citenamefont {Russo}\ \emph
	{et~al.}(2018{\natexlab{b}})\citenamefont {Russo}, \citenamefont {Akahane},\
	and\ \citenamefont {Tanaka}}]{russo2018water}%
\BibitemOpen
\bibfield  {author} {\bibinfo {author} {\bibfnamefont {J.}~\bibnamefont
		{Russo}}, \bibinfo {author} {\bibfnamefont {K.}~\bibnamefont {Akahane}}, \
	and\ \bibinfo {author} {\bibfnamefont {H.}~\bibnamefont {Tanaka}},\
}\href@noop {} {\bibfield  {journal} {\bibinfo  {journal} {Proc. Natl. Acad.
		Sci. USA}\ }\textbf {\bibinfo {volume} {115}},\ \bibinfo {pages} {E3333}
(\bibinfo {year} {2018}{\natexlab{b}})}\BibitemShut {NoStop}%
\bibitem [{\citenamefont {Mossa}\ and\ \citenamefont
	{Tarjus}(2003)}]{Tarjus2003}%
\BibitemOpen
\bibfield  {author} {\bibinfo {author} {\bibfnamefont {S.}~\bibnamefont
		{Mossa}}\ and\ \bibinfo {author} {\bibfnamefont {G.}~\bibnamefont {Tarjus}},\
}\href@noop {} {\bibfield  {journal} {\bibinfo  {journal} {J. Phys. Chem.}\
}\textbf {\bibinfo {volume} {119}},\ \bibinfo {pages} {8069} (\bibinfo {year}
{2003})}\BibitemShut {NoStop}%
\bibitem [{\citenamefont {Steinhardt}\ \emph {et~al.}(1983)\citenamefont
	{Steinhardt}, \citenamefont {Nelson},\ and\ \citenamefont
	{Ronchetti}}]{steinhardt1983bond}%
\BibitemOpen
\bibfield  {author} {\bibinfo {author} {\bibfnamefont {P.~J.}\ \bibnamefont
		{Steinhardt}}, \bibinfo {author} {\bibfnamefont {D.~R.}\ \bibnamefont
		{Nelson}}, \ and\ \bibinfo {author} {\bibfnamefont {M.}~\bibnamefont
		{Ronchetti}},\ }\href@noop {} {\bibfield  {journal} {\bibinfo  {journal}
		{Phys. Rev. B}\ }\textbf {\bibinfo {volume} {28}},\ \bibinfo {pages} {784}
	(\bibinfo {year} {1983})}\BibitemShut {NoStop}%
\bibitem [{\citenamefont {Tarjus}\ \emph {et~al.}(2005)\citenamefont {Tarjus},
	\citenamefont {Kivelson}, \citenamefont {Nussinov},\ and\ \citenamefont
	{Viot}}]{Tarjus2005}%
\BibitemOpen
\bibfield  {author} {\bibinfo {author} {\bibfnamefont {G.}~\bibnamefont
		{Tarjus}}, \bibinfo {author} {\bibfnamefont {S.~A.}\ \bibnamefont
		{Kivelson}}, \bibinfo {author} {\bibfnamefont {Z.}~\bibnamefont {Nussinov}},
	\ and\ \bibinfo {author} {\bibfnamefont {P.}~\bibnamefont {Viot}},\
}\href@noop {} {\bibfield  {journal} {\bibinfo  {journal} {J. Phys.: Condens.
		Matter}\ }\textbf {\bibinfo {volume} {17}},\ \bibinfo {pages} {R1143}
(\bibinfo {year} {2005})}\BibitemShut {NoStop}%
\bibitem [{\citenamefont {Quigley}\ \emph {et~al.}(2014)\citenamefont
	{Quigley}, \citenamefont {Alf{\`e}},\ and\ \citenamefont
	{Slater}}]{quigley2014communication}%
\BibitemOpen
\bibfield  {author} {\bibinfo {author} {\bibfnamefont {D.}~\bibnamefont
		{Quigley}}, \bibinfo {author} {\bibfnamefont {D.}~\bibnamefont {Alf{\`e}}}, \
	and\ \bibinfo {author} {\bibfnamefont {B.}~\bibnamefont {Slater}},\
}\href@noop {} {\enquote {\bibinfo {title} {Communication: On the stability
		of ice 0, ice i, and i h},}\ } (\bibinfo {year} {2014})\BibitemShut {NoStop}%
\bibitem [{\citenamefont {Matsumoto}\ \emph {et~al.}(2007)\citenamefont
	{Matsumoto}, \citenamefont {Baba},\ and\ \citenamefont
	{Ohmine}}]{Matsumoto2007}%
\BibitemOpen
\bibfield  {author} {\bibinfo {author} {\bibfnamefont {M.}~\bibnamefont
		{Matsumoto}}, \bibinfo {author} {\bibfnamefont {A.}~\bibnamefont {Baba}}, \
	and\ \bibinfo {author} {\bibfnamefont {I.}~\bibnamefont {Ohmine}},\
}\href@noop {} {\bibfield  {journal} {\bibinfo  {journal} {J. of Chem.
		Phys.}\ }\textbf {\bibinfo {volume} {127}},\ \bibinfo {pages} {134504}
(\bibinfo {year} {2007})}\BibitemShut {NoStop}%
\bibitem [{\citenamefont {Reinhardt}\ and\ \citenamefont
	{Doye}(2012)}]{reinhardt2012free}%
\BibitemOpen
\bibfield  {author} {\bibinfo {author} {\bibfnamefont {A.}~\bibnamefont
		{Reinhardt}}\ and\ \bibinfo {author} {\bibfnamefont {J.~P.}\ \bibnamefont
		{Doye}},\ }\href@noop {} {\bibfield  {journal} {\bibinfo  {journal} {The
			Journal of chemical physics}\ }\textbf {\bibinfo {volume} {136}},\ \bibinfo
	{pages} {054501} (\bibinfo {year} {2012})}\BibitemShut {NoStop}%
\bibitem [{\citenamefont {Hudait}\ \emph {et~al.}(2016)\citenamefont {Hudait},
	\citenamefont {Qiu}, \citenamefont {Lupi},\ and\ \citenamefont
	{Molinero}}]{hudait2016free}%
\BibitemOpen
\bibfield  {author} {\bibinfo {author} {\bibfnamefont {A.}~\bibnamefont
		{Hudait}}, \bibinfo {author} {\bibfnamefont {S.}~\bibnamefont {Qiu}},
	\bibinfo {author} {\bibfnamefont {L.}~\bibnamefont {Lupi}}, \ and\ \bibinfo
	{author} {\bibfnamefont {V.}~\bibnamefont {Molinero}},\ }\href@noop {}
{\bibfield  {journal} {\bibinfo  {journal} {Physical Chemistry Chemical
			Physics}\ }\textbf {\bibinfo {volume} {18}},\ \bibinfo {pages} {9544}
	(\bibinfo {year} {2016})}\BibitemShut {NoStop}%
\bibitem [{\citenamefont {Cheng}\ \emph {et~al.}(2018)\citenamefont {Cheng},
	\citenamefont {Dellago},\ and\ \citenamefont
	{Ceriotti}}]{cheng2018theoretical}%
\BibitemOpen
\bibfield  {author} {\bibinfo {author} {\bibfnamefont {B.}~\bibnamefont
		{Cheng}}, \bibinfo {author} {\bibfnamefont {C.}~\bibnamefont {Dellago}}, \
	and\ \bibinfo {author} {\bibfnamefont {M.}~\bibnamefont {Ceriotti}},\
}\href@noop {} {\bibfield  {journal} {\bibinfo  {journal} {Physical Chemistry
		Chemical Physics}\ }\textbf {\bibinfo {volume} {20}},\ \bibinfo {pages}
{28732} (\bibinfo {year} {2018})}\BibitemShut {NoStop}%
\bibitem [{\citenamefont {Romano}\ \emph {et~al.}(2014)\citenamefont {Romano},
	\citenamefont {Russo},\ and\ \citenamefont {Tanaka}}]{romano2014novel}%
\BibitemOpen
\bibfield  {author} {\bibinfo {author} {\bibfnamefont {F.}~\bibnamefont
		{Romano}}, \bibinfo {author} {\bibfnamefont {J.}~\bibnamefont {Russo}}, \
	and\ \bibinfo {author} {\bibfnamefont {H.}~\bibnamefont {Tanaka}},\
}\href@noop {} {\bibfield  {journal} {\bibinfo  {journal} {Physical Review
		B}\ }\textbf {\bibinfo {volume} {90}},\ \bibinfo {pages} {014204} (\bibinfo
{year} {2014})}\BibitemShut {NoStop}%
\bibitem [{\citenamefont {Wedekind}\ and\ \citenamefont
	{Reguera}(2008)}]{wedekind2008kinetic}%
\BibitemOpen
\bibfield  {author} {\bibinfo {author} {\bibfnamefont {J.}~\bibnamefont
		{Wedekind}}\ and\ \bibinfo {author} {\bibfnamefont {D.}~\bibnamefont
		{Reguera}},\ }\href@noop {} {\bibfield  {journal} {\bibinfo  {journal} {The
			Journal of Physical Chemistry B}\ }\textbf {\bibinfo {volume} {112}},\
	\bibinfo {pages} {11060} (\bibinfo {year} {2008})}\BibitemShut {NoStop}%
\bibitem [{\citenamefont {Russo}\ \emph {et~al.}(2013)\citenamefont {Russo},
	\citenamefont {Maggs}, \citenamefont {Bonn},\ and\ \citenamefont
	{Tanaka}}]{russo2013interplay}%
\BibitemOpen
\bibfield  {author} {\bibinfo {author} {\bibfnamefont {J.}~\bibnamefont
		{Russo}}, \bibinfo {author} {\bibfnamefont {A.~C.}\ \bibnamefont {Maggs}},
	\bibinfo {author} {\bibfnamefont {D.}~\bibnamefont {Bonn}}, \ and\ \bibinfo
	{author} {\bibfnamefont {H.}~\bibnamefont {Tanaka}},\ }\href@noop {}
{\bibfield  {journal} {\bibinfo  {journal} {Soft Matter}\ }\textbf {\bibinfo
		{volume} {9}},\ \bibinfo {pages} {7369} (\bibinfo {year} {2013})}\BibitemShut
{NoStop}%
\bibitem [{\citenamefont {Lechner}\ and\ \citenamefont
	{Dellago}(2008)}]{lechner2008accurate}%
\BibitemOpen
\bibfield  {author} {\bibinfo {author} {\bibfnamefont {W.}~\bibnamefont
		{Lechner}}\ and\ \bibinfo {author} {\bibfnamefont {C.}~\bibnamefont
		{Dellago}},\ }\href@noop {} {\bibfield  {journal} {\bibinfo  {journal} {The
			Journal of chemical physics}\ }\textbf {\bibinfo {volume} {129}},\ \bibinfo
	{pages} {114707} (\bibinfo {year} {2008})}\BibitemShut {NoStop}%
\bibitem [{\citenamefont {Ronceray}\ and\ \citenamefont
	{Harrowell}(2017)}]{ronceray2017suppression}%
\BibitemOpen
\bibfield  {author} {\bibinfo {author} {\bibfnamefont {P.}~\bibnamefont
		{Ronceray}}\ and\ \bibinfo {author} {\bibfnamefont {P.}~\bibnamefont
		{Harrowell}},\ }\href@noop {} {\bibfield  {journal} {\bibinfo  {journal}
		{Physical Review E}\ }\textbf {\bibinfo {volume} {96}},\ \bibinfo {pages}
	{042602} (\bibinfo {year} {2017})}\BibitemShut {NoStop}%
\bibitem [{\citenamefont {Wei}\ \emph {et~al.}(2018)\citenamefont {Wei},
	\citenamefont {Yang}, \citenamefont {Jiang}, \citenamefont {Dai},
	\citenamefont {Wang}, \citenamefont {Dyre}, \citenamefont {Douglass},\ and\
	\citenamefont {Harrowell}}]{wei2018assessing}%
\BibitemOpen
\bibfield  {author} {\bibinfo {author} {\bibfnamefont {D.}~\bibnamefont
		{Wei}}, \bibinfo {author} {\bibfnamefont {J.}~\bibnamefont {Yang}}, \bibinfo
	{author} {\bibfnamefont {M.-Q.}\ \bibnamefont {Jiang}}, \bibinfo {author}
	{\bibfnamefont {L.-H.}\ \bibnamefont {Dai}}, \bibinfo {author} {\bibfnamefont
		{Y.-J.}\ \bibnamefont {Wang}}, \bibinfo {author} {\bibfnamefont
		{J.}~\bibnamefont {Dyre}}, \bibinfo {author} {\bibfnamefont {I.}~\bibnamefont
		{Douglass}}, \ and\ \bibinfo {author} {\bibfnamefont {P.}~\bibnamefont
		{Harrowell}},\ }\href@noop {} {\bibfield  {journal} {\bibinfo  {journal}
		{arXiv preprint arXiv:1809.08589}\ } (\bibinfo {year} {2018})}\BibitemShut
{NoStop}%
\end{thebibliography}
%

\end{document}